\documentclass[12pt,preprint]{aastex}

\begin{document}

\title{A {\it Chandra} Study of the Rosette Star-Forming
Complex. II. Clusters in the Rosette Molecular Cloud}

\author{Junfeng Wang,\altaffilmark{1} Eric D.  
Feigelson,\altaffilmark{1} Leisa
K. Townsley,\altaffilmark{1} Carlos G. 
Rom\'an-Z\'u\~niga,\altaffilmark{2} Elizabeth Lada,\altaffilmark{3} 
and Gordon Garmire\altaffilmark{1}}

\altaffiltext{1}{Department of Astronomy \& Astrophysics, The
Pennsylvania State University, 525 Davey Lab, University Park, PA
16802; {\tt jwang@astro.psu.edu, edf@astro.psu.edu, townsley@astro.psu.edu}}

\altaffiltext{2}{Centro Astron\'{o}mico Hispano Alem\'{a}n, Camino Bajo 
Hu\'{e}tor 50, Granada, Spain 18008}

\altaffiltext{3}{Department of Astronomy, University of Florida, 211
Bryant Space Science Center, Gainesville, FL 32611}

\begin{abstract}

We explore here the young stellar populations in the Rosette Molecular
Cloud (RMC) region with high spatial resolution X-ray images from the
{\em Chandra X-ray Observatory}, which are effective in locating
weak-lined T Tauri stars as well as disk-bearing young stars. A total
of 395 X-ray point sources are detected, 299 of which (76\%) have an
optical or near-infrared (NIR) counterpart identified from deep
FLAMINGOS images.  From X-ray and mass sensitivity limits, we infer a
total population of $\sim 1700$ young stars in the survey
region. Based on smoothed stellar surface density maps, we investigate
the spatial distribution of the X-ray sources and define three
distinctive structures and substructures within them. Structures B and
C are associated with previously known embedded IR clusters, while
structure A is a new X-ray-identified unobscured cluster. A high mass
protostar RMCX \#89 = IRAS 06306+0437 and its associated sparse
cluster is studied. The different subregions are not coeval but do not
show a simple spatial-age pattern. Disk fractions vary between
subregions and are generally $\la 20$\% of the total stellar
population inferred from the X-ray survey. The data are consistent
with speculations that triggered star formation around the HII region
is present in the RMC, but do not support a simple sequential
triggering process through the cloud interior. While a significant
fraction of young stars are located in a distributed population
throughout the RMC region, it is not clear they originated in
clustered environments.

\end{abstract}


\section{Introduction}

In the last two decades, developments in infrared (IR) technologies,
such as $JHK$ arrays and the mid-IR {\em Spitzer Space Telescope}, 
have enabled large-scale studies of star formation processes in molecular clouds
\citep[e.g.,][]{Lada91,Lada95,Carpenter00,Evans03a,Churchwell04,Young05,Harvey06}. One
major discovery from such surveys of nearby clouds is that a large
fraction ($\sim$70\%-90\%) of stars form in embedded clusters
\citep{Lada03}.  The spatial distribution and hierarchical
structure of embedded clusters in giant molecular clouds (GMCs)
should reveal physical processes in cluster formation
\citep[e.g.,][]{Elmegreen00,Bonnell01,Lada03,Tan06,Ballesteros-Paredes07}.
It is particularly important to understand the influence of initial clusters 
on later star formation in adjacent cloud material.  The expansion of the 
HII region around massive stars in young stellar clusters can, in principle, 
sequentially trigger formation of second-generation OB
clusters, as described by \citet{Elmegreen77} \citep[see
also][]{Whitworth94,Dale07,Preibisch07}.  However, the operation
of triggering in specific cases of OB associations in GMCs is often
unclear. 

As one of the most massive Galactic GMCs, the Rosette Molecular 
Cloud (RMC) is ideal for studying the formation and possible triggering 
of embedded star clusters.  With a total extent of $\sim$150~pc 
\citep{Cox90} and  $\sim 10^5 M_{\odot}$ of molecular gas \citep{Blitz80}, 
it lies only 1.4~kpc from the Sun \citep{Hensberge00}. \citet[hereafter TFM03]{Townsley03} 
 and \citet[hereafter RL08]{RL08} review past and
present research on this well-studied and important star formation region.
Its orientation is perpendicular to the line-of-sight with a large, annular HII
region at its tip known as the Rosette Nebula. This HII region is
powered by the $\sim 2$~Myr old OB cluster NGC 2244 
\citep[hereafter Paper I] {Wang08}. The molecular gas exhibits the strongly 
clumped structure characteristic of turbulent GMCs \citep{Williams98, Heyer06}. 

\citet[henceforth, PL97]{PL97} identified seven embedded 
clusters in the dense molecular cloud cores from stellar concentrations
in near-IR (NIR) $JHK$ images which are associated with IRAS 
mid-IR sources (see Figure 1a in Paper I).  They suggested that both 
triggered and spontaneous star formation may be contributing to the 
forming clusters in this cloud.  The recent deeper FLAMINGOS NIR
imaging survey of Rosette \citep[hereafter REFL08]{RomanZuniga08} reveal four more clusters
across the complex. They find an increasing fraction of younger stars
(identified from $K$-band excesses associated with hot circumstellar disks)
with increasing distance from NGC 2244, implying a temporal sequence
of star formation across the complex.  Four new clusters and over 700
stars with dusty disks are located in mid-IR maps obtained with the
IRAC and MIPS detectors on the {\it Spitzer Space Telescope} 
\citep{Poulton08}.  Both the FLAMINGOS and $Spitzer$ studies argue
that the embedded star formation is not principally produced by
sequential triggering from the NGC~2244 OB cluster. 

Lying 2$^\circ$ from the Galactic Plane, IR imaging surveys in the
Rosette region are unavoidably contaminated by the overwhelming field
star populations.  Therefore, identification of young embedded populations
has relied on selecting stars with IR color excesses produced by dusty
circumstellar disks. Because the many diskfree pre-main sequence stars
(PMS) (weak-lined T Tauri stars, WTTS) that are cluster members cannot be identified in this way, 
statistical subtraction of the contaminating field population 
is needed to estimate the full embedded population from IR images.  
Even in very young clusters, the WTTS may outnumber stars with
prominent disks \citep[e.g.,][]{Walter88,Neuhaeuser95}.  

PMS stars both with and without disks are well-known to emit X-rays 
that can penetrate the heavy extinction of GMCs \citep{Feigelson07}.
The level of X-ray emission from young stellar
populations is elevated $10-10^4$ times above the levels of X-ray
emission from the old Galactic disk population \citep{Preibisch05}. As
a consequence, X-ray observations, particularly with the high spatial 
resolution {\it Chandra X-ray Observatory}, are highly efficient in selecting
PMS members in stellar clusters.   $Chandra$
images provide remarkably complete and unbiased PMS populations
with subarcsecond positions.  X-ray surveys thus complement IR surveys
in establishing the census of embedded PMS populations in GMCs which
in turn can address issues relating to triggered star formation.  $Chandra$
has made recent advances in the study of star formation triggering in the
vicinity of HII regions produced by the Cep OB3b, Trumpler 37 (IC 1396), 
NGC 6618 (M~17), and Trumpler 16 (Carina) OB associations
and a high latitude cluster  \citep{Getman06, Getman07, Broos07, 
Sanchawala07, Getman08a}. 

A previous X-ray study of the RMC was presented by \citet{Gregorio98}
based on an image from the {\em ROSAT} satellite.  They report faint
X-ray sources associated with embedded IR-emitting low mass star
clusters. However the poor spatial resolution and soft energy band of
{\em ROSAT} prevents identification of individual embedded stars. The
diffuse and stellar emission from the central Rosette Nebula cluster
NGC 2244 were studied in detail in TFM03 and Paper I.

We now examine a mosaic of four {\em Chandra} images in the RMC,
presented without detailed analysis in TFM03, to identify the PMS
population and study the cluster formation process in molecular
clouds, complementing the NIR studies (PL97, RL08, REFL08).  Figure~1
outlines the location of our images with respect to the molecular gas
and dust emission and the NIR clusters.  Section~\ref{sec:obs.sec}
describes the observations with \S~\ref{sec:counterpart.sec} outlining
the stellar counterparts of the X-ray sources.  Important individual
stars are discussed in \S~\ref{sec:previous}.  Characterization of the
embedded clusters, and non-clustered (distributed) population, follows
in \S~\ref{sec:cluster}. In particular we discover a previously
unidentified older cluster (\S~\ref{sec:XA}) and investigate clustered
star formation in the particular environment of the Rosette Complex
(\S~\ref{sec:distribSF}). In \S~\ref{sec:discuss} we summarize our
findings and discuss the implications for star formation modes in the
RMC.

\section{X-ray Sources in the RMC}\label{sec:obs.sec}

\subsection{{\it Chandra} Data Analysis}

The RMC was observed with the Imaging Array of {\em Chandra}'s
Advanced CCD Imaging Spectrometer (ACIS-I) in a sequence of 20~ks
exposures obtained in January 2001 (ObsIDs 1875-77). The observatory
and instrument are described by \citet{Weisskopf02}.  Details on the
data filtering, point source detection and extraction, variability
analysis, and spectral fitting procedures using the {\em ACIS Extract}
(AE) software package\footnote{The ACIS Extract software package and
  User's Guide are available online at
  \url{http://www.astro.psu.edu/xray/docs/TARA/ae\_users\_guide.html}.
}  \citep{ae} are given in Paper I.  This paper also defines the
quantities presented in the source tables below.

The resulting RMC source list contains 395 X-ray detections which we 
designate RMCX 1 through 395.  We 
divide these into a primary list of 347 highly reliable sources 
(Table~\ref{tbl:ch5_primary}) and a secondary list of 48 tentative 
sources (Table~\ref{tbl:ch5_tentative}) with $P_b \ge 0.001$ likelihood of 
being spurious background fluctuations based on Poisson statistics.  This 
approach has been adopted in our previous papers on {\em Chandra} 
observations of star forming regions \citep[e.g.][]{Townsley06, Wang07a, 
Broos07}; it allows the reader to evaluate the validity of
faint sources individually. Table~\ref{tbl:ch5_primary} and
Table~\ref{tbl:ch5_tentative} have a format that is identical to
Tables 1 and 2 in Paper I. A {\em Chandra} mosaic of the RMC
region (0.5-8 keV) overlaid with source extraction regions (calculated
with AE) is shown in Figure~\ref{fig:mosaic}.  The image here is shown
at reduced resolution (2\arcsec\ per pixel).

During the $\sim 20$ ks observations, eleven sources display
significant variability ($P_{KS} < 0.005$, shown in column 15 of
Tables~\ref{tbl:ch5_primary} and ~\ref{tbl:ch5_tentative}, where 
$P_{KS}$ is the significance of the statistic for 1-sided 
Kolmogorov-Smirnov test comparing a uniform  
count rate model to the distribution of source event time stamps).  These are  
associated with magnetic reconnection flares which are responsible for
most of PMS X-ray emission \citep{Feigelson07}.  Only one
of the variable sources (source \#119) has more than 100 net 
counts due to the short exposure; its light curve shows a slow 3-fold
rise in emission over 2 hours followed by a decay over the subsequent
3 hours.  Such slow-rise X-ray flares are not infrequently seen from PMS 
stars \citep{Getman08b}.  

Spectral analysis results for 164 sources with photometric 
significance {\em Signif} $\ge
2.0$ ({\em Signif} defined as the net counts value divided by the error on that 
value; see Table~\ref{tbl:ch5_primary} footnote) are presented in Table~\ref{tbl:ch5_thermal_spectroscopy}.   
The table entries are described in the table footnotes.  ACIS spectral
distributions were fit to thermal plasma models 
\citep[``apec;'',][]{Smith01} with interstellar soft
X-ray absorption \citep[``wabs;'',][]{Morrison83} convolved with the telescope mirror and detector  
spectral response.  Best-fit absorbing column densities range from 
negligible to $\log N_H \sim 23.5$~cm$^{-2}$, equivalent to a visual 
absorption of $A_V \sim$~200~mag \citep{Vuong03}. Temperatures 
range from $kT \sim 0.3$~keV to the hardest measurable by ACIS 
which we truncate at $kT = 15$ keV.  The range of
total band ($0.5-8$~keV) absorption corrected luminosities ($L_{t,c}$) derived
from spectral modeling is $29.6 \lesssim \log L_{t,c} \lesssim 32.4$
ergs~s$^{-1}$. Assuming a 2 keV plasma temperature and an average
$A_V=5$~mag visual extinction ($\log N_H \sim 21.9$~cm$^{-2}$), 
the faintest on-axis detection in Table~\ref{tbl:ch5_tentative} 
corresponds to a limiting luminosity of $\log L_{t,c} \sim
29.3$~ergs~s$^{-1}$.  Note however that the patchy extinction
throughout our fields will change the limiting sensitivity of the observation
somewhat (\S~\ref{sec:sensitivity}). The brightest source in the field is 
source \#119 with 369 ACIS counts in the 20~ks exposure. This count rate
is not high enough to cause photon pile-up in the detector. 

\subsection{Identification of Stellar Counterparts}\label{sec:counterpart.sec}

The X-ray sources, which are generally located with precision better than
0.5\arcsec\ in the 2MASS/Hipparcos reference frame, are associated with 
counterparts at other wavelengths to establish their stellar properties. 
Optical coverage of the RMC in the literature is poor because of the 
high visual extinction towards the molecular cloud and the contamination 
by foreground stars.  Few spectral types are measured.  We therefore use
optical and IR photometric catalogs including: Whole-Sky USNO-B1.0 Catalog
\citep[hereafter USNO]{Monet03}; 2MASS All-Sky Catalog of Point Sources
\citep[2MASS]{Cutri03}; and the University of Florida FLAMINGOS
Survey of Giant Molecular Clouds (PI: E. Lada).  The FLAMINGOS
observations of the Rosette Complex fields and IR data reduction are
described by Rom\'an-Z\'u\~niga (2006) and REFL08. We have
also associated our sources with the coordinates of $>$50,000 stars seen in 
$Spitzer$ IRAC 3.6$\mu$m images (Poulton et al.\ 2008, private 
communication).  Positional coincidence criteria are used to associate ACIS X-ray sources 
with optical and NIR (ONIR) sources as described in the Appendix of
\citet{Broos07}
\footnote{Software implementing the matching algorithm is available in
the TARA package at \url{http://www.astro.psu.edu/xray/docs/TARA/}. }.

Associations between ACIS sources and ONIR sources are reported in
Table~\ref{tbl:ch5_counterparts}.  Of the 395 ACIS sources in the RMC,
299 (76\%) have an ONIR counterpart identified.  $JHK$ magnitudes from
FLAMINGOS photometry are reported if available; we adopt 2MASS
photometry for six bright stars (RMCX~\#25, 65, 164, 185, 209, 342)
that are saturated in FLAMINGOS ($H<11$ mag, Rom\'an-Z\'u\~niga 2006).
NIR counterparts of 279 ACIS sources have measurements in all three
$JHK$ bands. The SIMBAD and VizieR catalog services were examined for
complementary information; these results are reported in the table
footnotes.  Useful sources include the TASS Mark IV Photometric Survey
of the Northern Sky \citep{Droege06} and two large optical surveys of
the Rosette \citep{MJD95,BC02} that partly covered our {\em Chandra}
fields. Source \#65 is too bright ($J=9.83$) for its spectral type
(G0) at the RMC's distance, leading to its classification as a
foreground star. The rest of the stars with known spectral types show
colors and magnitudes consistent with being intermediate- to high-mass
cloud members.

We have examined the spatial distribution of the 96 X-ray sources that
do not have associated ONIR counterparts in the context of Digital
Sky Survey optical plates and CO emission maps \citep{Heyer06}. Over 20 of these
sources are located in the brightest nebulosity of the
photodissociation region (PDR) of the Rosette Nebula, where source
detection becomes extremely difficult at ONIR wavelengths. We suspect
that most of these are cloud members.  About 50 others
are concentrated in regions where the dense molecular cores are
located and show X-rays with harder median energy ($MedE$).
Most of these are probably heavily obscured new members of the embedded
clusters, although a few may be extragalactic sources coincidentally lying
behind the embedded clusters.  

Only a small fraction of the X-ray sources, perhaps 10 of the 96
that do not have ONIR counterparts,  seem to be randomly distributed 
across the fields with a surface density consistent with the extragalactic
background of active galactic nuclei.  Extragalactic contaminants are seen 
in all $Chandra$ fields of young stellar clusters \citep[e.g.,][]
{Getman06,Wang07a,Broos07}.  Considering the contamination analysis of
these earlier studies, the absorption by the RMC gas, and the short 20~ks 
exposures considered here, we estimate a contamination of $\sim$3\% 
from extragalactic sources and $\sim 1$\% from older Galactic field 
star populations among the 395 ACIS sources, and even fewer among the
sources with ONIR counterparts listed in Table~\ref{tbl:ch5_counterparts}. 

\subsection{Limiting sensitivity and completeness}\label{sec:sensitivity}

The sensitivity limits of our X-ray observations are spatially uneven
due to the patchy absorption in the RMC and to degraded point spread
function (PSF) of the {\em Chandra} mirrors at large off-axis angles.
Based on previous experience with wavelet source detection and the
ACIS Extract software \citep{Getman06,Wang07a,Broos07}, we estimate
that the far-off-axis sensitivity is approximately 2 times worse than
on-axis (see \S~\ref{sec:new_sensitivity}). Sources with 8 or more net
counts would be detected anywhere in the three fields; over 60\% of
the sources in Table~\ref{tbl:ch5_primary} lie in this complete
regime.

The typical extinction is $A_V \sim 5-10$ in the RMC regions with
molecular concentrations and $A_V \sim 1$ in less obscured regions
(REFL08).  Assuming a $kT=2$ keV thermal plasma as the spectral model
of a typical PMS star \citep{Preibisch05}, the absorption-corrected
luminosity limit of 8 ACIS counts in 20~ks in the hard band (2--8 keV)
is $\log L_{h,c} \sim 29.6$ erg s$^{-1}$ in obscured regions and $\sim
30.0$ erg s$^{-1}$ in less obscured regions of the RMC (the observed
luminosity in the 0.5--8 keV band $\log L_{t} \sim 29.7$ erg s$^{-1}$
and $\sim 30.1$ erg s$^{-1}$, respectively).  Due to the strong
empirical correlation between $L_x$ and PMS stellar mass
\citep{Telleschi07}, these luminosity limits correspond to a mass
limit around $\sim 0.5M_{\odot}$.

Adopting the X-ray luminosity function (XLF) of the {\em Chandra} Orion 
Ultradeep Project (COUP) observation of the Orion Nebula Cluster as a 
calibrator for other PMS stellar populations, we can estimate the 
completeness of our RMC observations with respect to the full underlying 
PMS population.  Using the XLF of \citet{Feigelson05}, we find that the
RMC observations probe $\sim 20$\% ($\sim 30$\%) of the PMS population
down to the stellar limit of $M = 0.08$~M$_\odot$ in the obscured
(lightly obscured) regions of the RMC.  These fractions are adopted to 
obtain estimates of the total stellar populations in various RMC structures
below (\S~\ref{sec:cluster} and Table~\ref{tbl:cls_counts}).

\section{X-rays from Previously Studied Stars}\label{sec:previous}

\subsection{The O7 Star HD 46485} \label{sec:HD46485}

The only previously known O star in our fields, HD 46485
\citep[=BD+04$^{\circ}$1318, O7V;][]{Walborn73,Maiz-Apellaniz04} is
detected with 310 counts in the 20~ks exposure (RMCX~\#164). A
single-temperature thermal plasma model gives an acceptable fit to its
X-ray spectrum with $kT\sim 0.7$~keV and a low absorbing column
$N_{H}\sim 2.1\times 10^{21}$ cm$^{-2}$.  The fit can be improved with
a two-temperature thermal plasma model (Figure~\ref{fig:spectra}),
with a soft component ($kT_1\sim 0.3$ keV), a second harder component
($kT_2\sim 0.9$ keV), and an absorbing column of $N_H\sim 4.1\times
10^{21}$ cm$^{-2}$ ($A_V\sim 2.5$).  The soft plasma and the derived
X-ray luminosity, $\log L_{t,c}\sim 32.1$ erg s$^{-1}$, are similar to
the NGC 2244 O stars (e.g., HD 46223, HD 46202; Paper I), consistent
with being associated with instabilities in the radiatively
accelerated wind.  Note that some O stars exhibit a much harder X-ray
component; for example, $\theta^1$ Ori C (O7V) ionizing the Orion
Nebula has a component with $kT\sim 3$ keV \citep[e.g.][]{Gagne05}.
This hard component is attributed to a magnetically channeled wind
shock (MCWS). These differences in the X-ray properties of O stars of
similar spectral types may be attributed to the various magnetic field
strength in different stars; in this case, HD~46485 likely does not
have a strong field.  Measurements of magnetic properties of massive
stars \citep[e.g.,][]{Alecian08,Petit08} will be able to further
evaluate the link between X-ray emission and the presence of magnetic
fields in OB stars.

HD~46485 does not appear to have any relationship to the embedded
clusters of the RMC.  It is much less absorbed and lies outside all of
the clusters discussed here or in the infrared literature.  The
reported proper motion of HD~46485 is small \citep[$\mu_{\alpha}$ =
  -0.26 mas/yr, $\mu_{\delta}$ = -1.55 mas/yr, The HIPPARCOS
  Catalogue;][]{Perryman97}, and the radial velocity is $\sim 15$ km
s$^{-1}$ \citep{Evans67}. It might be a runaway O star from NGC~2244,
or a somewhat older member of the open clusters in the larger OB
association \citep{Perez91}.

\subsection{RMCX \#89 = IRAS 06306+0437: A new massive protostar}\label{sec:RMCX89} 

This star in PL97's cluster \#2 (PL2) exhibits unusually red colors in
the NIR, with $J-H=3.22$ and $H-K=2.41$.  These colors imply both
unusually strong reddening ($A_V\sim 30$) and a strong $K$-band
excess.  It is also a bright MSX point source (MSX6C
G206.7804-01.9395). The $Chandra$ source has only 21 net counts but
the X-ray properties are all consistent with a highly absorbed source:
the median photon energy is very hard (4.6 keV) which, based on the
Orion Nebula calibration between $MedE$ and $N_H$ \citep{Feigelson05},
corresponds to absorption $\log N_H=23.2$ cm$^{-2}$ or $A_V \sim 100$.
The observed X-ray luminosity in the hard band ($2-8$~keV) is $\log
L_h \simeq 31.2$ erg s$^{-1}$.  While not impossibly high for a lower
mass PMS star, this is at the top $\sim 1$\% of the PMS X-ray
luminosity function \citep{Feigelson05}.  Among protostars, only one
of several dozen Class~I in the Orion Nebula region flares up to a
comparable X-ray level \citep[e.g., COUP \#554 in OMC~1S,][]{Grosso05,
  Prisinzano08}.

Although the evolutionary stage of PMS stars cannot be directly established
from hard-band X-ray properties because magnetic reconnection flares 
are present throughout the Class I-III phases,  our past experience suggests 
that sources with absorption exceeding $\log N_H \sim 22.3$ cm$^{-2}$ are
protostars where most of the absorption arises in local envelopes rather than 
the larger-scale molecular cloud \citep{Getman06}.   
A protostellar classification is also indicated by the rapidly rising 
flux (high $24/8$~$\mu$m flux ratio) and roughly flat $25-100$~$\mu$m spectral
energy distribution measured with the $Spitzer$ and $IRAS$ satellites
\citep{Poulton08}.  

RMCX \#89 = IRAS 06306+0437 is also a massive system.  With a
dereddened $K$ magnitude around $K \sim 7$, it has a bolometric
luminosity associated with an early B or late O star (see the NIR
color magnitude diagram in Paper I). As O stars generally have softer
X-ray spectra than PMS stars, the correction from hard band to the
total $0.5-8$~keV band is larger; for the O6-O7 star $\theta^1$~C~Ori
in the Orion Nebula Cluster, this correction is a factor of $\sim
6$. If we assume the same correction, the estimated
absorption-corrected total band X-ray luminosity is $\log L_{t,c} \sim
32.0$ erg s$^{-1}$ for RMCX~\#89.  This source is similar to a
candidate massive YSO found in the $Chandra$ observation of the
Pismis~24 cluster in the NGC~6357 star forming complex
\citep{Wang07a}. In fact, IRAS 06306+0437 is a known water maser
source \citep{Brand94}. It was included in a CS(2-1) survey of IRAS
sources with far-IR color characteristics of ultra compact HII
regions, but not detected \citep{Bronfman96}. \citet{Wu06} observed
IRAS 06306+0437 among other water masers not associated with known HII
regions or low mass YSOs and detected NH$_{3}$(1,1) emission
coincident with the IRAS/MSX infrared peak.  They suggested the
ammonia core may be a high mass protostellar candidate.

The $IRAS$ catalog notes that the MIR source is extended.  We examine
the dusty environment of IRAS 06306+0437 in Figure~\ref{fig:source89}
which we extracted from the archived IRAC and MIPS data from the {\em
  Spitzer} satellite.  The star lies within a bright region of heated
dust surrounded by an elongated complex of infrared filaments
extending over several arcminutes which is probably associated with
the larger photodissociation region of the Rosette Nebula.  A
smaller-scale view of the dust emission around IRAS 06306+0437 is
shown by \citet{Poulton08}.  They note that this star does not have an
accompanying concentration of young Class I or II stellar companions;
i.e. that PL~2 is not a true stellar
cluster. Figure~\ref{fig:source89} shows 25 X-ray selected stars
within 1~pc of RMCX~\#89.  We have classified these as Class I, II or
III based on a preliminary examination of their NIR colors, including
two other Class~I protostars, three Class~II systems, and $\sim 20$
Class~III systems.  This is consistent with the concentration of NIR
$K$-band excess stars found in PL~2 (Figure 7 in REFL08).  Another
candidate protostar, RMCX~\#72 ($K=10.98$), is close to RMCX~\#89,
lying $\sim 40\arcsec$ to the east in a secondary dust
condensation. This further indicates the youth of this region.

We thus conclude that RMCX~\#89 is a massive proto-O star lying in a sparse
cluster of PMS stars, most of which are older Class~III systems.  The implied
age offset $-$ that the central O star is younger than the cluster of lower-mass
PMS stars $-$ is similar to that recently inferred from $Chandra$ study of the
embedded W3~Main cluster \citep{Feigelson08}.  

\subsection{Intermediate-mass X-ray Stars}\label{sec:intermediate}

\citet{Block93} list the five brightest $K$-band sources in the PL4 region. 
The two bright stars BG/IRS1 and BG/IRS2 (acronyms used in Block et 
al.\ 1993 = BG) form a close pair whose spectra 
show Br$\gamma$ and are likely Herbig Ae/Be (HAeBe) stars \citep{Hanson93}.  
RMCX~\#209 and \#207 are matched to BG/IRS2 and BG/IRS4, respectively. 
Both have $\sim$50 counts and the spectral fits yield $kT\sim 3$ keV with 
absorption $\log N_H\sim 22.2$ cm$^{-2}$ for RMCX~\#207 and $\log 
N_H\sim 22.5$ cm$^{-2}$ for RMCX~\#209. This absorption column 
($A_V\sim 15-20$) agrees with the $A_K\sim 1.5$ mag derived by
\citet{Hanson93}. The absorption corrected luminosities are $\log
L_{t,c}\sim 31.5$ erg s$^{-1}$. Their hard X-ray spectra are consistent
with the high X-ray temperatures ($\sim$20 MK) seen in X-ray surveys 
of HAeBe stars \citep{Hamaguchi05, Stelzer06b}.  RMCX~\#210 coincides 
with BG/IRS 5 but has too few counts for spectral fitting. 

The brightest X-ray source in our $Chandra$ fields is RMCX~\#119 with
369 ACIS counts.  The X-ray spectrum can be roughly fit with a
standard two-temperature thermal plasma model but there seems to be
some abundance anomaly around 1~keV. The derived X-ray luminosity is
$\log L_{t,c}\sim 31.9$ erg s$^{-1}$, too high for most lower-mass PMS
stars but consistent with the range of HAeBe systems.  The stellar
counterpart is bright, with $V = 13.9$, and was found not to have
strong H$\alpha$ emission in a follow-up study of $ROSAT$ sources in
the Rosette region \citep{BC02}.  Its location in the NIR
color-magnitude diagram indicates an intermediate mass star, although
possibly with weak or absent accretion as suggested by the weak
H$\alpha$ emission \citep{Calvet00}.

\subsection{Protostellar sources of Herbig-Haro outflows}

Five features of excited gas seen in the optical band [SII] survey 
by \citet{Ybarra04} lie in our $Chandra$ field.  These are probably 
Herbig-Haro outflows from protostars.   We have carefully
examined the X-ray fields in the vicinity of these sources.  In a region
6\arcmin\/ southwest of PL~2, the X-ray emitting star RMCX~\#51 is found
near the extended [SII] feature RMC-F.  Its IR counterpart is a 
heavily reddened star with $K=12.94$, which has no significant 
$K$-band excess.  This star may be responsible for the [SII] emission 
feature. 

RMC-H is an arc-shaped feature of excited gas outlining a globule
of gas northwest of PL~2 with a possible HH outflow originating in the
globule. Three X-ray sources are seen coincidently tracing the edge of the
globule. One of them is the Class I protostar RMCX~\#58 which may be
powering the jet-like outflow.  

\citet{Ybarra04} suggests that RMC-L is 
an HH outflow driven by a star in PL~4 associated with IRAS 06314+0427. 
In this context, the X-ray detected Class I protostar RMCX~ \#315 may be
the exciting source as it lies eastwards on the axis of symmetry of the 
bow-shock feature. The 3.5~pc separation between the star and shock is 
rather large, but comparable large-scale collimated flows have been detected 
in other star forming regions (e.g., HH80/81, $\ge 5$ pc long; Marti et al.\
1993). Other [SII] features such as RMC-E, RMC-G, and RMC-M do not 
have an obvious correlation with our X-ray sources.

\section{Stellar Clusters in the Chandra Population} \label{sec:cluster}

Visual examination of the spatial distribution of X-ray sources in Figure 
\ref{fig:mosaic} shows a non-random clustering.  This is also evident in the 
earlier presentation of the field shown in Figure~6a of TFM03.  We identify
the X-ray source concentrations, which can be considered to be `star 
clusters,' using two methods in \S \ref{sec:smooth}-\ref{sec:neighbor}.
The newly identified XA cluster is describe in \S \ref{sec:XA} and the 
previously established embedded XB and XC clusters are discussed in 
\S \ref{sec:XBXC}.  Tables~\ref{tbl:cls_comparison}-\ref{tbl:cls_counts} 
provide quantitative results.  

\subsection{Clusters from the Smoothed X-ray Source Distribution 
\label{sec:smooth}}

\citet{Wang07a} and \citet{Broos07} found that smoothed maps of the
stellar distribution derived from $Chandra$ source lists are effective
in revealing the structure of young stellar clusters.  For example, in
M~17, the X-ray stellar density maps highlighted the region's complex
structure, including the compact NGC~6618 cluster, a triggered
population along the edge of the HII region, several small embedded
clusters, and a previously unknown subcluster.  A similar approach
applied to 2MASS star distributions revealed the NGC~2237 subcluster
on the western side of the Rosette Nebula \citep{Li05}.  Compared to
infrared maps, X-ray maps are less affected by contamination from
Galactic field stars and show richer populations than disky IR-excess
samples.

Smoothed maps of the $Chandra$ sources in the RMC are shown in
Figure~\ref{fig:ch5_ssd}\footnote{Note that each star has the same
weight in these maps, so they do not accurately reflect the distribution of 
X-ray emission.  A smoothed map of the X-ray emission is shown by
TFM03.}.  The three ACIS fields have closely matched 20~ks
exposures, only a small sensitivity bias is present between the on-axis and
off-axis areas of each field, and contamination by extraneous sources
is small (\S \ref{sec:counterpart.sec}).  The three rows of 
panels in Figure~\ref{fig:ch5_ssd} show the total source sample, lightly obscured sources with 
$MedE\le 2.0$ keV, and heavily obscured sources with $MedE >
2.0$ keV.  The left and right panels show maps with 3\arcmin\/ and
2\arcmin\/ radius (1.2 and 0.8~pc) smoothing kernels, respectively. 

Three large-scale structures are seen as overdensities in the total sample,
irrespective of the adopted smoothing kernel size. We define these 
regions as XA, XB, and XC and trace their outline from a contour of the 
3\arcmin\/ kernel.  We note seven small-scale overdensities within the
three major structures in the 2\arcmin\/ map of the total population.  We
label these XA1-XA3, XB1-XB3, and XC1.  In addition, a distinct substructure
of heavily obscured stars extending southward from XC1 is seen in panel
Figure~\ref{fig:ch5_ssd}$f$ which we call XC2.  Note that the substructures 
are defined from smoothed contours on different maps and are not
contiguous; the star totals in the main structures are thus larger than the
sum of the stars in the substructures.  We label stars lying outside all
overdensities ``distributed'' stars in the RMC.

While the regions are defined based on subjectively chosen surface
density contours, we are confident that these structures are not
artifacts caused by the slightly degraded X-ray sensitivities at large
off-axis angles on the ACIS-I CCDs. The overdensities are not located
at the centers of the fields (where the telescope response is
maximized) and represent a factor of $2-20$ difference in stellar
density which cannot be solely attributed to the instrumental
sensitivity gradients.

\subsection{Clusters from the Nearest Neighbors Method \label{sec:neighbor}}

A second technique for identifying stellar clusters is analysis of the
distribution of nearest neighbors \citep{Casertano85,Diggle03}.  This
was used by REFL08 to establish the content of the seven PL97 clusters
in the $JHK$ catalog of the RMC. We apply the same nearest neighbor
method used by REFL08 based on the local density derived from the
distances to the 10 nearest neighbors to each star.

Contours of the resulting nearest neighbor maps for our {\em Chandra}
RMC observations are shown in Figure~\ref{fig:chandra_contour} for the
total, unobscured, and obscured populations. The locations and
morphologies of the resulting concentrations closely resemble the
structures that we defined using the smoothed surface density maps for
X-ray sources.  This indicates the robustness of the inferred
clustering irrespective of the methodology used to smooth the star
distribution.

\subsection{Evaluating the Sensitivity Across the Mosaic}\label{sec:new_sensitivity}

The maps shown in Figure~\ref{fig:ch5_ssd} used all detected X-ray
sources. However, the sensitivity across the field is not uniform;
there is a bias towards detecting faint sources in the area close to
the individual field aim points and in the overlapping regions with
deeper exposure.  It is possible that this sensitivity variations can
create spurious structure in an intrinsically smooth distribution of
sources. We evaluate the importance of this effect by creating a
smoothed source map similar to Figure~\ref{fig:ch5_ssd} which
removes the sensitivity degradation due to telescope vignetting,
effective exposure and background across the fields.

Following \citet{Lehmer05} and \citet{Luo08}, we estimate the
sensitivity across the X-ray imaging mosaic assuming a Poisson
model. The resulting relation can be approximately represented by
$\log N_{lim} = 0.558+0.322(\log b) +0.13(\log b)^2+0.037(\log b)^3$,
where $N_{lim}$ is the required number of counts for a source required
to be detected, and $b$ is the background counts measured in a local
region. The coefficients are obtained by constructing a lower envelope
in the plot of {\it Net Full} counts against {\it Bkgd Full} counts
tabulated in Table~\ref{tbl:ch5_primary}.  We then construct a
sensitivity map across the mosaic by dividing a smoothed map of
$N_{lim}$ by the telescope exposure maps in units of cm$^{2}$ s (shown
in Figure~\ref{fig:expmap}, see also Appendix of TFM03).  The
resulting sensitivity map is typically $6 \times 10^{-7}$ counts
cm$^{-2}$ s$^{-1}$ near the aimpoint and $1 \times 10^{-6}$ near the
edge of a typical 20~ks field. It reaches $2 \times 10^{-7}$ in a
small region around subcluster XA1 where the field overlaps the much
deeper $\sim$100~ks exposure of NGC~2244 discussed in Paper~I.

This corresponds to a $\sim$8~counts limit across the 20~ks exposures
and $\sim$40~counts in the overlapping region around XA1.  Applying
this spatially varying counts threshold removes the sensitivity bias
from telescope degradation and overlapping fields, and provides a
complete sub-sample of our X-ray detections.  This procedure removes
30\% of the weaker sources in Table~1 (the remaining sources with
number of counts above the detection threshold are marked in Table~1,
see table footnote). The smoothed stellar surface density map of this
reduced, spatially uniform sample is shown in
Figure~\ref{fig:uniform_contour}.

The three major X-ray clusters, XA, XB, and XC, are present as
prominent overdensities. Some substructures are less significant than
in the full sample map; XA2 and XB1 are weakest with densities $\sim
2$ times above the background level. Additional possible substructures
emerge: an enhancement at ($6^h 33^m 02^s$, $4\arcdeg 47\arcmin$) east
of XA3 and at ($6^h34^m05^s$, $4\arcdeg 33\arcmin$) northwest of
XC1. The reality of these features must be confirmed with deeper X-ray
observations.

\subsection{Identification of the new XA cluster \label{sec:XA}}

The three main stellar enhancements and seven secondary enhancements
are listed in Table \ref{tbl:cls_comparison} with approximate sizes.
Most are around 3\arcmin\/ or 1.2~pc in extent.  Four of the seven
substructures $-$ XB2, XB3, XC1, and XC2 $-$ are associated with
clusters identified from $JHK$ and $Spitzer$ images (PL97, RL08,
Poulton et al. 2008).  Figure~\ref{fig:co_ir} shows the outline of the
three main structures on the molecular CO map of \citet{Heyer06}.  It
shows that these same four substructures--XB2, XB3, XC1, and XC2--are
associated with dense molecular regions.

X-ray cluster XA, on the other hand, lies in the evacuated region on
the eastern side of the H$\alpha$-bright Rosette PDR. Located in the
north-west side of the north-westernmost ACIS field where molecular
material has been removed by the expanding HII region, its stars are
less obscured compared to the XB and XC concentrations embedded in the
molecular cloud.  This is shown in columns $2-4$ of
Table~\ref{tbl:cls_counts}: 30\% of XA stars have $MedE > 2$ keV,
corresponding to absorption $\log N_H \ga 22.0$ cm$^{-2}$ or $A_V \ga
6$, compared to $41-46$\% in concentrations XB and XC.

Most XA stars are concentrated in subcluster XA3 centered at
($\alpha$, $\delta$) = (06:32:38,+04:46:18).  XA1 and XA2 represent
small extensions to the northwest and southwest,
respectively. Sixty-one stars are in the XA X-ray sample.  Scaling by
the incompleteness factor for lightly obscured regions given in
\S~\ref{sec:sensitivity}, the estimated total stellar population in XA
is 300 stars.  This is about 15\% of the total population estimated
for the central NGC~2244 cluster using the same XLF-based method
($\sim$2000; Paper I).

The lightly absorbed XA cluster was not clearly identified in previous
IR surveys of the cloud.  The location of substructure XA1 is
separated by $\sim 20$\arcsec\ from IRAS 06297+0453. However a close
inspection of IRAS, DSS, and 2MASS images led us to conclude that this
is a spatial coincidence: IRAS 06297+0453 is associated with the very
bright foreground K0V star BD +04 1304 ($V=5.8$, $K=3.69$) and does
not belong to the RMC. The XA cluster was not covered in the earlier
PL97 survey. Six stars with $K$-band excesses are present in the
FLAMINGOS survey (Table~ \ref{tbl:cls_counts}), insufficient to
highlight a cluster using the 10th-nearest neighbor method. However,
it does appear as an enhancement in the smoothed surface density map
of all 2MASS sources shown by \citet[][Figures 1a and 2a]{Li05},
appearing as a stellar enhancement between XB and the main NGC~2244
cluster.  A group of 32 stars with disky MIR spectral energy
distributions is found within our XA contour in the $Spitzer$
IRAC/MIPS, but again these were not identified as a distinct cluster
by \citet[][see their Figures 6b]{Poulton08}.  Rather they appear as
an eastward extension of the central NGC 2244 cluster. The cluster is
more spatially distinct from the main cluster in the ACIS X-ray mosaic
of $\sim$20ks pointings towards NGC 2244 and RMC (see Figure 6a in
TFM03), which is not sensitive to obscuration.

The reason that the XA cluster emerges more clearly in our X-ray
survey than in infrared surveys is probably due to presence of nebular
contamination.  It lies on a bright ridge of enhanced H
Brackett-$\alpha$ and spatially variable PAH emission \citep[Figures 1
  and 4 of][]{Poulton08}.  These contaminants limit the detection or
photometric accuracy of stars in certain infrared bands which then
inhibits identification of disk emission from multiband spectral
energy distributions.

\subsection{The Embedded Clusters XB and XC \label{sec:XBXC}}

As noted above and in Table~\ref{tbl:cls_comparison}, the XB and XC
concentrations of X-ray stars 
shown in Figures~\ref{fig:ch5_ssd} and ~\ref{fig:chandra_contour} correspond to 
embedded clusters established from $IRAS$, 2MASS, FLAMINGOS,  
and $Spitzer$ studies.  Their stellar populations inferred from NIR 
studies have been most thoroughly described by RL08.   The XB and
XC clusters have higher fractions of obscured stars than the XA
cluster (46\% and 41\% $vs.$ 30\%; Table~\ref{tbl:cls_counts}).  This,
combined with their spatial relationship to dense molecular structures
(Figure~\ref{fig:co_ir}) confirms that they are at least partly embedded
in cloud material.  

The main advantage of our X-ray survey is that disk-free (PMS Class~III) 
members are readily detected, subject only to the sensitivity limitations
given in \S~\ref{sec:sensitivity}.  In most infrared surveys, Class~III
systems cannot be readily distinguished from the strong contamination
of Galactic field stars.  Infrared surveys are thus often limited to counting
disky (PMS Class~I and II) systems.  The X-ray samples are thus
usually larger than the IR-excess samples, and provide a less biased
way to estimate disk fractions which is relevant to stellar ages.  

The XB cluster is the sparsest of the three X-ray concentrations with
52 detected stars and an estimated total population of 200 stars
(Table~\ref{tbl:cls_counts}).  It has an ``L''-shaped structure with
three subcomponents of similar ($\sim 1$~pc) sizes.  Subcluster
XB2 associated with $Spitzer$ cluster C \citep{Poulton08}
appears more embedded than the other subclusters. Subcluster XB3 
is associated with NIR cluster PL~2 and MIR source IRAS 06306+0437.  
As discussed in \S~\ref{sec:RMCX89} above, this structure is 
dominated by the embedded massive star, RMCX~\#89.  Both XB1 
and XB2 contain bright MSX mid-IR sources (see footnotes to 
Table~\ref{tbl:ch5_counterparts}). 

The XC cluster is the richest of the three main X-ray structures in the
cloud with 160 detected stars.  In the principal concentration of stars,
XC1, roughly 2/3 are lightly obscured and 1/3 are heavily obscured.  
This rich cluster was first noted in $K^{\prime}$-band imaging by 
\citet{Block93}, studied spectroscopically in \citet{Hanson93},
and noted as a NIR cluster by PL97, \citet{Bica03}, and \citet{Li05b}. 
The sparser XC2 component is only evident in the stellar surface
density maps for the obscured population (Figure~
\ref{fig:ch5_ssd}$e$ and $f$) indicating it is deeply embedded. This 
region also appears much darker in the DSS optical image. It appears 
as a curved southern extension of the rich XC1 cluster in the X-ray 
maps, and is associated with the NIR clusters PL~5, REFL~8, and
the $Spitzer$ cluster E (Table~\ref{tbl:cls_comparison}).  The stellar
structure may extend beyond the southeastern edge of our ACIS field.
In contrast to XC1, only 1/4 of the stars in XC2 are lightly obscured as
measured by the $MedE$ indicator of soft X-ray absorption.  XC2 is thus
the most deeply embedded stellar cluster in the RMC regions we 
examine here. 

\subsection{Evolutionary Stages of the Principal Clusters} \label{sec:clusevol}

X-ray properties cannot discriminate very young clusters, dominated
by accreting Class I-II systems, from older clusters, dominated by
non-accreting Class III systems.   The reason for this is well-established:
the X-ray luminosity function and flaring properties do not greatly change
as low mass stars evolve from their Class I through III phases
\citep[e.g.][]{Preibisch05, Telleschi07, Stelzer07}.   X-ray selected
samples are thus relatively unbiased with respect to age, and their ages
can then be estimated from their rapidly evolving infrared properties.  Here
we focus on the high-quality $JHK$ photometry available from the
FLAMINGOS  survey (REFL08) for X-ray sources within the boundaries of
clusters XA, XB, and XC presented above.

Figure~\ref{fig:ch5_ccd} shows the $J-H$ vs.\ $H-K$ color-color
diagrams for the three clusters. The region between the left two dashed lines
is associated with Class~III objects (diskless WTTS) reddened
by interstellar extinction. To the right of this band are
sources exhibiting significant $K$-band excess; here we require $E
(H-K)> 2\sigma (H-K)$ based on FLAMINGOS photometric errors for the 
sources to be K-band excess stars. The
region between the middle and right-most dashed lines is occupied by
Class~II objects (PMS stars with circumstellar accretion disks), and
protostars still possessing thick envelopes (Class~I objects) lie beyond
the right-most dashed line.  Star counts in various regions are summarized
in the middle columns of Table~\ref{tbl:cls_counts}.

The colors of most sources in cluster XA are consistent with
PMS stars with small reddening, $A_V\sim 1$, typical of stars in the
central NGC~2244 cluster illuminating the Rosette Nebula (Paper~I).
A few Class II sources are present showing higher $A_V$; these may
lie within XA or behind XA in the molecular cloud.

Sources in cluster XB show a wide range of absorptions; some sources
are reddened with $A_V>5$ and the $K$-excess sources are among the
most reddened.  This may suggest the XB subclusters are distributed
along the line of sight, at different cloud depths. Three Class~I
sources are found in this region.  Despite the smaller number of
sources in region XB, it has more IR-excess sources than the more
populous XA region.  Subcluster XB1 has 11 detected X-ray stars with
an inferred total population around 40 stars, but has no IR-excess
systems.  XB2 has more MIR-excess than NIR-excess stars, while XB3 has
more NIR-excess than MIR-excess stars. We thus tentatively infer that
the XB region is not homogeneous and coeval.  Subcluster XB3, which
includes RMCX~\#89 = IRAS 06306+0437 discussed in \S~\ref{sec:RMCX89},
is clearly the youngest structure in the region as the presence of NIR
excess in the stars indicates inner disks have not been cleared
yet. 

X-ray sources in cluster XC are mostly lightly obscured PMS stars with
$A_V\sim 2$ and a dozen heavily obscured sources with $A_V> 10$.  The
XC cluster is very populous in the X-ray map, and only 5-10\% of these
stars have infrared excesses in the NIR or MIR bands.  This, together
with the presence of only one very young system in
Figure~\ref{fig:co_ir}, indicates that the region is not dominated by
active star formation today despite the presence of two $IRAS$ sources
resolved into several IR clusters (PL~4, PL~5, REFL~8 and $Spitzer$'s
E).  The very faint XC source RMCX~\#315 is associated with a faint
NIR source with unusual colors, $J-K > 3.8$; it is either a Class~I
system or a very heavily reddened Class~II system.  NIR $J$ vs.\ $J-H$
color-magnitude diagrams shown in Figure~\ref{fig:ch5_cmd} provide an
approximate range of the mass distribution and local absorptions for
the same stars shown in Figure~\ref{fig:ch5_ccd}.

No candidate new OB stars appear in the RMC NIR color-magnitude
diagram. This stands in marked contrast to $Chandra$ studies of other
prominent star forming regions where dozens of new obscured OB stars
were identified by virtue of their X-ray emission.  These include
Westerlund~1 \citep{Skinner06}, and RCW~38 \citep{Wolk06}, NGC~6357
\citep{Wang07a}, RCW~108 \citep{Wolk08}.  There is no reason to
believe that massive star formation is suppressed in the RMC; more
likely the population of the individual embedded cluster is simply too
small to contain massive stars as expected from the initial mass
function.

We conclude that cluster XA, showing less absorption and very few
IR-excess stars, appears older than clusters XB and XC.  The latter
structures are not unified, homogeneous clusters but rather
collections of smaller stellar concentrations.  Similar to the results
of REFL08, we find no clear spatial pattern of cluster ages in this
region.  XC is the most populous and, with only $5-10$\% fraction of
IR-excess sources and no more than one protostar, is not very young.
Active star formation is strongest in subcluster XB2 with several
protostars including the massive system RMCX~\#89 = IRAS 06306+0437.

\subsection{Relationship to the Molecular Gas Distribution \label{sec:moldist}}

Figure~\ref{fig:co_ir} shows the X-ray selected RMC stars, classified
by evolutionary stage from the NIR color-color diagram
(\S~\ref{sec:clusevol}), superposed on contours representing $^{12}$CO
emission \citep{Heyer06}.  The younger Class I and II sources are
mostly seen around the CO molecular ridge near cluster XB and the
molecular material near cluster XC.  Class III stars appear more
dispersed throughout the cloud and dominate the XA region where
molecular material is no longer present.  We can interpret this trend
as reflecting the drifting of older Class III stars away from their
birth sites \citep{Feigelson96}.  For example, with a characteristic
transverse speed of $\sim$1 km s$^{-1}$, a star would travel 1 pc
($\sim 2.3$\arcmin\/) in 1 Myr.  This is sufficient to disperse such
stars originally formed in the clusters throughout the $Chandra$
fields in a few Myrs, resulting in a ``dispersed'' population.

The relationship of cluster XA to molecular gas is clearly different
than for clusters XB and XC.  XA lies within the bright nebulosity of
the Rosette HII region where most of the molecular materials are
gone. This cluster of less embedded and older PMS stars (see
\S~\ref{sec:clusevol}) is likely an early population triggered by the
expansion of the HII region.  Its similarity to NGC~2237, a similar
lightly obscured cluster on the western side of the Rosette Nebula,
will be discussed in our Paper III (Wang et al., in preparation).  The
overdensity of stars in XB, including the NIR PL~2 cluster, is
situated on the `rim' of molecular material adjacent to the expanding
Rosette PDR.  It includes XB2, the youngest subcluster in our examined
portion of the RMC.  This cluster, and possibly the similar PL~1
cluster located south-west of our $Chandra$ fields in a molecular
clump, are reasonably interpreted as triggered by the expanding shell
of the NGC~2244 HII region.  The molecular clumps are probably
undergoing photoevaporation and will eventually become unobscured
clusters like region XA.

Region XC contains the large and most extended population of embedded
stars, including several infrared clusters.  Its relationship to a
dense molecular clump has been noted by \citet{Heyer06} and REFL08.
Our analysis above (\S~\ref{sec:clusevol}) suggests that subcluster
XC1 (PL~4) is partially emerging from the molecular core while XC2 is
younger and still embedded.  As discussed by REFL08 and
\citet{Poulton08}, it is not clear that star formation here has been
triggered by the expanding HII region.

Since the X-ray sample provides an estimate of the total stellar
population of each cluster, we can roughly estimate the star formation
efficiency (SFE) of molecular material in the RMC.  \citet{Williams95}
estimate the molecular masses of 70 clumps in the RMC using an
empirical CO-to-H$_{2}$ ratio $X_{RMC}=1.1\times 10^{{20}}$ $H_{2}$
molecules cm$^{-2}$ (K km s$^{-1}$)$^{-1}$.  Mass estimates are made
assuming either local thermodynamic equilibrium or gravitational
binding of gas motions within clumps.  If we assume an average stellar
mass of 0.4~M$_\odot$ \citep{Kroupa02}, the SFE
$M_{star}/M_{star+gas}$ is approximately 15\% in region XB and 11\% in
region XC, adopting the LTE clump masses. If the gravitational clump
masses in \citet{Williams95} are used, the SFE becomes 24\% and 32\%
in XB and XC, respectively. These values are similar to the SFEs
inferred for nearby embedded rich clusters: NGC~2071 (SFE$\sim $12\%)
and NGC~2024 (33\%) in the Orion Molecular Cloud \citep{Lada91},
$\rho$ Oph \citep[9\%;][]{Wilking83}, and Mon~R2
\citep[25\%;][]{Wolf90}. As noted by RL08, the XC ($\sim$PL~4) cluster
lies at the high end of SFEs; this may indicate that on-going star
formation has significantly depleted gas in the region.

\subsection{Clustered $vs.$ Distributed Star Formation \label{sec:distribSF}}

Infrared observations have established that $70-90$\% of stars in 
GMCs form in embedded clusters \citep{Lada03}. 
Unclustered PMS stars have been established mainly in the
Orion Molecular Cloud complex, but even there it has been difficult
to quantify their population and star formation processes
\citep{Strom93, Allen96, Carpenter00}.  \citet{Allen07} present new 
survey results from {\em Spitzer} maps of nearby star forming clouds, 
and many show a significant distributed population. Distributed star
formation in more distant clouds like the RMC is largely unstudied
due to the heavy contamination from Galactic field stars outside of
dense clusters.  Modern astrophysical models of star formation in 
GMCs suggests that individual stars may 
commonly form asynchronously in supersonically turbulent cloud
clumps \citep{MacLow04}.

Because our X-ray identified young stellar population suffers little
contamination, we are able to examine the fraction of stars that are
formed both in clusters and at low stellar densities in the RMC.  The
total number of sources in the three fields is 553, of which 303 lie
within the contours we use to define clusters XA, XB, and XC
(Table~\ref{tbl:cls_counts}). Among the 250 stars distributed outside
the clusters, at most 10\% of the total detections could be Galactic
and extragalactic contaminants given the short exposure time and the
presence of the molecular cloud (\S~\ref{sec:counterpart.sec}).  This
implies that $\sim$35\% of stars could be formed in a distributed
fashion throughout the RMC region, consistent with estimates from
infrared studies \citep[e.g.][]{Carpenter00, Lada03}.  After
correcting the sensitivity variation across the fields (\S~\ref{sec:new_sensitivity}), the fraction
of distributed stars is $\sim$45\%, which implies that clustered star
formation would then be the governing mechanism in the RMC but
distributed star formation is rather important.

However, further interpretation of this fraction should proceed with
caution. There are at least three sources of complexity here. First,
our definition of cluster boundaries based on smoothed stellar density
maps (Figure~\ref{fig:ch5_ssd}) is arbitrary, which are about 0.8
stars arcmin$^{-2}$ for the X-ray-sampled stellar density. In
comparison, in REFL08 the cluster boundaries are defined by 0.2 stars
arcmin$^{-2}$ contour levels in the nearest neighbor density map of
IR-excess sources. They find that 60\% stars in the complex are in
clusters. The clustering fraction reaches 86\% if only the stars
within the molecular cloud are considered.  Besides the differences in
the field coverage of the X-ray and IR surveys (see also
\S~\ref{sec:discuss_XIR}), we could get different distributed
fractions with different, reasonable choices of cluster boundaries.

Second, a significant number of lower mass PMS stars ($M \la
0.5$~M$_\odot$) is missing from our survey due to short X-ray
exposures (\S~\ref{sec:sensitivity}).  While there may be no
preference for lower-mass stars to be clustered or distributed, it is
possible that higher-mass stars preferentially form in denser clusters
from primordial mass segregation \citep[e.g.,][]{Bonnell98}. 

Third, the distributed stars may have been born in a clustered
environment but have moved into a wider region.  Numerical simulations
show that the formation process of a cluster of stars is highly
dynamic and chaotic; stellar encounters violently eject some of the
young stars from multiple systems \citep{Bate03}.  Sparse clusters in
particular can lose $\approx 50\%$ of their stars over a few Myrs.
Thus the observed distributed populations may not have formed {\it in
  situ}.

\section{Discussion}\label{sec:discuss}

\subsection{Comparing X-ray and infrared surveys}\label{sec:discuss_XIR}

An underlying theme to this and similar studies concerns the
relationships between X-ray and infrared surveys of star forming
molecular clouds to uncover young stellar populations.  Each as
advantages and disadvantages.  With instruments like FLAMINGOS, NIR
surveys can cover large cloud areas to faint magnitudes.  These
catalogs can include virtually the entire cloud population at low to
moderate extinctions ($A_V \la 20$), but often suffer bad (often
factors $> 10$) contamination by foreground and background field
stars.  Used by themselves, PMS samples are restricted to stars with
NIR-bright inner disks, usually Class~II and lightly obscured Class~I
systems.  The resulting NIR-excess samples can constitute only a small
fraction ($\la 10$\%) of the total stellar populations which are often
dominated by systems with cool or absent disks.  With space-based
instruments like $Spitzer$'s IRAC, MIR surveys also cover large cloud
areas with very high sensitivity and penetration to $A_V \sim 100$.
As with NIR studies, MIR observations are dominated by field star
populations and samples are restricted to MIR-excess systems.  The
NIR-excess stars are recovered and new subpopulations emerge: embedded
Class~I and some Class~0 systems, low luminosity PMS brown dwarf
systems, and a few older systems with transitional MIR-excess disks
\citep[e.g.,][]{Bouwman06,Furlan07}.  Diskfree Class~III members are
catalogued but cannot be distinguished from contaminants.

X-ray surveys have very different characteristics because they trace
magnetic reconnection flaring rather than stellar and disk bolometric
emission.  Older Galactic field stars with much lower flaring levels
are nearly absent, and most extragalactic contaminants are readily
removed by the absence of stellar counterparts.  X-rays can penetrate
$A_V > 100$ but with reduced sensitivity.  X-ray studies also give
estimates of the total stellar populations, and measurements of
absorption for each star individually.  While some Class~I and II
systems are recovered, the main advantage of X-ray observations is
that disk-free Class~III PMS members are readily detected, subject to
the sensitivity limitations which correlate with stellar mass.  Since
many clusters are dominated by Class~III stars, the X-ray populations
can considerably exceed the NIR- or MIR-excess populations.  A
significant drawback is areal coverage: the $XMM-Newton$ telescope has
a $\sim 30$\arcmin\/ field but suffers confusion in crowded fields,
while the {\it Chandra X-ray Observatory} has a 17\arcmin\/ field of
view with subarcsecond resolution.  This limitation can be somewhat
mitigated with mosaics as described here.  The combination of X-ray
and infrared studies are more effective at uncovering PMS populations
in molecular clouds than any one method.  We find here that the X-ray
samples are quite free from contamination and can be 10 times richer
than the NIR- or MIR-excess samples (Table~\ref{tbl:cls_counts}).
This greatly assists in tracing the spatial structure of star
formation in the cloud.  Subsequent infrared photometry then gives the
evolutionary classification, absorption and approximate mass of each
X-ray-selected star without bias towards luminous disks
(Figures~\ref{fig:ch5_ccd}-\ref{fig:ch5_cmd}).

\subsection{RMC Young Stellar Clusters}

There is good agreement between our X-ray survey and REFL08 NIR survey
(and other surveys). Taking this combined X-ray/infrared approach, we
emerge with a number of results on the PMS population in the portion
of the RMC examined with $Chandra$.  First, we confirm past findings
that star formation is not spatially uniform in the cloud, and we
estimate that about 1/3 of the stars lie in unclustered environments
(\S~\ref{sec:distribSF}).  This supports a large body of
infrared-based survey work giving similar results in other clouds
\citep{Lada03}.

Second, we estimate the total population in the examined portion of
the cloud to be about 1700 stars down to the substellar limit
(Table~\ref{tbl:cls_counts}).  This estimate is based on an
extrapolation of the empirical PMS $L_x$-mass correlation
\citep{Telleschi07} and knowledge of our X-ray sensitivity limits
(\S~\ref{sec:sensitivity}).  This is similar to the total population
of $\sim 2000$ stars found for the central NGC~2244 cluster which
illuminates the Rosette Nebula HII region (Paper I).

Third, in contrast to the single concentrated OB cluster at the center
of the Rosette Nebula, the RMC population is divided into three
smaller clusters which themselves can be divided into sparse
subclusters.  Portions of X-ray clusters XB and XC are associated with
well-established infrared clusters PL~2 and PL~4 (\S~\ref{sec:XBXC}).
The X-ray stellar distributions and absorptions provide more detail on
these structures.  XC is richest with a total population around 800
stars; subcluster XC2 is more deeply embedded with younger stars than
XC1.  Similarly in cluster XB, group XB2 has more protostars than XB1.
These structures are thus inhomogeneous and not coeval.  We caution
that, our current identifications of structural components are from
Figure~\ref{fig:ch5_ssd} (and Figure~\ref{fig:chandra_contour}) rather
than Figure~\ref{fig:uniform_contour}, as the addition of more sources
is critically needed for substructure identification.  While some
substructures were cross-identified in the infrared (see
Table~\ref{tbl:cls_comparison}), the weak substructures certainly
require further confirmation regarding the details of their locations
and sizes.

Subcluster XB2 = PL~2 is sparse but contains the highest-mass star in
our examined region of the RMC, the heavily obscured, infrared-excess
O9/B0 star RMCX~\#89 = IRAS 06306+0437 (\S~\ref{sec:RMCX89}).  There
is only one lower-mass protostar, RMCX~\#72, lies at a projected
distance of 0.3~pc to RMCX~\#89.  The absence of a substantial cluster
surrounding this massive protostar has interesting implications for
star formation at the edge of HII regions where triggering may be
active.  The system cannot realistically be dynamically ejected from a
more crowded region, as it is still embedded in its dense dusty
envelope.  One possibility is that the massive star formed in
isolation.  We recall that HD~46223, one of the two O4 stars in the
unobscured NGC~2244 cluster lies in a puzzling location near the
southeast boundary of the nebula and, based on our $Chandra$ study,
has no lower mass companions (Paper I).  IRAS 06306+0437 may be like
HD~46223 seen at an earlier phase.  Another possibility is that the
massive star has formed before (except for RMCX~\#72) an accompanying
cluster of lower mass stars.  If this scenario applies, the star
formation sequence is opposite to the one \citet{Feigelson08} inferred
for the embedded W3~Main clusters, where the OB stars appeared to have
formed after, not before, an extended cluster of lower-mass PMS stars.

While it is reasonable that the density of obscured X-ray sources
decreases significantly towards the HII region ionized by the NGC 2244
cluster, it was surprising that the new lightly obscured cluster XA
appeared in the X-ray survey in a region with no molecular material
(\S~\ref{sec:XA}). Infrared surveys likely missed this structure due
to the paucity of infrared-excess stars and confusion by the bright
PAH and atomic nebular emission, a stellar enhancement associated with
XA can be discerned in 2MASS, FLAMINGOS, and $Spitzer$ stellar
distributions.  Its location and low fraction of IR-excess sources
suggest that it may have formed through an early epoch of triggered
star formation produced by the expanding Rosette Nebula HII region.
It appears similar to the small NGC 2237 cluster on the western side
of the Rosette Nebula \citep[REFL08]{Li05}.  A detailed $Chandra$
study of NGC~2237 is presented in Paper III.

\subsection{Star Formation Modes in the RMC}

While Paper III will present a more comprehensive discussion of star
formation in the Rosette complex, some immediate inferences can be
made concerning the RMC structures presented here.  NGC~2244 is the
youngest OB cluster among the sub-associations \citep{Blaauw64}.
\citet{Cox90} argue that sequential star formation extends into the
RMC following the compression and triggering process described by
\citet{Elmegreen77}.  However, neither REFL08, \citet{Poulton08} nor
the present study support this model in the XB and XC region of the
RMC.  No clear spatial sequence of stellar ages is seen.  It is
plausible, however, that cluster XA was triggered by the expanding
Rosette HII region.  At present it lies in (or, at least, projected
upon) the ionized region rather than in the molecular cloud.  It has
the smallest fraction of infrared-excess stars of the clusters
examined here, and hence was likely the earliest formed.

We thank Bruce Elmegreen and Zhiyun Li for helpful discussions on
cluster structure and formation. We thank Travis Rector and Mark Heyer
for kindly providing the KPNO MOSAIC images of the Rosette Nebula and
the CO emission maps of the Rosette Complex, respectively. J.W. thanks
C. Poulton for providing the full source catalog of Spitzer/IRAC
3.6$\mu$m detections. This work was supported by {\it Chandra X-ray
  Observatory} grants GO1-2008X and GO3-4010X awarded to
L.K.T. FLAMINGOS was designed and constructed by the IR
instrumentation group (PI: R. Elston) at the University of Florida,
Department of Astronomy with support from NSF grant AST97-31180 and
Kitt Peak National Observatory.  The data were collected under the
NOAO Survey Program, ``Towards a Complete Near-Infrared Spectroscopic
Survey of Giant Molecular Clouds'' (PI: E. Lada) and supported by NSF
grants AST97-3367 and AST02-02976 to the University of
Florida. E.A.L. also acknowledges support from NASA LTSA NNG05D66G.
This publication makes use of data products from the Two Micron All
Sky Survey, which is a joint project of the University of
Massachusetts and the Infrared Processing and Analysis
Center/California Institute of Technology, funded by NASA and the
National Science Foundation.  This research has made use of the SIMBAD
database and the VizieR catalogue access tool, operated at CDS,
Strasbourg, France.


\begin{thebibliography}{93}

\bibitem[{{Alecian} {et~al.}(2008){Alecian}, {Wade}, {Catala}, {Bagnulo},
  {Boehm}, {Bohlender}, {Bouret}, {Donati}, {Folsom}, {Grunhut}, \&
  {Landstreet}}]{Alecian08}
{Alecian}, E., {Wade}, G.~A., {Catala}, C., {Bagnulo}, S., {Boehm}, T.,
  {Bohlender}, D., {Bouret}, J.-C., {Donati}, J.-F., {Folsom}, C.~P.,
  {Grunhut}, J., \& {Landstreet}, J.~D. 2008, \aap, 481, L99

\bibitem[{{Allen}(1996)}]{Allen96}
{Allen}, L. 1996, PhD thesis, , University of Massachusetts, Amherst, MA, USA

\bibitem[Allen et al.(2007)]{Allen07} Allen, L., et al.\ 2007, 
Protostars and Planets V, ed. B.~{Reipurth}, D.~{Jewitt}, \& K.~{Keil}, 361 

\bibitem[{{Ballesteros-Paredes} {et~al.}(2007){Ballesteros-Paredes}, {Klessen},
  {Mac Low}, \& {Vazquez-Semadeni}}]{Ballesteros-Paredes07}
{Ballesteros-Paredes}, J., {Klessen}, R.~S., {Mac Low}, M.-M., \&
  {Vazquez-Semadeni}, E. 2007, Protostars and Planets V, ed. 
  B.~{Reipurth}, D.~{Jewitt}, \& K.~{Keil}, 63

\bibitem[{{Bate} {et~al.}(2003){Bate}, {Bonnell}, \& {Bromm}}]{Bate03}
{Bate}, M.~R., {Bonnell}, I.~A., \& {Bromm}, V. 2003, \mnras, 339, 577

\bibitem[{{Bergh{\"o}fer} \& {Christian}(2002)}]{BC02}
{Bergh{\"o}fer}, T.~W. \& {Christian}, D.~J. 2002, \aap, 384, 890

\bibitem[Bessell \& Brett(1988)]{Bessell88}
Bessell, M.~S., \& Brett, J.~M.\ 1988, \pasp, 100, 1134

\bibitem[{{Bica} {et~al.}(2003){Bica}, {Dutra}, {Soares}, \& {Barbuy}}]{Bica03}
{Bica}, E., {Dutra}, C.~M., {Soares}, J., \& {Barbuy}, B. 2003, \aap, 404, 223

\bibitem[{{Blaauw}(1964)}]{Blaauw64}
{Blaauw}, A. 1964, \araa, 2, 213

\bibitem[{{Blitz} \& {Thaddeus}(1980)}]{Blitz80}
{Blitz}, L. \& {Thaddeus}, P. 1980, \apj, 241, 676

\bibitem[{{Block} {et~al.}(1993){Block}, {Geballe}, \& {Dyson}}]{Block93}
{Block}, D.~L., {Geballe}, T.~R., \& {Dyson}, J.~E. 1993, \aap, 273, L41+

\bibitem[{{Bonnell} {et~al.}(2001){Bonnell}, {Bate}, {Clarke}, \&
  {Pringle}}]{Bonnell01}
{Bonnell}, I.~A., {Bate}, M.~R., {Clarke}, C.~J., \& {Pringle}, J.~E. 2001,
  \mnras, 323, 785

\bibitem[{{Bonnell} \& {Davies}(1998)}]{Bonnell98}
{Bonnell}, I.~A. \& {Davies}, M.~B. 1998, \mnras, 295, 691

\bibitem[Bouwman et al.(2006)]{Bouwman06} Bouwman, J., Lawson, 
W.~A., Dominik, C., Feigelson, E.~D., Henning, T., Tielens, A.~G.~G.~M., 
\& Waters, L.~B.~F.~M.\ 2006, \apjl, 653, L57 

\bibitem[{{Brand} {et~al.}(1994){Brand}, {Cesaroni}, {Caselli}, {Catarzi},
  {Codella}, {Comoretto}, {Curioni}, {Curioni}, {di Franco}, {Felli},
  {Giovanardi}, {Olmi}, {Palagi}, {Palla}, {Panella}, {Pareschi}, {Rossi},
  {Speroni}, \& {Tofani}}]{Brand94}
{Brand}, J., {Cesaroni}, R., {Caselli}, P., {Catarzi}, M., {Codella}, C.,
  {Comoretto}, G., {Curioni}, G.~P., {Curioni}, P., {di Franco}, S., {Felli},
  M., {Giovanardi}, C., {Olmi}, L., {Palagi}, F., {Palla}, F., {Panella}, D.,
  {Pareschi}, G., {Rossi}, E., {Speroni}, N., \& {Tofani}, G. 1994, \aaps, 103,
  541

\bibitem[{{Bronfman} {et~al.}(1996){Bronfman}, {Nyman}, \& {May}}]{Bronfman96}
{Bronfman}, L., {Nyman}, L.-A., \& {May}, J. 1996, \aaps, 115, 81

\bibitem[{{Broos} {et~al.}(2002){Broos}, {Townsley}, {Getman}, \& {Bauer}}]{ae}
{Broos}, P., {Townsley}, L.~K., {Getman}, K.~V., \& {Bauer}, F. 2002, ACIS
  Extract, An ACIS Point Source Extraction Package, Pennsylvania State
  University,
  \url{http://www.astro.psu.edu/xray/docs/TARA/ae\_users\_guide.html}

\bibitem[{{Broos} {et~al.}(2007){Broos}, {Feigelson}, {Townsley}, {Getman},
  {Wang}, {Garmire}, {Jiang}, \& {Tsuboi}}]{Broos07}
{Broos}, P.~S., {Feigelson}, E.~D., {Townsley}, L.~K., {Getman}, K.~V., {Wang},
  J., {Garmire}, G.~P., {Jiang}, Z., \& {Tsuboi}, Y. 2007, \apjs, 169, 353

\bibitem[{{Calvet} {et~al.}(2000){Calvet}, {Hartmann}, \& {Strom}}]{Calvet00}
{Calvet}, N., {Hartmann}, L., \& {Strom}, S.~E. 2000, Protostars and Planets
  IV, 377

\bibitem[{{Carpenter}(2000)}]{Carpenter00}
{Carpenter}, J.~M. 2000, \aj, 120, 3139

\bibitem[{{Casertano} \& {Hut}(1985)}]{Casertano85}
{Casertano}, S. \& {Hut}, P. 1985, \apj, 298, 80

\bibitem[{{Churchwell} {et~al.}(2004){Churchwell}, {Whitney}, {Babler},
  {Indebetouw}, {Meade}, {Watson}, {Wolff}, {Wolfire}, {Bania}, {Benjamin},
  {Clemens}, {Cohen}, {Devine}, {Dickey}, {Heitsch}, {Jackson}, {Kobulnicky},
  {Marston}, {Mathis}, {Mercer}, {Stauffer}, \& {Stolovy}}]{Churchwell04}
{Churchwell}, E., {Whitney}, B.~A., {Babler}, B.~L., {Indebetouw}, R., {Meade},
  M.~R., {Watson}, C., {Wolff}, M.~J., {Wolfire}, M.~G., {Bania}, T.~M.,
  {Benjamin}, R.~A., {Clemens}, D.~P., {Cohen}, M., {Devine}, K.~E., {Dickey},
  J.~M., {Heitsch}, F., {Jackson}, J.~M., {Kobulnicky}, H.~A., {Marston},
  A.~P., {Mathis}, J.~S., {Mercer}, E.~P., {Stauffer}, J.~R., \& {Stolovy},
  S.~R. 2004, \apjs, 154, 322

\bibitem[{{Cox} {et~al.}(1990){Cox}, {Deharveng}, \& {Leene}}]{Cox90}
{Cox}, P., {Deharveng}, L., \& {Leene}, A. 1990, \aap, 230, 181

\bibitem[{{Cutri} {et~al.}(2003){Cutri}, {Skrutskie}, {van Dyk}, {Beichman},
  {Carpenter}, {Chester}, {Cambresy}, {Evans}, {Fowler}, {Gizis}, {Howard},
  {Huchra}, {Jarrett}, {Kopan}, {Kirkpatrick}, {Light}, {Marsh}, {McCallon},
  {Schneider}, {Stiening}, {Sykes}, {Weinberg}, {Wheaton}, {Wheelock}, \&
  {Zacarias}}]{Cutri03}
{Cutri}, R.~M., {Skrutskie}, M.~F., {van Dyk}, S., {Beichman}, C.~A.,
  {Carpenter}, J.~M., {Chester}, T., {Cambresy}, L., {Evans}, T., {Fowler}, J.,
  {Gizis}, J., {Howard}, E., {Huchra}, J., {Jarrett}, T., {Kopan}, E.~L.,
  {Kirkpatrick}, J.~D., {Light}, R.~M., {Marsh}, K.~A., {McCallon}, H.,
  {Schneider}, S., {Stiening}, R., {Sykes}, M., {Weinberg}, M., {Wheaton},
  W.~A., {Wheelock}, S., \& {Zacarias}, N. 2003, {2MASS All Sky Catalog of
  point sources.} (The IRSA 2MASS All-Sky Point Source Catalog, NASA/IPAC
  Infrared Science Archive.~http://irsa.ipac.caltech.edu/applications/Gator/)

\bibitem[{{Dale} {et~al.}(2007){Dale}, {Bonnell}, \& {Whitworth}}]{Dale07}
{Dale}, J.~E., {Bonnell}, I.~A., \& {Whitworth}, A.~P. 2007, \mnras, 375, 1291


\bibitem[{{Diggle}(2003)}]{Diggle03}
{Diggle}, P. 2003, {Statistical Analysis of Spatial Point Patterns}, 2nd
  Edition, Hodder Arnold Publication

\bibitem[{{Droege} {et~al.}(2006){Droege}, {Richmond}, {Sallman}, \&
  {Creager}}]{Droege06}
{Droege}, T.~F., {Richmond}, M.~W., {Sallman}, M.~P., \& {Creager}, R.~P. 2006,
  \pasp, 118, 1666

\bibitem[{{Elmegreen}(2000)}]{Elmegreen00}
{Elmegreen}, B.~G. 2000, \apj, 530, 277

\bibitem[{{Elmegreen} \& {Lada}(1977)}]{Elmegreen77}
{Elmegreen}, B.~G. \& {Lada}, C.~J. 1977, \apj, 214, 725

\bibitem[{{Evans}(1967)}]{Evans67}
{Evans}, D.~S. 1967, in IAU Symposium, Vol.~30, Determination of Radial
  Velocities and their Applications, ed. A.~H. {Batten} \& J.~F. {Heard}, 57

\bibitem[{{Evans} {et~al.}(2003){Evans}, {Allen}, {Blake}, {Boogert}, {Bourke},
  {Harvey}, {Kessler}, {Koerner}, {Lee}, {Mundy}, {Myers}, {Padgett},
  {Pontoppidan}, {Sargent}, {Stapelfeldt}, {van Dishoeck}, {Young}, \&
  {Young}}]{Evans03a}
{Evans}, II, N.~J., {Allen}, L.~E., {Blake}, G.~A., {Boogert}, A.~C.~A.,
  {Bourke}, T., {Harvey}, P.~M., {Kessler}, J.~E., {Koerner}, D.~W., {Lee},
  C.~W., {Mundy}, L.~G., {Myers}, P.~C., {Padgett}, D.~L., {Pontoppidan}, K.,
  {Sargent}, A.~I., {Stapelfeldt}, K.~R., {van Dishoeck}, E.~F., {Young},
  C.~H., \& {Young}, K.~E. 2003, \pasp, 115, 965

\bibitem[{{Feigelson} {et~al.}(2007){Feigelson}, {Townsley}, {G{\"u}del}, \&
  {Stassun}}]{Feigelson07}
{Feigelson}, E., {Townsley}, L., {G{\"u}del}, M., \& {Stassun}, K. 2007, in
  Protostars and Planets V, ed. B.~{Reipurth}, D.~{Jewitt}, \& K.~{Keil},  313

\bibitem[{{Feigelson}(1996)}]{Feigelson96}
{Feigelson}, E.~D. 1996, \apj, 468, 306

\bibitem[{{Feigelson} {et~al.}(2005){Feigelson}, {Getman}, {Townsley},
  {Garmire}, {Preibisch}, {Grosso}, {Montmerle}, {Muench}, \&
  {McCaughrean}}]{Feigelson05}
{Feigelson}, E.~D., {Getman}, K., {Townsley}, L., {Garmire}, G., {Preibisch},
  T., {Grosso}, N., {Montmerle}, T., {Muench}, A., \& {McCaughrean}, M. 2005,
  \apjs, 160, 379

\bibitem[{{Feigelson} \& {Townsley}(2008)}]{Feigelson08}
{Feigelson}, E.~D. \& {Townsley}, L.~K. 2008, \apj, 673, 354

\bibitem[Furlan et al.(2007)]{Furlan07} Furlan, E., et al.\ 
2007, \apj, 664, 1176 

\bibitem[{{Gagn{\'e}} {et~al.}(2005){Gagn{\'e}}, {Oksala}, {Cohen}, {Tonnesen},
  {ud-Doula}, {Owocki}, {Townsend}, \& {MacFarlane}}]{Gagne05}
{Gagn{\'e}}, M., {Oksala}, M.~E., {Cohen}, D.~H., {Tonnesen}, S.~K.,
  {ud-Doula}, A., {Owocki}, S.~P., {Townsend}, R.~H.~D., \& {MacFarlane}, J.~J.
  2005, \apj, 628, 986

\bibitem[{{Getman} {et~al.}(2007){Getman}, {Feigelson}, {Garmire}, {Broos}, \&
  {Wang}}]{Getman07}
{Getman}, K.~V., {Feigelson}, E.~D., {Garmire}, G., {Broos}, P., \& {Wang}, J.
  2007, \apj, 654, 316

\bibitem[{Getman} {et~al.}(2008a)]{Getman08a}
{Getman}, K.~V., {Feigelson}, E.~D., {Lawson}, W.~A., {Broos}, P.~S., \&
  {Garmire}, G.~P. 2008, \apj, 673, 331

\bibitem[{Getman} {et~al.}(2008b)]{Getman08b} {Getman}, K.~V., 
{Feigelson}, E.~D., {Broos}, P.~S., {Micela}, G., 
\& {Garmire}, G.~P.\ 2008, \apj, 688, 418 

\bibitem[{{Getman} {et~al.}(2006){Getman}, {Feigelson}, {Townsley}, {Broos},
  {Garmire}, \& {Tsujimoto}}]{Getman06}
{Getman}, K.~V., {Feigelson}, E.~D., {Townsley}, L., {Broos}, P., {Garmire},
  G., \& {Tsujimoto}, M. 2006, \apjs, 163, 306

\bibitem[{{Gregorio-Hetem} {et~al.}(1998){Gregorio-Hetem}, {Montmerle},
  {Casanova}, \& {Feigelson}}]{Gregorio98}
{Gregorio-Hetem}, J., {Montmerle}, T., {Casanova}, S., \& {Feigelson}, E.~D.
  1998, \aap, 331, 193

\bibitem[{{Grosso} {et~al.}(2005){Grosso}, {Feigelson}, {Getman}, {Townsley},
  {Broos}, {Flaccomio}, {McCaughrean}, {Micela}, {Sciortino}, {Bally}, {Smith},
  {Muench}, {Garmire}, \& {Palla}}]{Grosso05}
{Grosso}, N., {Feigelson}, E.~D., {Getman}, K.~V., {Townsley}, L., {Broos}, P.,
  {Flaccomio}, E., {McCaughrean}, M.~J., {Micela}, G., {Sciortino}, S.,
  {Bally}, J., {Smith}, N., {Muench}, A.~A., {Garmire}, G.~P., \& {Palla}, F.
  2005, \apjs, 160, 530

\bibitem[{{Hamaguchi} {et~al.}(2005){Hamaguchi}, {Yamauchi}, \&
  {Koyama}}]{Hamaguchi05}
{Hamaguchi}, K., {Yamauchi}, S., \& {Koyama}, K. 2005, \apj, 618, 360

\bibitem[{{Hanson} {et~al.}(1993){Hanson}, {Geballe}, {Conti}, \&
  {Block}}]{Hanson93}
{Hanson}, M.~M., {Geballe}, T.~R., {Conti}, P.~S., \& {Block}, D.~L. 1993,
  \aap, 273, L44

\bibitem[{{Harvey} {et~al.}(2006){Harvey}, {Chapman}, {Lai}, {Evans}, {Allen},
  {J{\o}rgensen}, {Mundy}, {Huard}, {Porras}, {Cieza}, {Myers}, {Mer{\'{\i}}n},
  {van Dishoeck}, {Young}, {Spiesman}, {Blake}, {Koerner}, {Padgett},
  {Sargent}, \& {Stapelfeldt}}]{Harvey06}
{Harvey}, P.~M., {Chapman}, N., {Lai}, S.-P., {Evans}, II, N.~J., {Allen},
  L.~E., {J{\o}rgensen}, J.~K., {Mundy}, L.~G., {Huard}, T.~L., {Porras}, A.,
  {Cieza}, L., {Myers}, P.~C., {Mer{\'{\i}}n}, B., {van Dishoeck}, E.~F.,
  {Young}, K.~E., {Spiesman}, W., {Blake}, G.~A., {Koerner}, D.~W., {Padgett},
  D.~L., {Sargent}, A.~I., \& {Stapelfeldt}, K.~R. 2006, \apj, 644, 307

\bibitem[{{Hensberge} {et~al.}(2000){Hensberge}, {Pavlovski}, \&
  {Verschueren}}]{Hensberge00}
{Hensberge}, H., {Pavlovski}, K., \& {Verschueren}, W. 2000, \aap, 358, 553

\bibitem[{{Heyer} {et~al.}(2006){Heyer}, {Williams}, \& {Brunt}}]{Heyer06}
{Heyer}, M.~H., {Williams}, J.~P., \& {Brunt}, C.~M. 2006, \apj, 643, 956

\bibitem[{{Kroupa}(2002)}]{Kroupa02}
{Kroupa}, P. 2002, Science, 295, 82



\bibitem[Lada \& Adams(1992)]{Lada92} Lada, C.~J., \& Adams, F.~C.\ 1992, \apj, 393, 278 

\bibitem[{{Lada} \& {Lada}(2003)}]{Lada03}
{Lada}, C.~J. \& {Lada}, E.~A. 2003, \araa, 41, 57

\bibitem[{{Lada} {et~al.}(1991){Lada}, {Evans}, {Depoy}, \& {Gatley}}]{Lada91}
{Lada}, E.~A., {Evans}, II, N.~J., {Depoy}, D.~L., \& {Gatley}, I. 1991, \apj,
  371, 171

\bibitem[{{Lada} \& {Lada}(1995)}]{Lada95}
{Lada}, E.~A. \& {Lada}, C.~J. 1995, \aj, 109, 1682

\bibitem[Lehmer et al.(2005)]{Lehmer05} Lehmer, B.~D., et al.\ 
2005, \apjs, 161, 21 

\bibitem[{{Li}(2005)}]{Li05}
{Li}, J.~Z. 2005, \apj, 625, 242

\bibitem[{{Li} \& {Smith}(2005)}]{Li05b}
{Li}, J.~Z. \& {Smith}, M.~D. 2005, \apj, 620, 816

\bibitem[{Luo} {et~al.}(2008)]{Luo08}
{Luo}, B., et al.\ 2008, \apjs, 179, 19 

\bibitem[{{Mac Low} \& {Klessen}(2004)}]{MacLow04}
{Mac Low}, M.-M. \& {Klessen}, R.~S. 2004, Reviews of Modern Physics, 76, 125

\bibitem[{{Ma{\'{\i}}z-Apell{\'a}niz}
  {et~al.}(2004){Ma{\'{\i}}z-Apell{\'a}niz}, {Walborn}, {Galu{\'e}}, \&
  {Wei}}]{Maiz-Apellaniz04}
{Ma{\'{\i}}z-Apell{\'a}niz}, J., {Walborn}, N.~R., {Galu{\'e}}, H.~{\'A}., \&
  {Wei}, L.~H. 2004, \apjs, 151, 103

\bibitem[{{Massey} {et~al.}(1995){Massey}, {Johnson}, \&
  {Degioia-Eastwood}}]{MJD95}
{Massey}, P., {Johnson}, K.~E., \& {Degioia-Eastwood}, K. 1995, \apj, 454, 151

\bibitem[Meyer et al.(1997)]{Meyer97} Meyer, M.~R., Calvet, N.,
\& Hillenbrand, L.~A.\ 1997, \aj, 114, 288

\bibitem[{{Monet} {et~al.}(2003){Monet}, {Levine}, {Canzian}, {Ables}, {Bird},
  {Dahn}, {Guetter}, {Harris}, {Henden}, {Leggett}, {Levison}, {Luginbuhl},
  {Martini}, {Monet}, {Munn}, {Pier}, {Rhodes}, {Riepe}, {Sell}, {Stone},
  {Vrba}, {Walker}, {Westerhout}, {Brucato}, {Reid}, {Schoening}, {Hartley},
  {Read}, \& {Tritton}}]{Monet03}
{Monet}, D.~G., {Levine}, S.~E., {Canzian}, B., {Ables}, H.~D., {Bird}, A.~R.,
  {Dahn}, C.~C., {Guetter}, H.~H., {Harris}, H.~C., {Henden}, A.~A., {Leggett},
  S.~K., {Levison}, H.~F., {Luginbuhl}, C.~B., {Martini}, J., {Monet},
  A.~K.~B., {Munn}, J.~A., {Pier}, J.~R., {Rhodes}, A.~R., {Riepe}, B., {Sell},
  S., {Stone}, R.~C., {Vrba}, F.~J., {Walker}, R.~L., {Westerhout}, G.,
  {Brucato}, R.~J., {Reid}, I.~N., {Schoening}, W., {Hartley}, M., {Read},
  M.~A., \& {Tritton}, S.~B. 2003, \aj, 125, 984

\bibitem[{{Morrison} \& {McCammon}(1983)}]{Morrison83}
{Morrison}, R. \& {McCammon}, D. 1983, \apj, 270, 119

\bibitem[{{Neuhaeuser} {et~al.}(1995){Neuhaeuser}, {Sterzik}, {Schmitt},
  {Wichmann}, \& {Krautter}}]{Neuhaeuser95}
{Neuhaeuser}, R., {Sterzik}, M.~F., {Schmitt}, J.~H.~M.~M., {Wichmann}, R., \&
  {Krautter}, J. 1995, \aap, 297, 391

\bibitem[{{Perez}(1991)}]{Perez91}
{Perez}, M.~R. 1991, Revista Mexicana de Astronomia y Astrofisica, 22, 99

\bibitem[{{Perryman} {et~al.}(1997){Perryman}, {Lindegren}, {Kovalevsky},
  {Hoeg}, {Bastian}, {Bernacca}, {Cr{\'e}z{\'e}}, {Donati}, {Grenon}, {van
  Leeuwen}, {van der Marel}, {Mignard}, {Murray}, {Le Poole}, {Schrijver},
  {Turon}, {Arenou}, {Froeschl{\'e}}, \& {Petersen}}]{Perryman97}
{Perryman}, M.~A.~C., {Lindegren}, L., {Kovalevsky}, J., {Hoeg}, E., {Bastian},
  U., {Bernacca}, P.~L., {Cr{\'e}z{\'e}}, M., {Donati}, F., {Grenon}, M., {van
  Leeuwen}, F., {van der Marel}, H., {Mignard}, F., {Murray}, C.~A., {Le
  Poole}, R.~S., {Schrijver}, H., {Turon}, C., {Arenou}, F., {Froeschl{\'e}},
  M., \& {Petersen}, C.~S. 1997, \aap, 323, L49

\bibitem[{{Petit} {et~al.}(2008){Petit}, {Wade}, {Drissen}, {Montmerle}, \&
  {Alecian}}]{Petit08}
{Petit}, V., {Wade}, G.~A., {Drissen}, L., {Montmerle}, T., \& {Alecian}, E.
  2008, \mnras, L56

\bibitem[{{Phelps} \& {Lada}(1997)}]{PL97}
{Phelps}, R.~L. \& {Lada}, E.~A. 1997, \apj, 477, 176

\bibitem[{{Poulton} {et~al.}(2008){Poulton}, {Robitaille}, {Greaves},
  {Bonnell}, {Williams}, \& {Heyer}}]{Poulton08}
{Poulton}, C.~J., {Robitaille}, T.~P., {Greaves}, J.~S., {Bonnell}, I.~A.,
  {Williams}, J.~P., \& {Heyer}, M.~H. 2008, \mnras, 132

\bibitem[{{Preibisch} {et~al.}(2005){Preibisch}, {Kim}, {Favata}, {Feigelson},
  {Flaccomio}, {Getman}, {Micela}, {Sciortino}, {Stassun}, {Stelzer}, \&
  {Zinnecker}}]{Preibisch05}
{Preibisch}, T., {Kim}, Y.-C., {Favata}, F., {Feigelson}, E.~D., {Flaccomio},
  E., {Getman}, K., {Micela}, G., {Sciortino}, S., {Stassun}, K., {Stelzer},
  B., \& {Zinnecker}, H. 2005, \apjs, 160, 401

\bibitem[{{Preibisch} \& {Zinnecker}(2007)}]{Preibisch07}
{Preibisch}, T. \& {Zinnecker}, H. 2007, in IAU Symposium, Vol. 237, IAU
  Symposium, ed. B.~G. {Elmegreen} \& J.~{Palous}, 270

\bibitem[{{Prisinzano} {et~al.}(2008){Prisinzano}, {Micela}, {Flaccomio},
  {Stauffer}, {Megeath}, {Rebull}, {Robberto}, {Smith}, {Feigelson}, {Grosso},
  \& {Wolk}}]{Prisinzano08}
{Prisinzano}, L., {Micela}, G., {Flaccomio}, E., {Stauffer}, J.~R., {Megeath},
  T., {Rebull}, L., {Robberto}, M., {Smith}, K., {Feigelson}, E.~D., {Grosso},
  N., \& {Wolk}, S. 2008, \apj, 677, 401 

\bibitem[{{Rom{\'a}n-Z{\'u}{\~n}iga}
  {et~al.}(2008{\natexlab{a}}){Rom{\'a}n-Z{\'u}{\~n}iga}, {Elston}, {Ferreira},
  \& {Lada}}]{RomanZuniga08}
{Rom{\'a}n-Z{\'u}{\~n}iga}, C.~G., {Elston}, R., {Ferreira}, B., \& {Lada},
  E.~A. 2008{\natexlab{a}}, \apj, 672, 861

\bibitem[{{Rom{\'a}n-Z{\'u}{\~n}iga}
    {et~al.}(2008{\natexlab{b}}){Rom{\'a}n-Z{\'u}{\~n}iga}, {Lada}, \&
    {Elston}}]{RL08} {Rom{\'a}n-Z{\'u}{\~n}iga}, C.~G., {Lada}, E.~A.,
  \& {Elston}, R.  2008{\natexlab{b}}, Handbook of Star Forming
  Regions, Vol. I: The Northern Hemisphere, ed. B.~{Reipurth}, in
  press, astro-ph/0910.0931

\bibitem[{{Sanchawala} {et~al.}(2007){Sanchawala}, {Chen}, {Lee}, {Chu},
  {Nakajima}, {Tamura}, {Baba}, \& {Sato}}]{Sanchawala07}
{Sanchawala}, K., {Chen}, W.-P., {Lee}, H.-T., {Chu}, Y.-H., {Nakajima}, Y.,
  {Tamura}, M., {Baba}, D., \& {Sato}, S. 2007, \apj, 656, 462

\bibitem[{{Siess} {et~al.}(2000){Siess}, {Dufour}, \& {Forestini}}]{Siess00}
{Siess}, L., {Dufour}, E., \& {Forestini}, M. 2000, \aap, 358, 593

\bibitem[{{Skinner} {et~al.}(2006){Skinner}, {G{\"u}del}, {Schmutz}, \&
  {Zhekov}}]{Skinner06}
{Skinner}, S., {G{\"u}del}, M., {Schmutz}, W., \& {Zhekov}, S. 2006, \apss,
  304, 97

\bibitem[{{Smith} {et~al.}(2001){Smith}, {Brickhouse}, {Liedahl}, \&
  {Raymond}}]{Smith01}
{Smith}, R.~K., {Brickhouse}, N.~S., {Liedahl}, D.~A., \& {Raymond}, J.~C.
  2001, \apjl, 556, L91

\bibitem[{{Stelzer} {et~al.}(2007){Stelzer}, {Flaccomio}, {Briggs}, {Micela},
  {Scelsi}, {Audard}, {Pillitteri}, \& {G{\"u}del}}]{Stelzer07}
{Stelzer}, B., {Flaccomio}, E., {Briggs}, K., {Micela}, G., {Scelsi}, L.,
  {Audard}, M., {Pillitteri}, I., \& {G{\"u}del}, M. 2007, \aap, 468, 463

\bibitem[{{Stelzer} {et~al.}(2006){Stelzer}, {Micela}, {Hamaguchi}, \&
  {Schmitt}}]{Stelzer06b}
{Stelzer}, B., {Micela}, G., {Hamaguchi}, K., \& {Schmitt}, J.~H.~M.~M. 2006,
  \aap, 457, 223

\bibitem[{{Strom} {et~al.}(1993){Strom}, {Strom}, \& {Merrill}}]{Strom93}
{Strom}, K.~M., {Strom}, S.~E., \& {Merrill}, K.~M. 1993, \apj, 412, 233

\bibitem[{{Tan} {et~al.}(2006){Tan}, {Krumholz}, \& {McKee}}]{Tan06}
{Tan}, J.~C., {Krumholz}, M.~R., \& {McKee}, C.~F. 2006, \apjl, 641, L121

\bibitem[{{Telleschi} {et~al.}(2007){Telleschi}, {G{\"u}del}, {Briggs},
  {Audard}, \& {Palla}}]{Telleschi07}
{Telleschi}, A., {G{\"u}del}, M., {Briggs}, K.~R., {Audard}, M., \& {Palla}, F.
  2007, \aap, 468, 425

\bibitem[{{Townsley} {et~al.}(2006){Townsley}, {Broos}, {Feigelson}, {Brandl},
  {Chu}, {Garmire}, \& {Pavlov}}]{Townsley06}
{Townsley}, L.~K., {Broos}, P.~S., {Feigelson}, E.~D., {Brandl}, B.~R., {Chu},
  Y.-H., {Garmire}, G.~P., \& {Pavlov}, G.~G. 2006, \aj, 131, 2140

\bibitem[{{Townsley} {et~al.}(2003){Townsley}, {Feigelson}, {Montmerle},
  {Broos}, {Chu}, \& {Garmire}}]{Townsley03}
{Townsley}, L.~K., {Feigelson}, E.~D., {Montmerle}, T., {Broos}, P.~S., {Chu},
  Y.-H., \& {Garmire}, G.~P. 2003, \apj, 593, 874

\bibitem[{{Vuong} {et~al.}(2003){Vuong}, {Montmerle}, {Grosso}, {Feigelson},
  {Verstraete}, \& {Ozawa}}]{Vuong03}
{Vuong}, M.~H., {Montmerle}, T., {Grosso}, N., {Feigelson}, E.~D.,
  {Verstraete}, L., \& {Ozawa}, H. 2003, \aap, 408, 581

\bibitem[{{Walborn}(1973)}]{Walborn73}
{Walborn}, N.~R. 1973, \aj, 78, 1067

\bibitem[{{Walter} {et~al.}(1988){Walter}, {Brown}, {Mathieu}, {Myers}, \&
  {Vrba}}]{Walter88}
{Walter}, F.~M., {Brown}, A., {Mathieu}, R.~D., {Myers}, P.~C., \& {Vrba},
  F.~J. 1988, \aj, 96, 297

\bibitem[{{Wang} {et~al.}(2007){Wang}, {Townsley}, {Feigelson}, {Getman},
  {Broos}, {Garmire}, \& {Tsujimoto}}]{Wang07a}
{Wang}, J., {Townsley}, L.~K., {Feigelson}, E.~D., {Getman}, K.~V., {Broos},
  P.~S., {Garmire}, G.~P., \& {Tsujimoto}, M. 2007, \apjs, 168, 100

\bibitem[{{Wang} {et~al.}(2008){Wang}, {Townsley}, {Feigelson}, {Getman},
  {Broos}, {Roman-Zuniga}, \& {Lada}}]{Wang08}
{Wang}, J., {Townsley}, L.~K., {Feigelson}, E.~D., {Getman}, K.~V., {Broos},
  P.~S., {Roman-Zuniga}, C.~G., \& {Lada}, E.~A. 2008, ApJ, 675, 464  

\bibitem[{{Weisskopf} {et~al.}(2002){Weisskopf}, {Brinkman}, {Canizares},
  {Garmire}, {Murray}, \& {Van Speybroeck}}]{Weisskopf02}
{Weisskopf}, M.~C., {Brinkman}, B., {Canizares}, C., {Garmire}, G., {Murray},
  S., \& {Van Speybroeck}, L.~P. 2002, \pasp, 114, 1

\bibitem[{{Whitworth} {et~al.}(1994){Whitworth}, {Bhattal}, {Chapman},
  {Disney}, \& {Turner}}]{Whitworth94}
{Whitworth}, A.~P., {Bhattal}, A.~S., {Chapman}, S.~J., {Disney}, M.~J., \&
  {Turner}, J.~A. 1994, \mnras, 268, 291

\bibitem[{{Wilking} \& {Lada}(1983)}]{Wilking83}
{Wilking}, B.~A. \& {Lada}, C.~J. 1983, \apj, 274, 698

\bibitem[{{Williams} \& {Blitz}(1998)}]{Williams98}
{Williams}, J.~P. \& {Blitz}, L. 1998, \apj, 494, 657

\bibitem[{{Williams} {et~al.}(1995){Williams}, {Blitz}, \&
  {Stark}}]{Williams95}
{Williams}, J.~P., {Blitz}, L., \& {Stark}, A.~A. 1995, \apj, 451, 252

\bibitem[{{Wolf} {et~al.}(1990){Wolf}, {Lada}, \& {Bally}}]{Wolf90}
{Wolf}, G.~A., {Lada}, C.~J., \& {Bally}, J. 1990, \aj, 100, 1892

\bibitem[{{Wolk} {et~al.}(2006){Wolk}, {Spitzbart}, {Bourke}, \&
  {Alves}}]{Wolk06}
{Wolk}, S.~J., {Spitzbart}, B.~D., {Bourke}, T.~L., \& {Alves}, J. 2006, \aj,
  132, 1100

\bibitem[{{Wolk} {et~al.}(2008){Wolk}, {Spitzbart}, {Bourke}, {Gutermuth},
  {Vigil}, \& {Comer{\'o}n}}]{Wolk08}
{Wolk}, S.~J., {Spitzbart}, B.~D., {Bourke}, T.~L., {Gutermuth}, R.~A.,
  {Vigil}, M., \& {Comer{\'o}n}, F. 2008, \aj, 135, 693

\bibitem[{{Wu} {et~al.}(2006){Wu}, {Zhang}, {Yu}, {Miller}, {Mao}, {Sun}, \&
  {Wang}}]{Wu06}
{Wu}, Y., {Zhang}, Q., {Yu}, W., {Miller}, M., {Mao}, R., {Sun}, K., \& {Wang},
  Y. 2006, \aap, 450, 607

\bibitem[{{Ybarra} \& {Phelps}(2004)}]{Ybarra04}
{Ybarra}, J.~E. \& {Phelps}, R.~L. 2004, \aj, 127, 3444

\bibitem[{{Young} {et~al.}(2005){Young}, {Harvey}, {Brooke}, {Chapman},
  {Kauffmann}, {Bertoldi}, {Lai}, {Alcal{\'a}}, {Bourke}, {Spiesman}, {Allen},
  {Blake}, {Evans}, {Koerner}, {Mundy}, {Myers}, {Padgett}, {Salinas},
  {Sargent}, {Stapelfeldt}, {Teuben}, {van Dishoeck}, \& {Wahhaj}}]{Young05}
{Young}, K.~E., {Harvey}, P.~M., {Brooke}, T.~Y., {Chapman}, N., {Kauffmann},
  J., {Bertoldi}, F., {Lai}, S.-P., {Alcal{\'a}}, J., {Bourke}, T.~L.,
  {Spiesman}, W., {Allen}, L.~E., {Blake}, G.~A., {Evans}, II, N.~J.,
  {Koerner}, D.~W., {Mundy}, L.~G., {Myers}, P.~C., {Padgett}, D.~L.,
  {Salinas}, A., {Sargent}, A.~I., {Stapelfeldt}, K.~R., {Teuben}, P., {van
  Dishoeck}, E.~F., \& {Wahhaj}, Z. 2005, \apj, 628, 283

\end{thebibliography}

\begin{deluxetable}{rcrrrrrrrrrrrrrccccc}
\centering \rotate
\tabletypesize{\tiny} \tablewidth{0pt}
\tablecolumns{20}

\tablecaption{ {\it Chandra} Catalog:  Primary Source Properties
\label{tbl:ch5_primary}}

\tablehead{
\multicolumn{2}{c}{Source} &
  &
\multicolumn{4}{c}{Position} &
  &
\multicolumn{5}{c}{Extracted Counts} &
  &
\multicolumn{6}{c}{Characteristics} \\
\cline{1-2} \cline{4-7} \cline{9-13} \cline{15-20}

\colhead{RMCX} & \colhead{CXOU J} &
  &
\colhead{$\alpha_{\rm J2000}$} & \colhead{$\delta_{\rm J2000}$} &
\colhead{Err}
& \colhead{$\theta$} &
  &
\colhead{Net} & \colhead{$\Delta$Net} & \colhead{Bkgd} &
\colhead{Net} & \colhead{PSF} &
  &
\colhead{Signif} & \colhead{$\log P_B$} & \colhead{Anom} &
\colhead{Var} &\colhead{EffExp} & \colhead{$MedE$}  \\

\colhead{\#} & \colhead{} &
  &
\colhead{(deg)} & \colhead{(deg)} & \colhead{(\arcsec)} &
\colhead{(\arcmin)} &
  &
\colhead{Full} & \colhead{Full} & \colhead{Full} & \colhead{Hard} &
\colhead{Frac} &
  &
\colhead{} & \colhead{} & \colhead{} & \colhead{} & \colhead{(ks)} &
\colhead{(keV)}
 \\

\colhead{(1)} & \colhead{(2)} &
  &
\colhead{(3)} & \colhead{(4)} & \colhead{(5)} & \colhead{(6)} &
  &
\colhead{(7)} & \colhead{(8)} & \colhead{(9)} & \colhead{(10)} &
\colhead{(11)}
&
  &
\colhead{(12)} & \colhead{(13)} & \colhead{(14)} & \colhead{(15)} &
\colhead{(16)} & \colhead{(17)}  }

\startdata
   1$\ast$ & 063232.43$+$043705.7 &  &   98.135158 &   4.618251 &  0.7 &
6.0 &  &     7.5 &   3.4 &   0.5 &     6.7 & 0.91 &  &   1.9 & $<$-5
& g... & \nodata &    9.8 & 3.9 \\
   3$\ast$ & 063236.58$+$043611.4 &  &   98.152431 &   4.603172 &  0.9 &
6.6 &  &     6.2 &   3.2 &   0.8 &     4.4 & 0.89 &  &   1.6 & -4.6 &
.... & a &   17.9 & 2.6 \\
   4$\ast$ & 063238.08$+$043250.1 &  &   98.158706 &   4.547265 &  1.2 &
9.9 &  &    16.8 &   4.9 &   2.2 &     0.6 & 0.91 &  &   3.1 & $<$-5
& g... & \nodata &   10.6 & 1.3 \\
   5$\ast$ & 063239.50$+$043628.9 &  &   98.164603 &   4.608042 &  0.6 &
6.3 &  &    12.3 &   4.1 &   0.7 &     3.5 & 0.90 &  &   2.6 & $<$-5
& .... & a &   18.0 & 1.8 \\
   6$\ast$ & 063239.51$+$043124.9 &  &   98.164639 &   4.523587 &  1.2 &
9.9 &  &    11.7 &   4.4 &   3.3 &     3.9 & 0.91 &  &   2.3 & $<$-5
& .... & a &   16.2 & 1.9 \\
   7$\ast$ & 063240.73$+$043653.3 &  &   98.169734 &   4.614833 &  0.5 &
5.9 &  &    14.4 &   4.4 &   0.6 &     9.6 & 0.90 &  &   2.9 & $<$-5
& .... & a &   18.1 & 3.1 \\
   8$\ast$ & 063242.33$+$043217.9 &  &   98.176413 &   4.538329 &  1.3 &
9.0 &  &     7.0 &   3.5 &   2.0 &     6.7 & 0.90 &  &   1.7 & -3.6 &
.... & a &   17.3 & 3.6 \\
   9$\ast$ & 063245.25$+$043206.5 &  &   98.188575 &   4.535152 &  0.7 &
8.4 &  &    21.4 &   5.3 &   1.6 &     1.9 & 0.90 &  &   3.6 & $<$-5
& .... & c &   17.5 & 0.8 \\
  10$\ast$ & 063246.11$+$043612.9 &  &   98.192141 &   4.603596 &  0.8 &
7.3 &  &    10.8 &   4.1 &   2.2 &     7.7 & 0.90 &  &   2.3 & $<$-5
& .... & a &   35.3 & 2.4 \\
  11$\ast$ & 063246.72$+$043831.4 &  &   98.194699 &   4.642082 &  0.5 &
4.5 &  &     5.8 &   3.0 &   0.2 &     0.8 & 0.89 &  &   1.6 & $<$-5
& .... & a &   18.5 & 1.4 \\
\enddata

\tablecomments{Table \ref{tbl:ch5_primary} is published in its
entirety in the
electronic edition of the {\it Astrophysical Journal}.  A portion is
shown here for guidance regarding its form and content.}

\tablecomments{{\bf Column 1:} X-ray source sequence number, sorted
by RA. $\ast$ denotes the sources included in the complete sample of X-ray detections after sensitivity correction (see \S~\ref{sec:new_sensitivity}).
{\bf Column 2:} IAU designation.
{\bf Columns 3,4:} Right ascension and declination for epoch J2000.0.
{\bf Column 5:} Estimated $1\sigma$ random position error computed as 
the standard deviation of the local PSF inside
extraction region divided by the square root of the extracted counts. 
{\bf Column 6:} Off-axis angle in arcminutes
{\bf Columns 7,8:} Estimated net counts extracted in the total energy
band (0.5\nodata8~keV), and the average of its Poisson upper and lower $1\sigma$ 
error (including background contribution). 
{\bf Column 9:} Background counts extracted (total band).
{\bf Column 10:} Estimated net counts extracted in the hard energy
band (2\nodata8~keV).
{\bf Column 11:} Fraction of the PSF (at 1.497 keV) enclosed within
the extraction region.  A PSF fraction significantly below 90\% usually
indicates that the source lies in a crowded region.
{\bf Column 12:} Photometric significance computed as the ratio of the
net counts to the upper error on net counts.
{\bf Column 13:} Log probability that extracted counts (total band)
are solely from background based on Poisson statistics.  Some sources 
have $P_B$ values above the 0.1\% threshold that defines the catalog 
because local background estimates can rise during the final extraction 
iteration. 
{\bf Column 14:}  Source anomalies:  g = fractional time that source
was on a detector (FRACEXPO from {\em mkarf}) is $<0.9$ ; e = source
on field edge; p = source piled up; s = source on readout streak.
{\bf Column 15:} Variability characterization based on the 
Kolmogorov-Smirnov statistic
(total band):  a = no evidence for variability ($0.05<P_{KS}$); b =
possibly variable ($0.005<P_{KS}<0.05$); c = definitely variable
($P_{KS}<0.005$).  No value is reported for sources with fewer than 4
counts or for sources in chip gaps or on field edges.
{\bf Column 16:} Effective exposure time: approximate time the source
would have to be observed on-axis to obtain the reported number of
counts.
{\bf Column 17:} Background-corrected median photon energy (total
band).}

\end{deluxetable}
\begin{deluxetable}{rcrrrrrrrrrrrrrccccc}
\centering \rotate
\tabletypesize{\tiny} \tablewidth{0pt}
\tablecolumns{20}

\tablecaption{ {\it Chandra} Catalog:  Tentative Source Properties
\label{tbl:ch5_tentative}}

\tablehead{
\multicolumn{2}{c}{Source} &
  &
\multicolumn{4}{c}{Position} &
  &
\multicolumn{5}{c}{Extracted Counts} &
  &
\multicolumn{6}{c}{Characteristics} \\
\cline{1-2} \cline{4-7} \cline{9-13} \cline{15-20}

\colhead{RMCX} & \colhead{CXOU J} &
  &
\colhead{$\alpha_{\rm J2000}$} & \colhead{$\delta_{\rm J2000}$} &
\colhead{Err}
& \colhead{$\theta$} &
  &
\colhead{Net} & \colhead{$\Delta$Net} & \colhead{Bkgd} &
\colhead{Net} & \colhead{PSF} &
  &
\colhead{Signif} & \colhead{$\log P_B$} & \colhead{Anom} &
\colhead{Var} &\colhead{EffExp} & \colhead{$MedE$}  \\

\colhead{\#} & \colhead{} &
  &
\colhead{(deg)} & \colhead{(deg)} & \colhead{(\arcsec)} &
\colhead{(\arcmin)} &
  &
\colhead{Full} & \colhead{Full} & \colhead{Full} & \colhead{Hard} &
\colhead{Frac} &
  &
\colhead{} & \colhead{} & \colhead{} & \colhead{} & \colhead{(ks)} &
\colhead{(keV)}
 \\

\colhead{(1)} & \colhead{(2)} &
  &
\colhead{(3)} & \colhead{(4)} & \colhead{(5)} & \colhead{(6)} &
  &
\colhead{(7)} & \colhead{(8)} & \colhead{(9)} & \colhead{(10)} &
\colhead{(11)}
&
  &
\colhead{(12)} & \colhead{(13)} & \colhead{(14)} & \colhead{(15)} &
\colhead{(16)} & \colhead{(17)}  }

\startdata
   2$\ast$ & 063232.91$+$043437.2 &  &   98.137138 &   4.577017 &  1.4 &
8.4 &  &     5.9 &   3.4 &   2.1 &     0.6 & 0.90 &  &   1.5 & -2.8 &
g... & \nodata &   13.0 & 0.7 \\
  12 & 063247.43$+$043150.2 &  &   98.197650 &   4.530632 &  1.3 &
7.9 &  &     4.5 &   3.0 &   1.5 &     4.0 & 0.90 &  &   1.2 & -2.3 &
.... & a &   17.6 & 2.1 \\
  16 & 063249.22$+$043425.9 &  &   98.205124 &   4.573882 &  1.1 &
7.8 &  &     4.9 &   3.4 &   3.1 &     3.7 & 0.90 &  &   1.2 & -1.8 &
.... & a &   33.1 & 7.8 \\
  26 & 063254.45$+$043724.6 &  &   98.226879 &   4.623523 &  0.9 &
6.3 &  &     4.6 &   3.0 &   1.4 &     3.1 & 0.90 &  &   1.3 & -2.6 &
.... & \nodata &   35.6 & 3.4 \\
  29 & 063255.82$+$043419.1 &  &   98.232618 &   4.571979 &  1.1 &
7.2 &  &     5.2 &   3.4 &   2.8 &     0.0 & 0.89 &  &   1.3 & -2.1 &
.... & a &   34.6 & 1.3 \\
  33 & 063258.15$+$043003.7 &  &   98.242303 &   4.501051 &  1.1 &
6.6 &  &     3.3 &   2.5 &   0.7 &     3.5 & 0.89 &  &   1.1 & -2.3 &
.... & a &   15.9 & 5.2 \\
  36 & 063300.61$+$043030.7 &  &   98.252556 &   4.508536 &  0.9 &
5.9 &  &     3.5 &   2.5 &   0.5 &     0.0 & 0.90 &  &   1.1 & -2.7 &
.... & a &   17.6 & 1.3 \\
  53 & 063308.34$+$044203.4 &  &   98.284762 &   4.700962 &  1.0 &
7.3 &  &     4.6 &   3.2 &   2.4 &     2.4 & 0.89 &  &   1.2 & -1.9 &
.... & a &   33.9 & 1.6 \\
  57 & 063308.52$+$043634.3 &  &   98.285532 &   4.609538 &  0.3 &
4.0 &  &     4.7 &   3.0 &   1.3 &     0.0 & 0.90 &  &   1.3 & -2.7 &
g... & \nodata &   23.3 & 0.9 \\
  58 & 063309.33$+$043804.3 &  &   98.288891 &   4.634532 &  1.1 &
6.1 &  &     6.9 &   3.7 &   3.1 &     5.9 & 0.89 &  &   1.6 & -2.8 &
.... & a &   35.4 & 3.3 \\
\enddata

\tablecomments{Table \ref{tbl:ch5_tentative} is published in its
entirety in the
electronic edition of the {\it Astrophysical Journal}.  A portion is
shown here for guidance regarding its form and content. The column descriptions are the same as in Table~1.}

\end{deluxetable}


\begin{deluxetable}{rcrrrcccrcccccrc}
\centering \rotate
\tabletypesize{\scriptsize} \tablewidth{0pt}
\tablecolumns{16}

\tablecaption{X-ray Spectroscopy for Brighter RMCX 
Sources \label{tbl:ch5_thermal_spectroscopy}}

\tablehead{
\multicolumn{4}{c}{RMCX\tablenotemark{a}} &
  &
\multicolumn{3}{c}{Spectral Fit\tablenotemark{b}} &
  &
\multicolumn{5}{c}{X-ray Luminosities\tablenotemark{c}} &
  &
\colhead{Notes\tablenotemark{d}} \\
\cline{1-4} \cline{6-8} \cline{10-14}

\colhead{Seq} & \colhead{CXOU J} & \colhead{Net} & \colhead{Signif} &
  &
\colhead{$\log N_H$} & \colhead{$kT$} & \colhead{$\log EM$} &
  &
\colhead{$\log L_s$} & \colhead{$\log L_h$} & \colhead{$\log
L_{h,c}$} & \colhead{$\log L_t$} & \colhead{$\log L_{t,c}$} &
  &
\colhead{}  \\

\colhead{} & \colhead{} & \colhead{Cts} & \colhead{} &
  &
\colhead{(cm$^{-2}$)} & \colhead{(keV)} & \colhead{(cm$^{-3}$)} &
  &
\multicolumn{5}{c}{(ergs s$^{-1}$)} &
  &
\colhead{} \\

\colhead{(1)} & \colhead{(2)} & \colhead{(3)} & \colhead{(4)} &
  &
\colhead{(5)} & \colhead{(6)} & \colhead{(7)} &
  &
\colhead{(8)} & \colhead{(9)} & \colhead{(10)} &\colhead{(11)} &
\colhead{(12)}
&
  &
\colhead{(13)}
}
\startdata
   5 & 063239.50$+$043628.9 &    12.3 &   2.6 &  &  {\tiny $-0.9$}
21.8 {\tiny $+0.3$} &  2.2  & {\tiny $-0.4$} 53.4 {\tiny $+0.3$} &  &
29.67 & 29.98 & 30.02 & 30.15 & 30.42 &  & \nodata \\
   6 & 063239.51$+$043124.9 &    11.7 &   2.3 &  &   22.1  & {\em
2.0 } & \phantom{{\tiny $+0.2$}} 53.7 {\tiny $+0.2$} &  & 29.72 &
30.14 & 30.21 & 30.28 & 30.65 &  & \nodata \\
   7 & 063240.73$+$043653.3 &    14.4 &   2.9 &  &  {\tiny $-0.5$}
22.2 {\tiny $+0.4$} &  7.0  & {\tiny $-0.3$} 53.6 {\tiny $+0.7$} &  &
29.48 & 30.48 & 30.55 & 30.52 & 30.75 &  & \nodata \\
   9 & 063245.25$+$043206.5 &    21.4 &   3.6 &  &  {\em  20.0 } &
{\tiny $-0.05$} 0.2 {\tiny $+0.09$} & {\tiny $-0.3$} 53.8 {\tiny
$+0.2$} &  & 30.45 & 27.47 & 27.47 & 30.45 & 30.48 &  & \nodata \\
  10 & 063246.11$+$043612.9 &    10.8 &   2.3 &  &   22.1  &  3.5  &
53.3  &  & 29.28 & 29.98 & 30.05 & 30.06 & 30.34 &  & \nodata \\
  14 & 063248.65$+$043404.4 &    14.6 &   2.7 &  &  {\em  20.0 } &
{\tiny $-0.3$} 1.2 {\tiny $+0.5$} & {\tiny $-0.2$} 52.9 {\tiny
$+0.2$} &  & 29.77 & 29.10 & 29.10 & 29.85 & 29.86 &  & \nodata \\
  17 & 063249.80$+$043641.9 &    10.2 &   2.2 &  &  {\tiny $-0.7$}
22.2 {\tiny $+0.4$} &  3.4  &  53.2  &  & 29.07 & 29.91 & 30.00 &
29.97 & 30.29 &  & \nodata \\
  18 & 063251.19$+$043412.1 &    29.9 &   4.4 &  &  {\tiny $-0.6$}
21.9 {\tiny $+0.4$} & {\tiny $-0.9$} 1.7 {\tiny $+2.9$} & {\tiny
$-0.4$} 53.7 {\tiny $+0.6$}
&  & 29.86 & 30.02 & 30.07 & 30.24 & 30.60 &  & \nodata \\
  21 & 063251.73$+$043621.4 &    10.4 &   2.3 &  &  {\tiny $-0.3$}
22.7 {\tiny $+0.2$} & {\em  2.0 } & {\tiny $-0.4$} 53.8 {\tiny
$+0.3$} &  &   \nodata & 30.06 & 30.32 & 30.08 &   \nodata &  &
\nodata \\
  22 & 063252.30$+$043744.3 &    15.1 &   2.9 &  &  {\tiny $-0.7$}
22.1 {\tiny $+0.4$} & {\tiny $-2.9$} 4.0 \phantom{{\tiny $-2.9$}} &
{\tiny $-0.4$} 53.3 {\tiny $+0.6$} &  & 29.36 & 30.10 & 30.16 & 30.17
& 30.43 &  & \nodata \\
\enddata

\tablecomments{Table \ref{tbl:ch5_thermal_spectroscopy} is published
in its entirety in the
electronic edition of the {\it Astrophysical Journal}.  A portion is
shown here for guidance regarding its form and content.}

\tablenotetext{a}{ For convenience {\bf columns 1\nodata4} reproduce the
source identification, net counts, and photometric significance data
from Table~\ref{tbl:ch5_primary}.  }

\tablenotetext{b}{
All spectral fits used the ``wabs(apec)'' model in {\it XSPEC} and assumed
0.3$Z_{\odot}$ abundances.
{\bf Columns 5 and 6} present the best-fit values for the column
density and plasma temperature parameters.
{\bf Column 7} presents the emission measure for the model spectrum,
assuming a distance of 1.4~kpc.
{\it Quantities in italics} were frozen in the fit.
{\it Uncertainties} represent 90\% confidence intervals.
More significant digits are used for uncertainties $<0.1$ in order to
avoid large rounding errors; for consistency, the same number of
significant digits is used for both lower and upper uncertainties.
Uncertainties are omitted when {\it XSPEC} was unable to compute them
or when their values were so large that the parameter is effectively
unconstrained.
Fits lacking uncertainties should be considered to merely be a spline
fit to the data to obtain rough estimates of luminosities; their spectral
parameter values are unreliable.
}

\tablenotetext{c}{ X-ray luminosities are presented in {\bf columns
8\nodata12}:  s = soft band (0.5\nodata2 keV); h = hard band (2\nodata8 keV); t =
total band (0.5\nodata8 keV). Absorption-corrected luminosities are
subscripted with a $c$; they are omitted when $\log N_H > 22.5$ 
cm$^{-2}$ since the soft band emission is essentially unmeasurable.}

\tablenotetext{d}{ {\bf 2T} means a two-temperature model was used.  Well-known
counterparts from Table~\ref{tbl:ch5_counterparts} are listed here for the
convenience of the reader. }
\end{deluxetable}

\begin{deluxetable}{rcccrrrccrrrcc}
\centering \rotate
\tabletypesize{\tiny} \tablewidth{0pt}
\tablecolumns{14}

\tablecaption{Stellar Counterparts \label{tbl:ch5_counterparts}}
\tablehead{
\multicolumn{2}{c}{X-ray Source} & & \multicolumn{11}{c}{Optical/Infrared Counterpart and Photometry} \\
\cline{1-2} \cline{4-14}

\colhead{RMCX} & \colhead{CXOU J} & \colhead{} & \colhead{USNO B1.0} &
\colhead{B} & \colhead{R} & \colhead{I} & \colhead{2MASS} &
\colhead{FLAMINGOS} & \colhead{J} & \colhead{H} & \colhead{K} &
\colhead{2MASS} & \colhead{Spitzer/IRAC}\\

\colhead{} & \colhead{} & \colhead{} & \colhead{} & \colhead{(mag)}
& \colhead{(mag)} & \colhead{(mag)} & \colhead{ID} & \colhead{ID} &
\colhead{(mag)} & \colhead{(mag)} & \colhead{(mag)} & 
\colhead{PhCcFlg} & \colhead{3.6$\mu$m} \\

\colhead{(1)} & \colhead{(2)} & \colhead{} & \colhead{(3)} &
\colhead{(4)} &
\colhead{(5)} & \colhead{(6)} & \colhead{(7)} &
\colhead{(8)} & \colhead{(9)} & \colhead{(10)} &\colhead{(11)} &
\colhead{(12)} & \colhead{(13)}
}

\startdata
  1 & 063232.43$+$043705.7 & & \nodata &  \nodata &  \nodata &  \nodata & \nodata & \nodata &  \nodata &  \nodata &  \nodata & \nodata\nodata & $\surd$ \\
  2 & 063232.91$+$043437.2 & & \nodata &  \nodata &  \nodata &  \nodata & \nodata & \nodata &  \nodata &  \nodata &  \nodata & \nodata\nodata & $\times$ \\
  3 & 063236.58$+$043611.4 & & \nodata &  \nodata &  \nodata &  \nodata & \nodata & \nodata &  \nodata &  \nodata &  \nodata & \nodata\nodata & $\times$ \\
  4 & 063238.08$+$043250.1 & & 0945$-$0095483 &  \nodata &  \nodata & 14.54 & \nodata & \nodata &  \nodata &  \nodata &  \nodata & \nodata\nodata & $\times$ \\
  5 & 063239.50$+$043628.9 & & \nodata &  \nodata &  \nodata &  \nodata & \nodata & 063239$+$043629 & 14.69 & 13.45 & 12.54 & \nodata\nodata & $\surd$ \\
  6 & 063239.51$+$043124.9 & & \nodata &  \nodata &  \nodata &  \nodata & \nodata & \nodata &  \nodata &  \nodata &  \nodata & \nodata\nodata & $\times$ \\
  7 & 063240.73$+$043653.3 & & \nodata &  \nodata &  \nodata &  \nodata & \nodata & \nodata &  \nodata &  \nodata &  \nodata & \nodata\nodata & $\times$ \\
  8 & 063242.33$+$043217.9 & & \nodata &  \nodata &  \nodata &  \nodata & \nodata & \nodata &  \nodata &  \nodata &  \nodata & \nodata\nodata & $\times$ \\
  9 & 063245.25$+$043206.5 & & 0945$-$0095525 & 18.41 & 16.59 & 14.86 & 06324526$+$0432074 & 063245$+$043207 & 13.68 & 12.99 & 12.79 & AAA000 & $\surd$ \\
 10 & 063246.11$+$043612.9 & & \nodata &  \nodata &  \nodata &  \nodata & \nodata & 063246$+$043610 & 14.69 & 14.29 & 14.06 & \nodata\nodata & $\surd$ \\
\enddata

\tablecomments{Table \ref{tbl:ch5_counterparts} with complete notes is
published in its entirety in the electronic edition of the {\it
Astrophysical Journal}.  A portion is shown here for guidance
regarding its form and content.}

\tablecomments{{\bf Columns 1\nodata2} reproduce the sequence number and
source identification from Table~\ref{tbl:ch5_primary} and
Table~\ref{tbl:ch5_tentative}. For convenience, [MJD95]=Massey, Johnson,
\& Degioia-Eastwood (1995), [BC02]=Bergh{\"o}fer \& Christian\ (2002),
[TASS4]=TASS Mark IV Photometric Survey of the Northern Sky
\citep{Droege06}, [VGK85]=Voroshilov et al. (1985),
[BGD93]=Block, Geballe, \& Dyson (1993), [MSX]=the complete MSX6C 
catalogue in the Galactic Plane.}

\tablenotetext{RMCX 24}{=[MJD95] 289}  
\tablenotetext{RMCX 25}{=[MJD95] 278=[TASS4] 712279=[MSX] G206.6836-01.9932; V=12.6}
\tablenotetext{RMCX 27}{=[MJD95] 265}
\tablenotetext{RMCX 32}{=[MJD95] 237=[MSX] G206.8124-02.0424}  
\tablenotetext{RMCX 42}{=[MJD95] 230}
\tablenotetext{RMCX 48}{=[MJD95] 259}
\tablenotetext{RMCX 54}{=[MJD95] 256}  
\tablenotetext{RMCX 57}{=[MJD95] 260}
\tablenotetext{RMCX 58}{=[MSX] G206.7216-01.9384}
\tablenotetext{RMCX 65}{=HD 259533=[TASS4] 663214; V=10.6, spectral
type G0}
\tablenotetext{RMCX 72}{=[BC02] 135}  
\tablenotetext{RMCX 89}{=IRAS 06306+0437=[MSX] G206.7804-01.9395}
\tablenotetext{RMCX 95}{=[TASS4] 712310=[VGK85] NGC 2244 +04 199;
V=13.2; spectral type A0V} 
\tablenotetext{RMCX 119}{=[BC02] 136}
\tablenotetext{RMCX 146}{=[MSX] G207.0058-01.9412}
\tablenotetext{RMCX 164}{=HD 46485=[BC02] 138=[TASS4] 663303;
spectral type O8}
\tablenotetext{RMCX 165}{=[MSX] G206.9965-01.8833}
\tablenotetext{RMCX 185}{=[TASS4] 663323}
\tablenotetext{RMCX 207}{=[BGD93] IRS 4}
\tablenotetext{RMCX 209}{=[BGD93] IRS 2}
\tablenotetext{RMCX 210}{=[BGD93] IRS 5}
\tablenotetext{RMCX 220}{=[MSX] G207.0302-01.8290}
\tablenotetext{RMCX 241}{=[MSX] G207.0070-01.8037}
\tablenotetext{RMCX 278}{=[MSX] G207.1335-01.8414}
\tablenotetext{RMCX 342}{=[TASS4] 663391=[VGK85] NGC 2244 +04 226;
spectral type B9V}
\tablenotetext{RMCX 352}{=[MSX] G207.1564-01.7804}
\tablenotetext{RMCX 357}{=[MSX] G207.1792-01.7878}
\tablenotetext{RMCX 381}{=[TASS4] 1472087=[VGK85] NGC 2244 +04 229;
spectral type A0:}
\tablenotetext{RMCX 387}{=[MSX] G207.1118-01.7013}

\end{deluxetable}

\begin{deluxetable}{lccccccc}
\centering 
\tablewidth{0pt}
\tablecolumns{8}

\tablecaption{X-ray Stellar Clusters and Infrared Counterparts \label{tbl:cls_comparison}}
\tablehead{
\multicolumn{4}{c}{X-ray structure} & &
\multicolumn{3}{c}{IR structure} \\  
\cline{1-4} \cline{6-8}
\colhead{RMC} & \colhead{R.A.} & \colhead{Dec.} & \colhead{Extent} & 
& \colhead{$JHK$}  &  \colhead{$Spitzer$}  & \colhead{$IRAS$} \\  
\colhead{Struc} & \colhead{(J2000)} & \colhead{(J2000)} & &  & &  \\  
}

\startdata
XA\hfill & 06:32:38 & $+$04:46:18 & $9^{\prime} \times 6^{\prime}$ & & \nodata & \nodata&  \nodata\\
\hfill XA1 & 06:32:24 & +04:50:30 & $1.5^{\prime}\times 3.5^{\prime}$&& \nodata & \nodata & 06297$+$0453 \\
\hfill XA2 & 06:32:25 & +04:39:00 & $3^{\prime}\times 1.5^{\prime}$ && \nodata & \nodata & \nodata \\
\hfill XA3 & 06:32:38 & +04:46:18 & $6^{\prime}\times 3.5^{\prime}$ && \nodata &\nodata & \nodata \\
&&&&&&& \\
XB\hfill   & 06:33:15 & +04:35:11 & $4^{\prime} \times 8^{\prime}$ && \nodata &\nodata  &\nodata  \\
\hfill XB1 & 06:32:54 & +04:37:00 & $3^{\prime}\times 3^{\prime}$ && \nodata & \nodata & \nodata \\
\hfill XB2 & 06:33:14 & +04:31:30 & $2^{\prime}\times 2^{\prime}$ && \nodata & C & \nodata \\
\hfill XB3 & 06:33:15 & +04:35:11 & $3.5^{\prime}\times 3.5^{\prime}$ && PL 2 & \nodata & 06306$+$0437\\
&&&&&&& \\
XC\hfill   & 06:34:12 & +04:28:15 & $8^{\prime}\times 8^{\prime}$ && \nodata & \nodata & \nodata \\
\hfill XC1 & 06:34:12 & +04:28:15 & $5^{\prime}\times 6^{\prime}$ && PL 4 & E & 06314$+$0427 \\
\hfill XC2 & 06:34:24 & +04:22:00 & $4^{\prime}\times 4^{\prime}$ && PL 5, REFL 8 & E & 06317$+$0426 \\
\enddata

\tablecomments{ $JHK$ clusters are from \citet[PL][]{PL97} and \citet[REFL][]{RomanZuniga08}. $Spitzer$ clusters are from \citet{Poulton08}.} 
  
\end{deluxetable}

\begin{deluxetable}{lcccccccccccc}
\centering
\tabletypesize{\small}
\tablewidth{0pt}
\tablecolumns{13}

\tablecaption{Star counts in X-ray clusters \label{tbl:cls_counts}}

\tablehead{

\multicolumn{5}{c}{X-ray population} &&
\multicolumn{4}{c}{NIR pop} &&
\multicolumn{2}{c}{MIR pop} \\ 
\cline{1-5} \cline{7-10} \cline{12-13}

\colhead{RMC} & \colhead{$N_X$} & \colhead{$N_{X,h}$}  &  
\colhead{$f_{X,h}$}  &\colhead{$N_{tot}$} &&
\colhead{$N_{NIR}$} & \colhead{$N_{d,NIR}$}&  \colhead{$f^{NIR}_{d,NIR}$} & \colhead{$f^{NIR}_{d,tot}$} &&
\colhead{$N_{MIR}$} & \colhead{$f^{MIR}_{d,tot}$} \\

\colhead{Struc} &&&& \colhead{Total} && &&& && & \\

\colhead{1} & \colhead{2} & \colhead{3} & \colhead{4} & \colhead{5} && 
\colhead{6} & \colhead{7} & \colhead{8} & \colhead{9} && 
\colhead{10} & \colhead{11} 

}

\startdata
XA          &    91 & 27 &    30\% & 300 && 61 &  6 &     10\% &    ~2\% &&  32 &    11\% \\
\hfill XA1 &   ~6 & ~2 &    33\% & ~20 && ~3 &  0 &    ~0\% &    ~0\% &&  ~2 &    ~6\% \\
\hfill XA2 &   ~4 & ~1 &    25\% & ~15 && ~3 &  0 &    ~0\% &    ~0\% &&  ~0 &    ~0\% \\
\hfill XA3 &   29 & ~5 &    17\% &  100 && 23 &  4 &    17\% &    ~4\% &&  11 &    11\% \\
&&&&&&&&&&&& \\
XB          &   52 &  24 &    46\% &  200 && 37 &  8 &    22\% &   ~4\%  &&  16 &    ~8\% \\
\hfill XB1 &   11 & ~6 &    55\% &  ~40 && ~5 &  0 &   ~0\% &    ~0\% &&  ~0 &    ~0\% \\
\hfill XB2 &   ~5 & ~4 &    80\% &  ~20 && ~3 &  1 &   33\% &    ~5\% &&  ~8 &    40\% \\
\hfill XB3 &   14 & ~7 &    50\% &  ~20 && 11 &  4 &    36\% &    20\% &&  ~1 &   ~5\% \\
&&&&&&&&&&&& \\
XC          & 160 & 66 &    41\% &  800 &&131 &  9 &    ~7\% &    ~1\% &&  62 &   ~8\% \\
\hfill XC1 &   67 & 22 &   33\% &   340 &&  57 &  5 &   ~9\% &    ~1\% &&  17 &   ~5\% \\
\hfill XC2 &   17 & 13 &   76\% &   ~90 &&  14 &  1 &   ~7\% &    ~1\% &&  16 &   18\% \\
&&&&&&&&&&&& \\
Distributed & 250 & 105 & 42\% & 1100 && \nodata & \nodata & \nodata & \nodata && \nodata & \nodata \\
\enddata

\tablecomments{
{\bf Column 1}: X-ray region defined in Table \ref{tbl:cls_comparison}.  The Distributed values represent the total sources in the X-ray fields minus those in the X-ray regions.  \\
{\bf Column 2}: $N_{X}$ = Number of X-ray sources in the region 
defined by the contours in Figure \ref{fig:ch5_ssd}$a$ and $b$ \\
{\bf Column 3}: $N_{X,h}$ = Number of hard X-ray sources ($MedE>2.0$ keV) contained in the region\\
{\bf Column 4}: $f_{X,h} = N_{X,h}/N_{X}$ = Fraction of hard X-ray stars with respect to all X-ray stars.  \\
{\bf Column 5}: $N_{tot}$ = Estimated total stellar population in the region based on the X-ray sample; see \S \ref{sec:sensitivity} \\
{\bf Column 6}: $N_{NIR}$ = Number of NIR stars in the X-ray region 
from \citet{RomanZuniga08} \\
{\bf Column 7}: $N_{d,NIR}$ = Number of $K$-band excess disky stars in the X-ray region from \citet{RomanZuniga08} \\
{\bf Column 8}: $f^{NIR}_{d,NIR} = N_{d,NIR}/N_{NIR}$ = Fraction of $K$-band excess disky stars with respect to the NIR sample in the X-ray region \\
{\bf Column 9}: $f^{NIR}_{d,tot} = N_{d,NIR}/N_{tot}$ = Fraction of $K$-band excess disky stars with respect to the estimated total stellar population in the X-ray region \\
{\bf Column 10}: $N_{MIR}$ = Number of MIR-band excess disky stars in the X-ray region from \citet{Poulton08} \\
{\bf Column 11}:  $f^{MIR}_{d,NIR}$ = $N_{MIR}/N_{tot}$ = Fraction of MIR-band excess disky stars with respect to the estimated total stellar population in the X-ray region  
}

\end{deluxetable}

\begin{figure}[h]
\centering 
\plotone{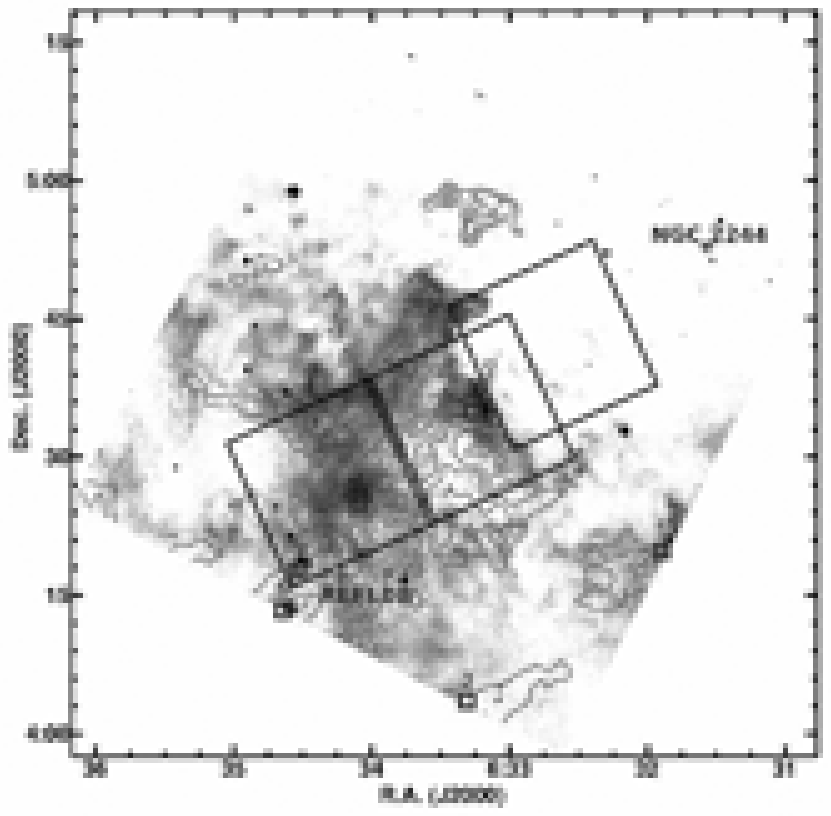}
\caption{A $1^{\circ}\times 1^{\circ}$ MSX 8.3~$\mu$m image
overlaid with $^{12}$CO(J=1-0) emission contours \citep{Heyer06},
outlining the distribution of the ISM in the RMC. The small squares with
sequence numbers denote the locations of \citet{PL97} IR clusters
(except cluster PL7, which is off the MSX image). The large boxes
outline the ACIS-I 17\arcmin$\times$17\arcmin\/ pointings. The embedded clusters are clearly
associated with the CO clumps and the dust emission
peaks. The center of the Rosette Nebula's ionizing cluster NGC 2244 
is marked. \label{fig:msx}}
\end{figure}
\begin{figure}[h]
\centering
\plotone{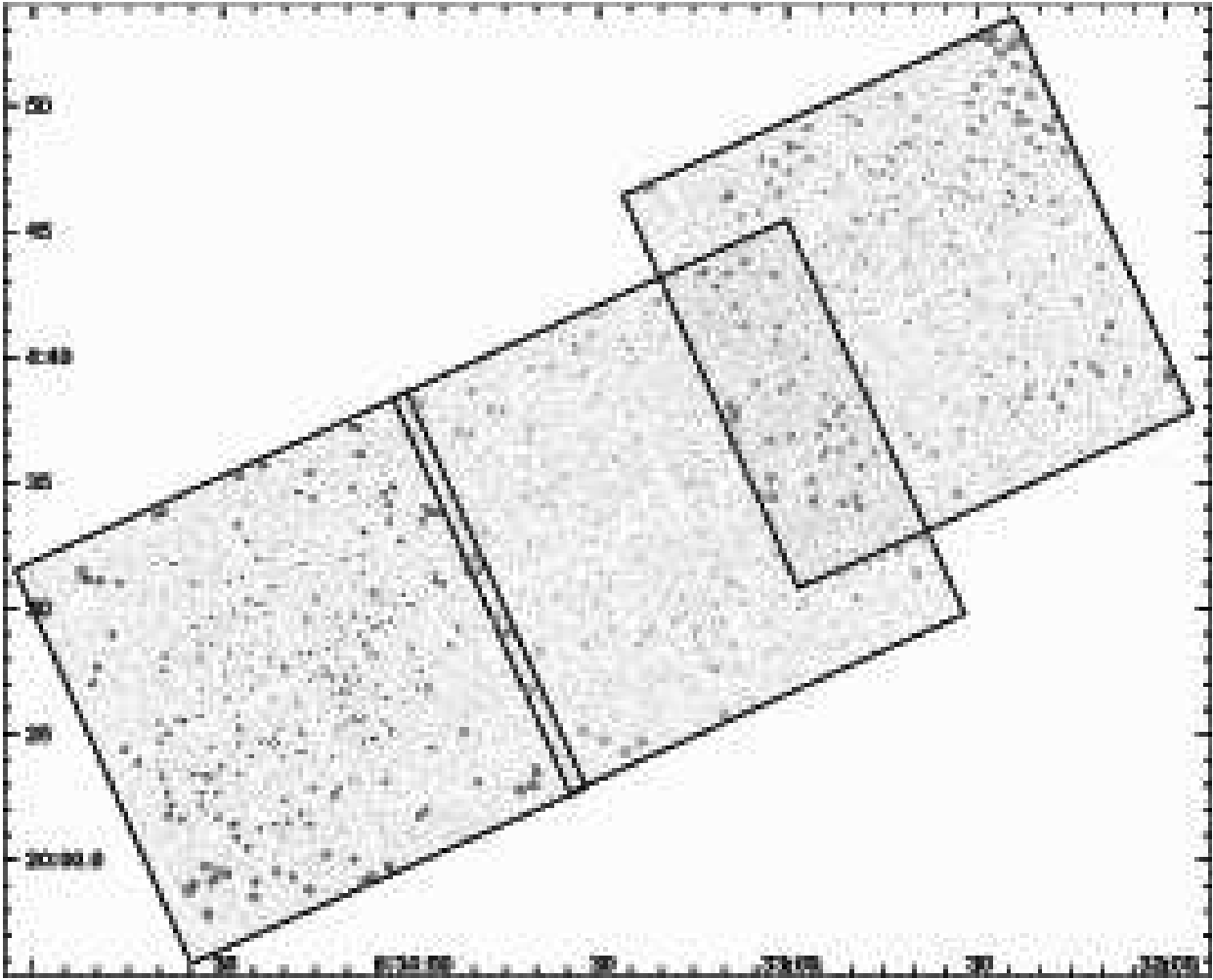}
\caption{A {\em Chandra}
mosaic of the RMC region (0.5-8 keV) overlaid with source extraction
regions derived by {\it ACIS Extract} from appropriate {\em Chandra} PSFs. Note
that these PSFs grow larger in size with increasing off-axis
angles. Sources covered by different pointings are represented by
polygons of different color (red: ObsID 1875; green: ObsID 1876; blue:
ObsID 1877). Note that some sources have two extraction regions as
they lie in the overlapping regions of two
observations.\label{fig:mosaic}}
\end{figure}
\begin{figure}[h]
\centering
\plotone{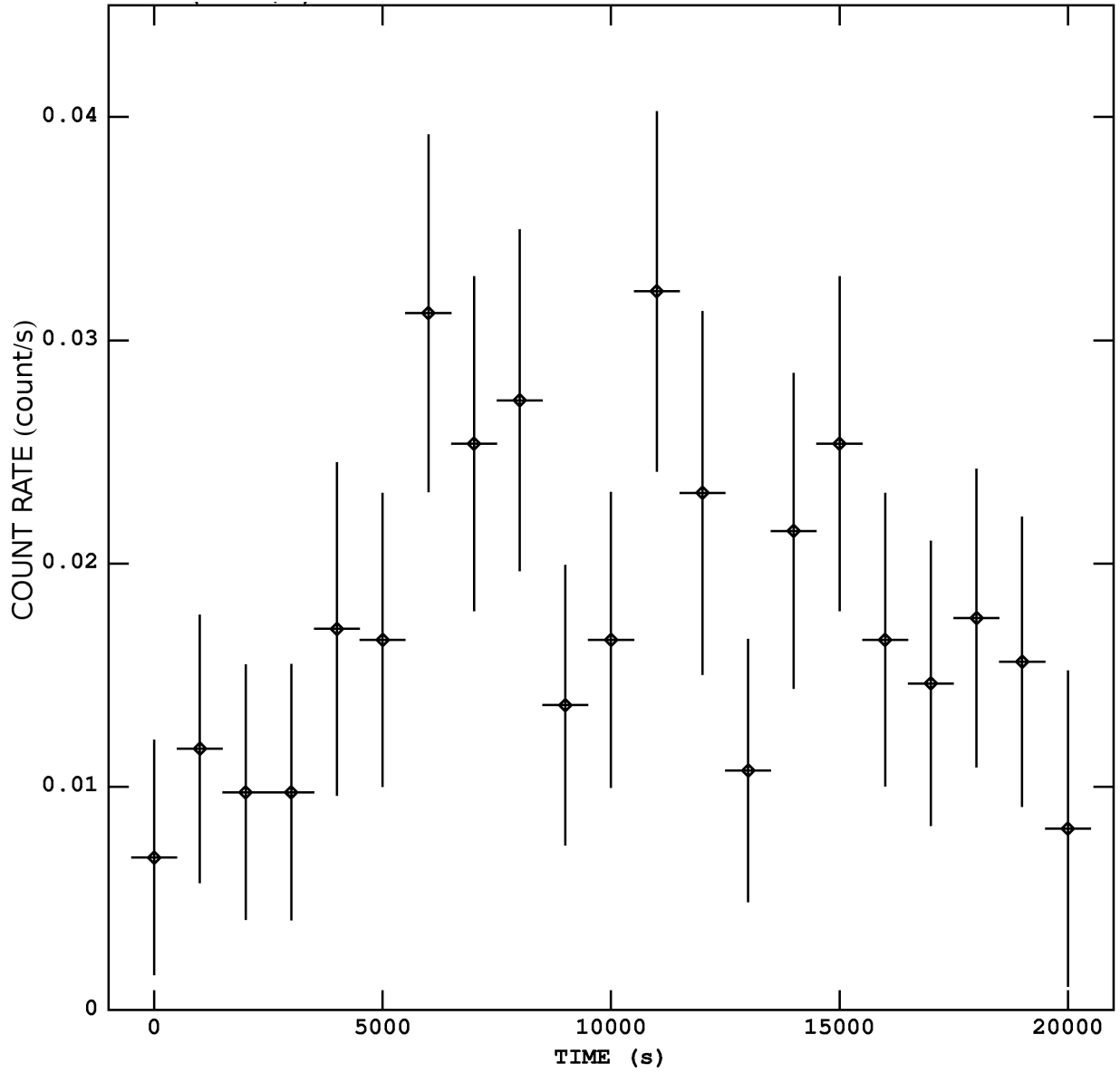}
\caption{The variable X-ray
light curve of ACIS source \#119. The $K-S$ test finds a significant
variation ($P_{KS}\le 0.005$), but no flaring characterized by fast
rise and slow decay as commonly seen in PMS stars is
present. \label{fig:lcurve}}
\end{figure}
\begin{figure}[h]
\centerline{\includegraphics[angle=-90,width=0.8\textwidth]{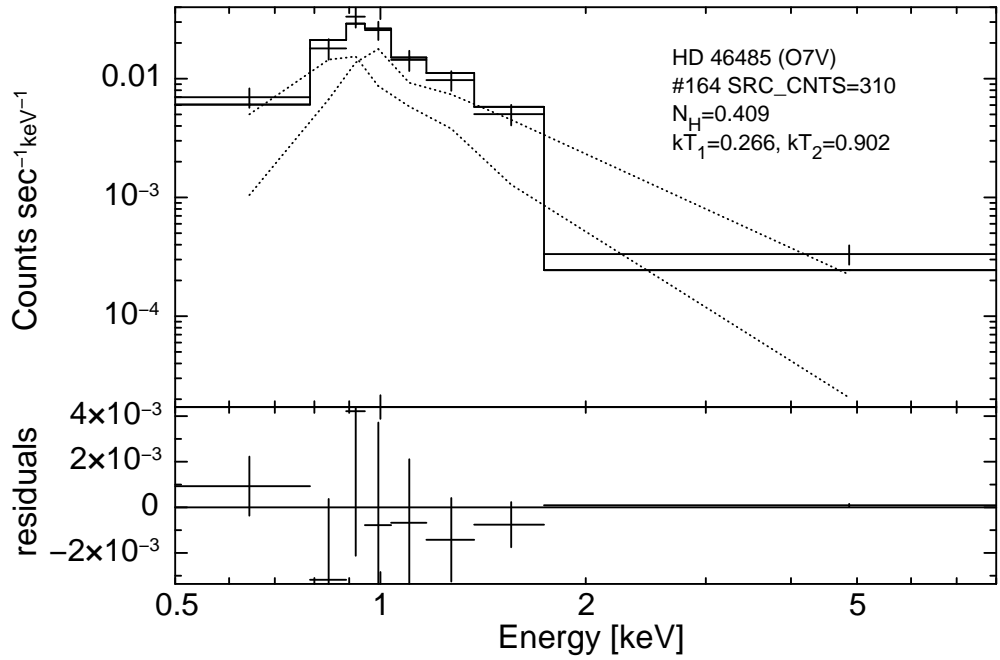}}
\caption{The X-ray spectrum and the
spectral fit for the O7 star (\#164). The best fit adopts a
two-temperature thermal plasma model, with soft $kT$ and a low
absorbing column $N_H\sim 4.1\times 10^{21}$ cm$^{-2}$. 
\label{fig:spectra}} 
\end{figure}
\begin{figure}[h]
\centering 
\plotone{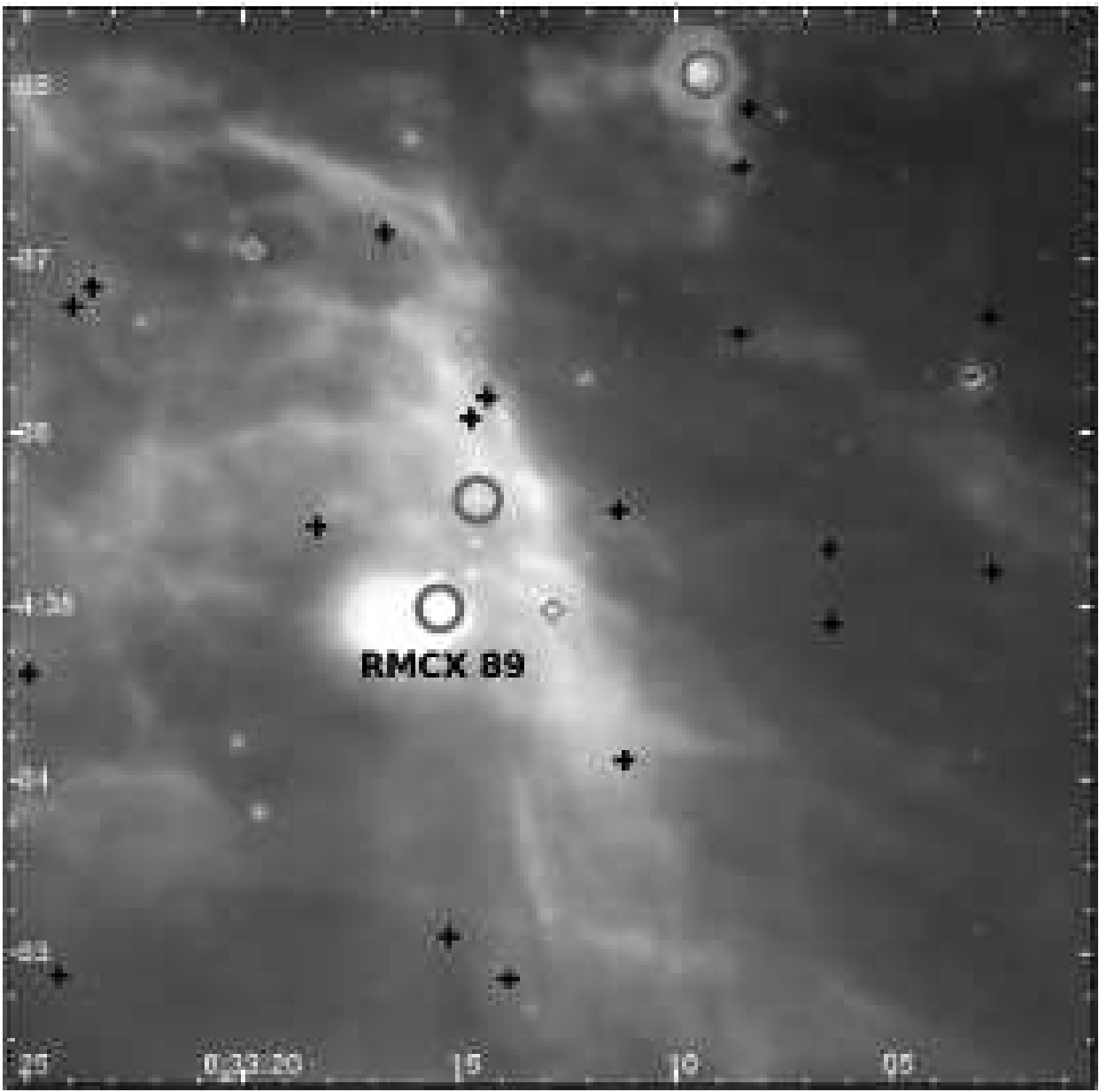}
\caption{Three-color {\em Spitzer}/IRAC+MIPS image of the
neighborhood region of source \#89. Red, green, and blue represents
24$\mu$m, 8$\mu$m, and 3.6$\mu$m emission, respectively. Overlaid
symbols are the X-ray-selected stars that are Class I (magenta
circles), Class II (green diamonds), and Class III (black crosses) based on
their NIR colors. \label{fig:source89}}
\end{figure}

\begin{figure}[h]
\centering 
\includegraphics[width=0.4\textwidth]{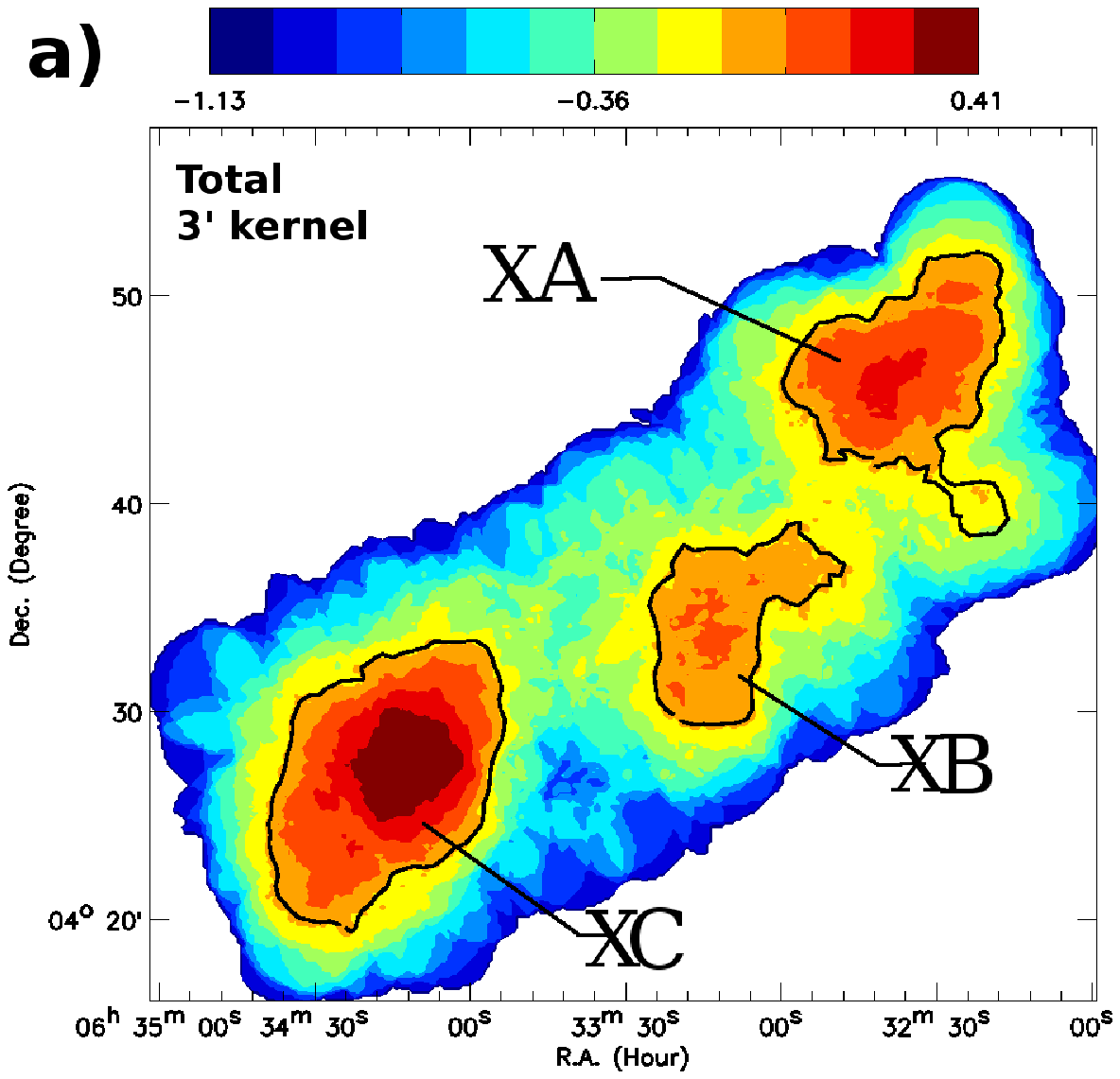}
\includegraphics[width=0.4\textwidth]{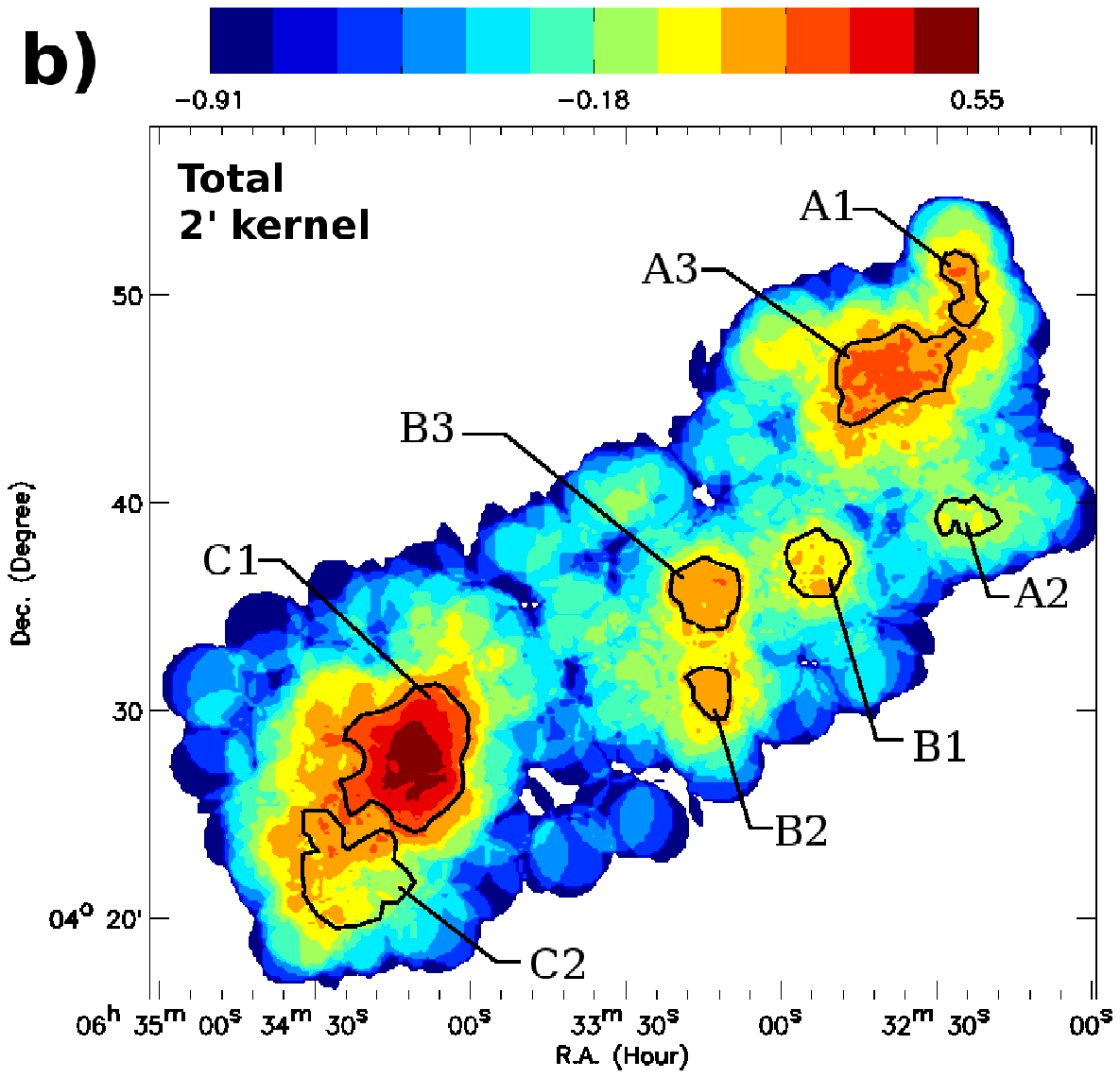}
\includegraphics[width=0.4\textwidth]{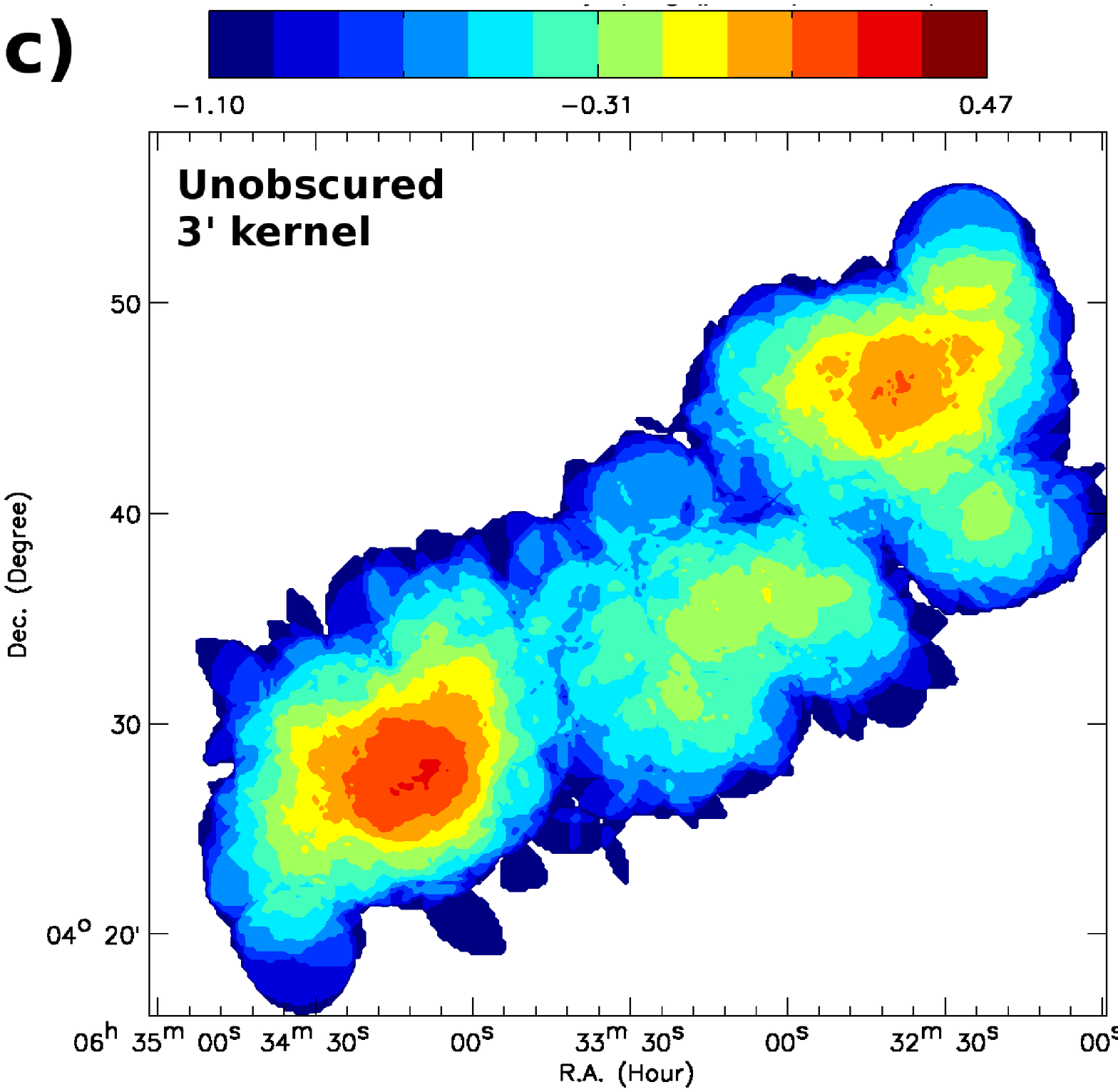}
\includegraphics[width=0.4\textwidth]{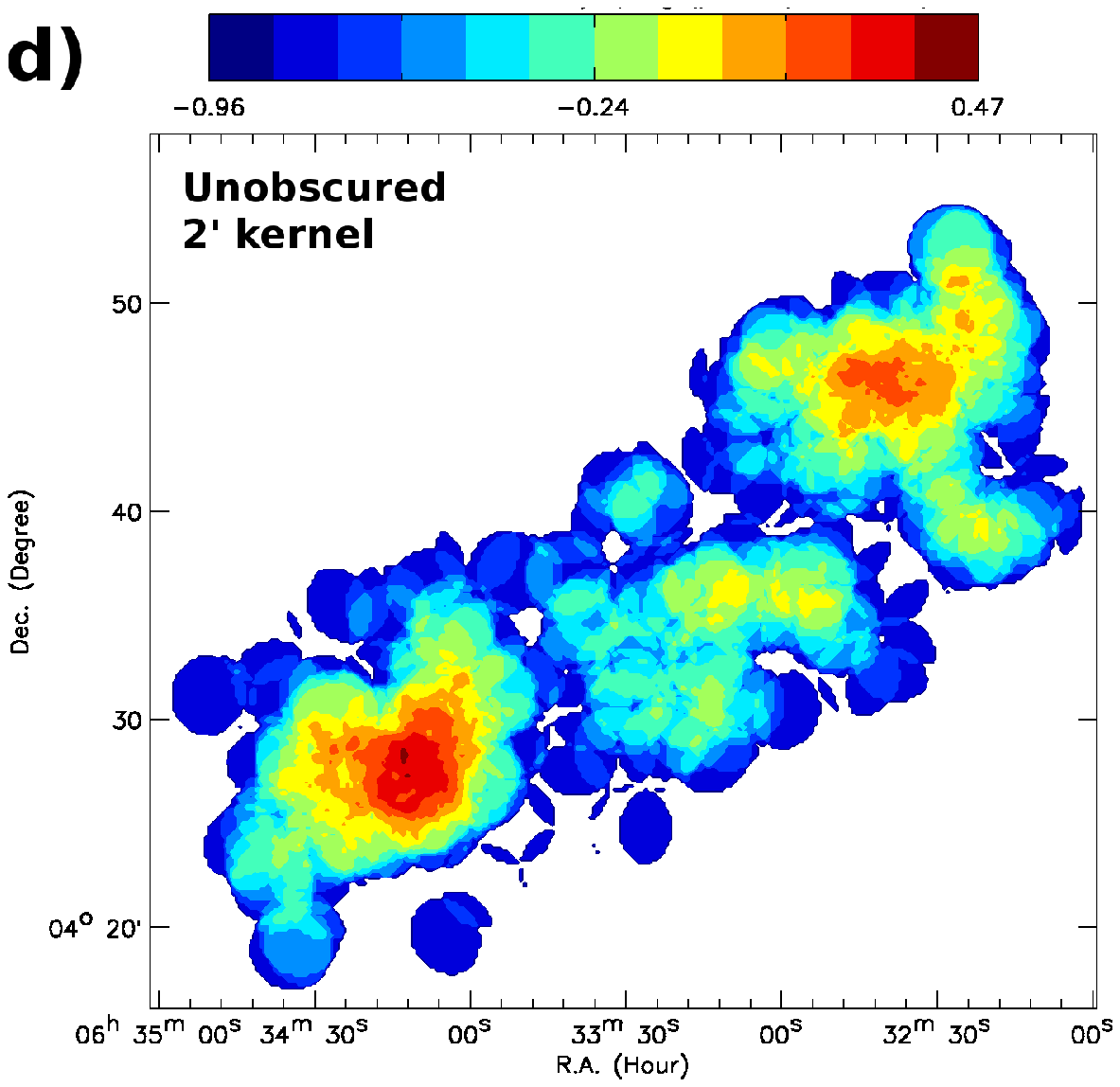}
\includegraphics[width=0.4\textwidth]{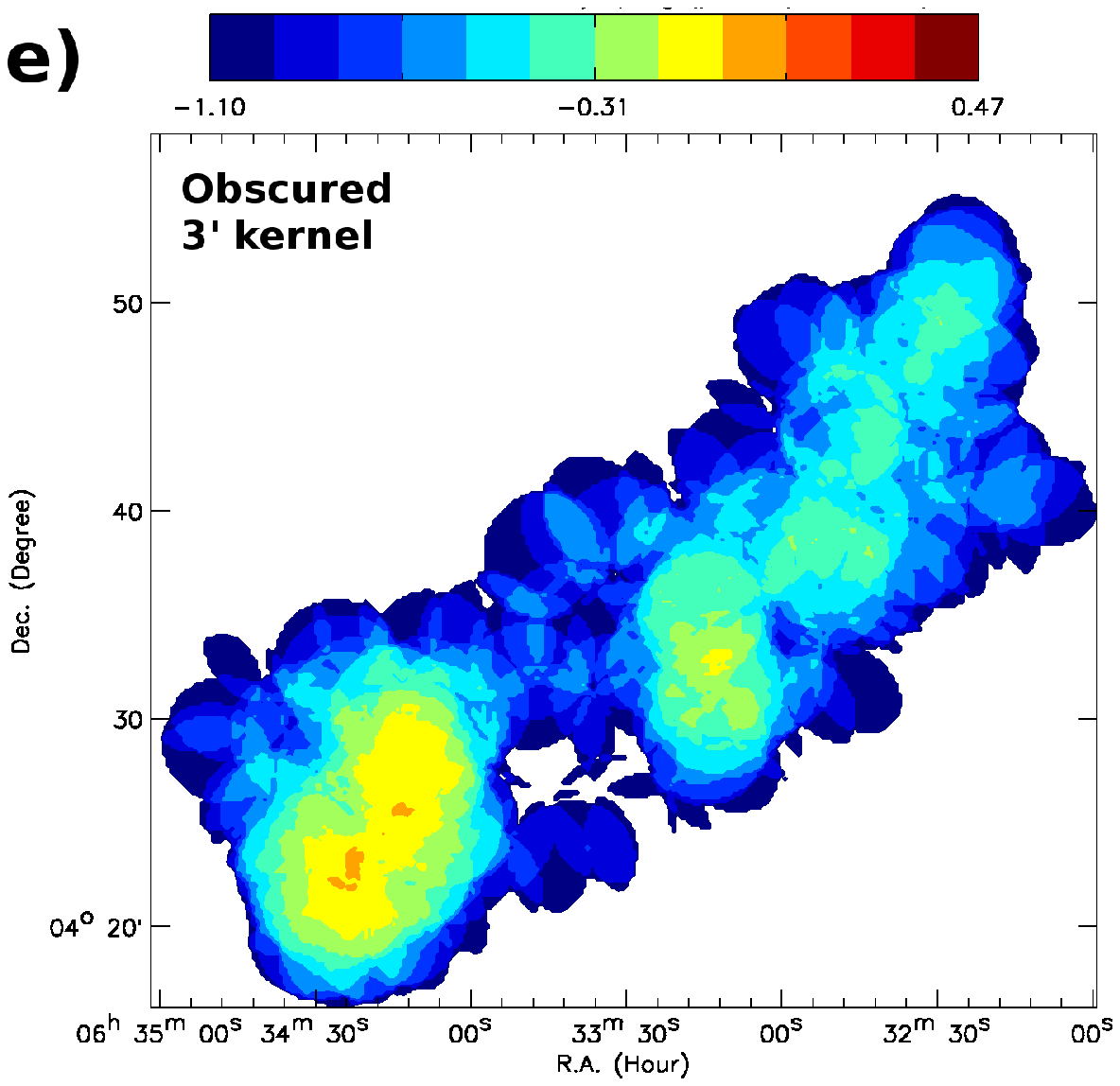}
\includegraphics[width=0.4\textwidth]{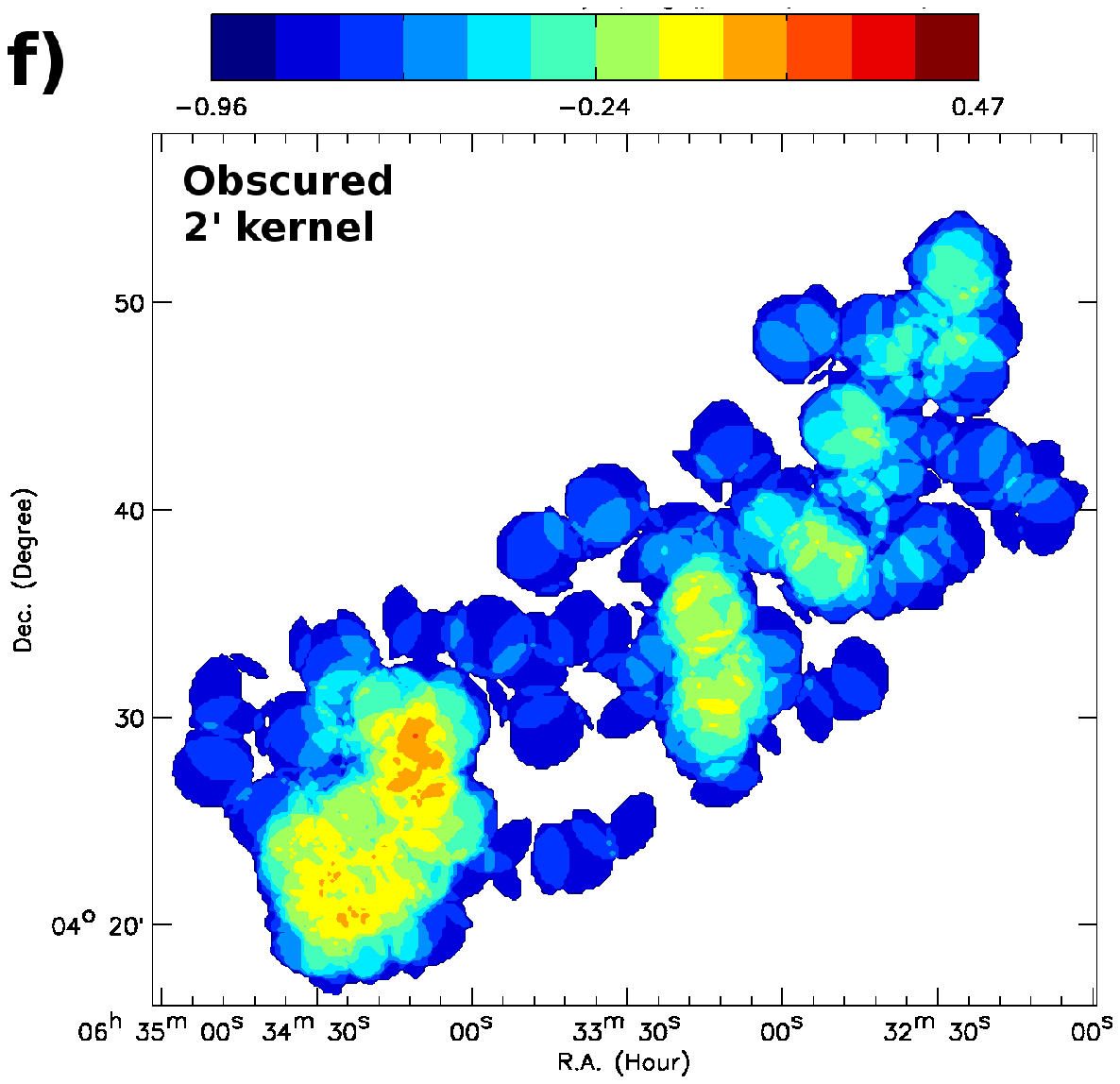}

\caption{(a) The stellar surface density ($\log$\#stars arcmin$^{-2}$) map for all RMC
sources smoothed with a 3\arcmin\/ radius kernel. (b) The same as (a)
but using a 2\arcmin\/ radius kernel. (c) The same as (a) but for the
unobscured population. (d) The same as (b) but for the unobscured
population. (e) The same as (a) but for the obscured population. (f)
The same as (b) but for the obscured population.  The density scaling
is the same between maps of the unobscured population and obscured
population for a fair comparison.\label{fig:ch5_ssd}}
\end{figure}
\begin{figure}[h]
\centering 
\plotone{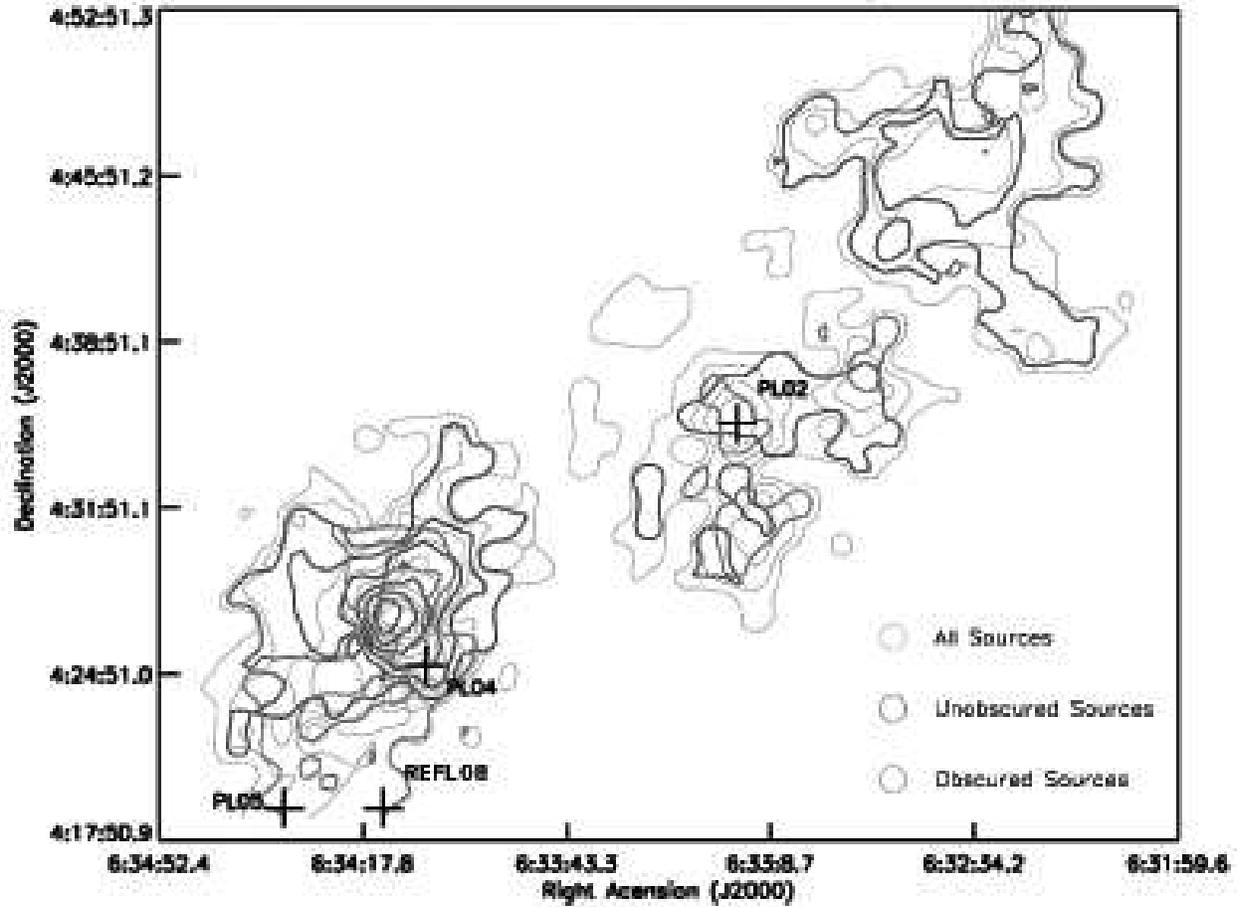}
\caption{Surface density contours based on
10th neighbor distances for the {\em Chandra} RMC sources. The green
contours are densities for all candidates. The blue lines are contours
for unobscured sources ($medE\le 2.0$keV), and the red lines are
contours for obscured sources ($medE > 2.0$keV). All contours follow
the levels of density for obscured sources, starting at 0.057 stars
arcmin$^{-2}$ with steps of 0.45 stars
arcmin$^{-2}$.\label{fig:chandra_contour}}
\end{figure}
\begin{figure}[h]
\centering 
\plotone{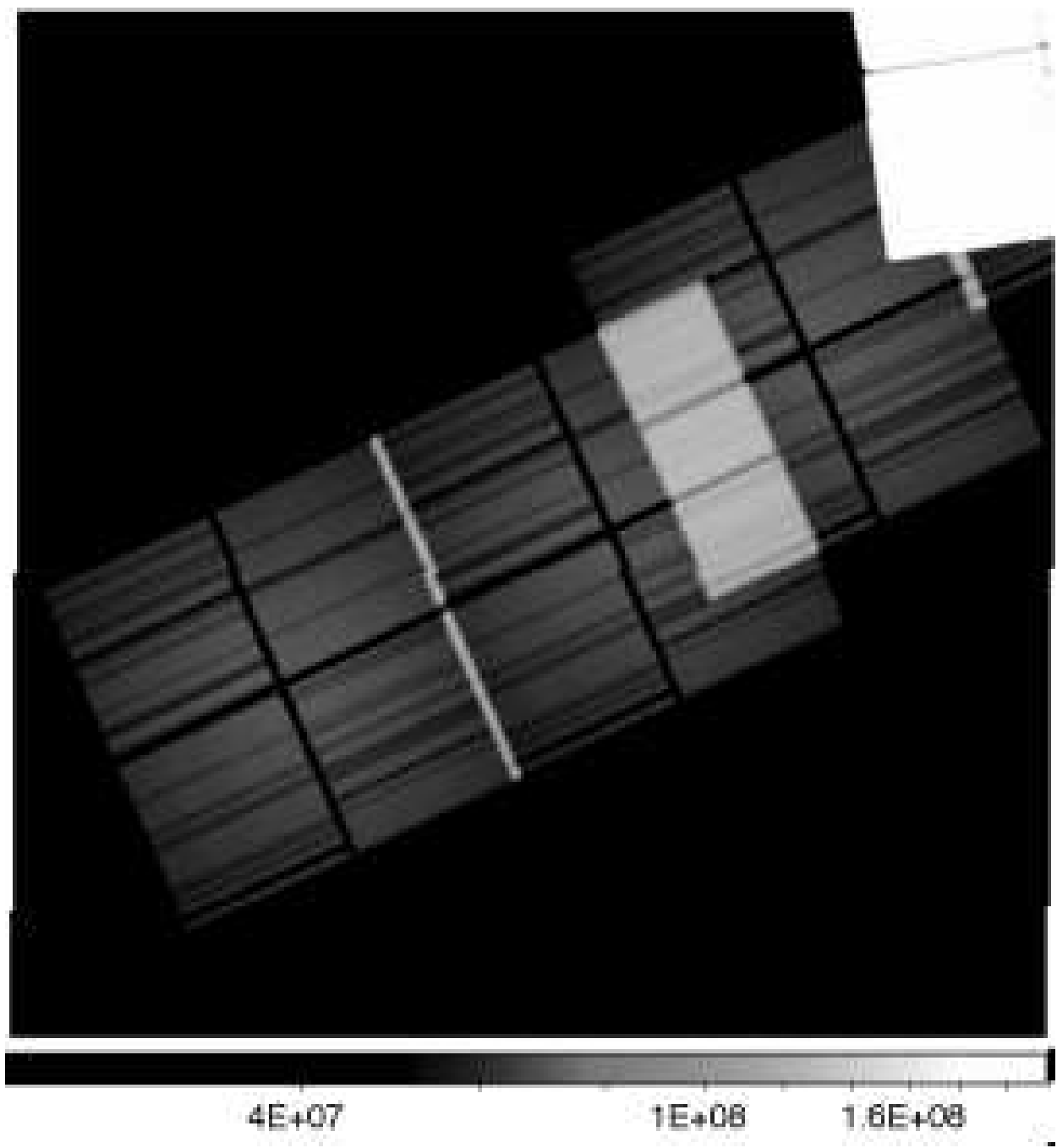}
\caption{The combined {\em Chandra} exposure maps for the RMC observations. The unit is cm$^2$ s. Note the enhanced sensitivity in the regions where individual observations overlap. \label{fig:expmap}}
\end{figure}
\begin{figure}[h]
\centering 
\includegraphics[width=0.54\textwidth]{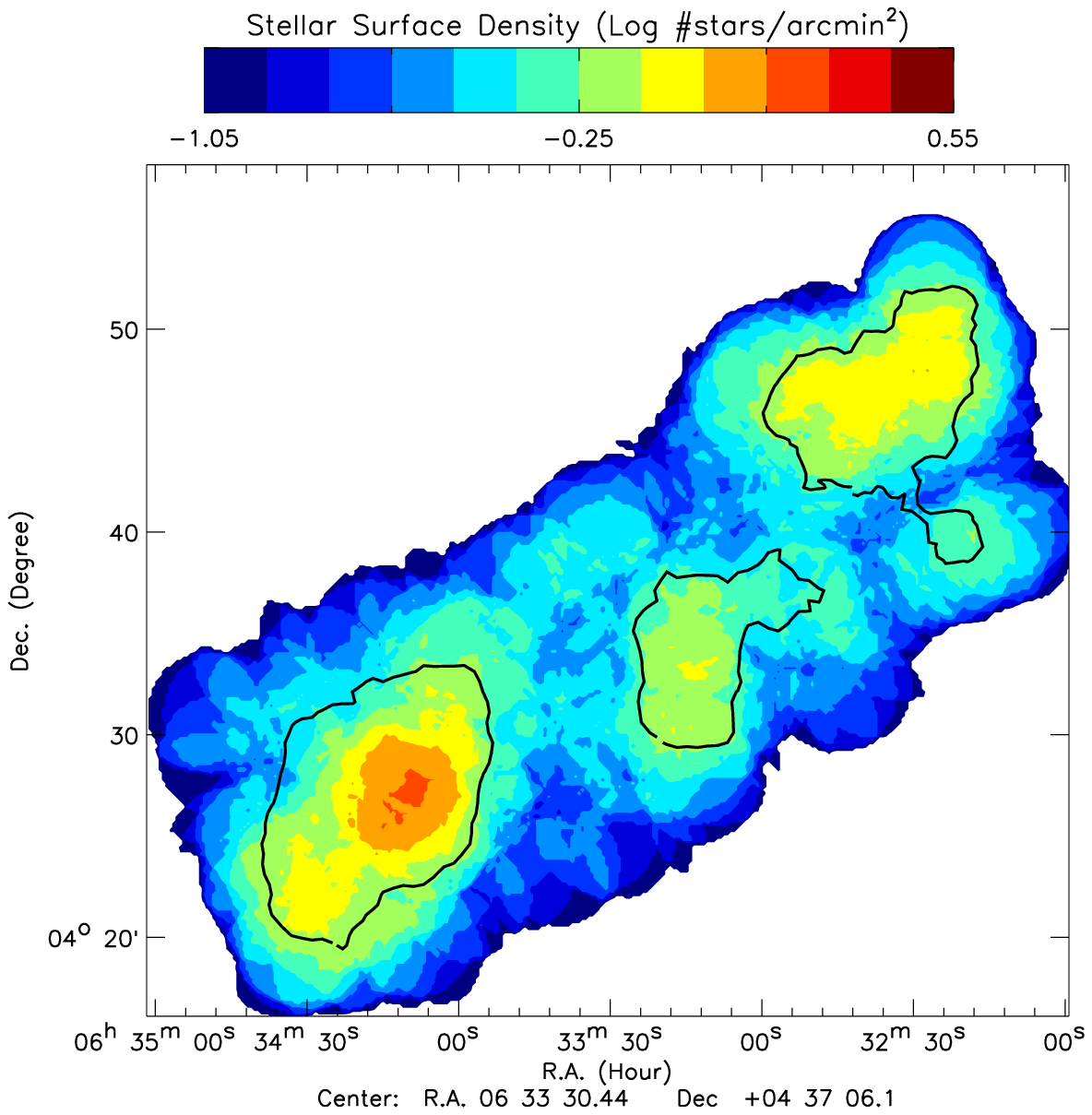}
\includegraphics[width=0.54\textwidth]{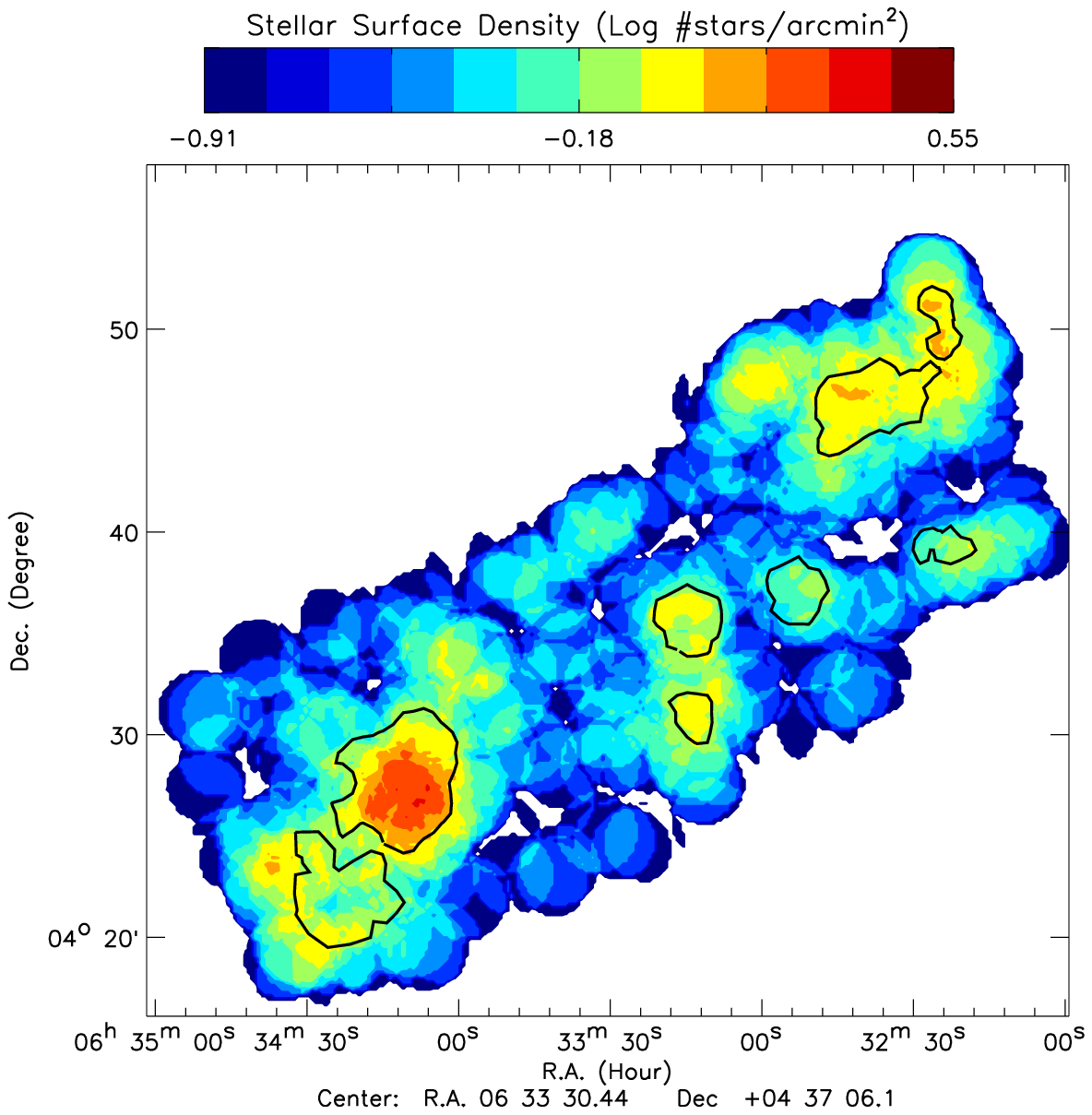}
\caption{The stellar surface density ($\log$\#stars arcmin$^{-2}$) map for a sub-sample of RMC sources (photon flux $> 1\times 10^{-6}$ count cm$^{-2}$ s$^{-1}$) smoothed with a 3\arcmin\/ radius kernel (top panel) and a 2\arcmin\/ radius kernel (lower panel). The cluster regions and substructure regions from Figure~\ref{fig:ch5_ssd} are shown.\label{fig:uniform_contour}}
\end{figure}
\begin{figure}[h]
\centering 
\includegraphics[width=2.6in]{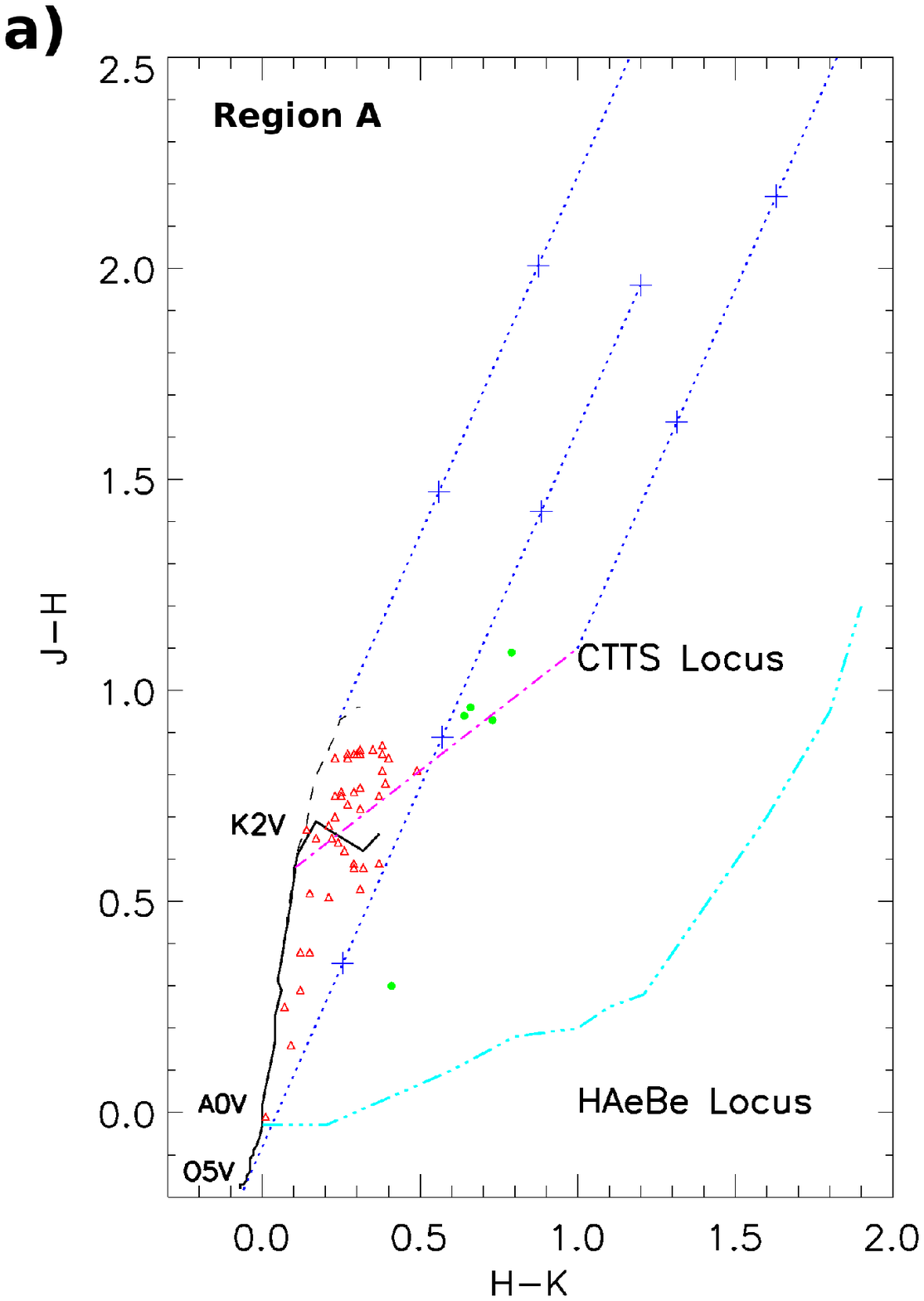}
\includegraphics[width=2.4in]{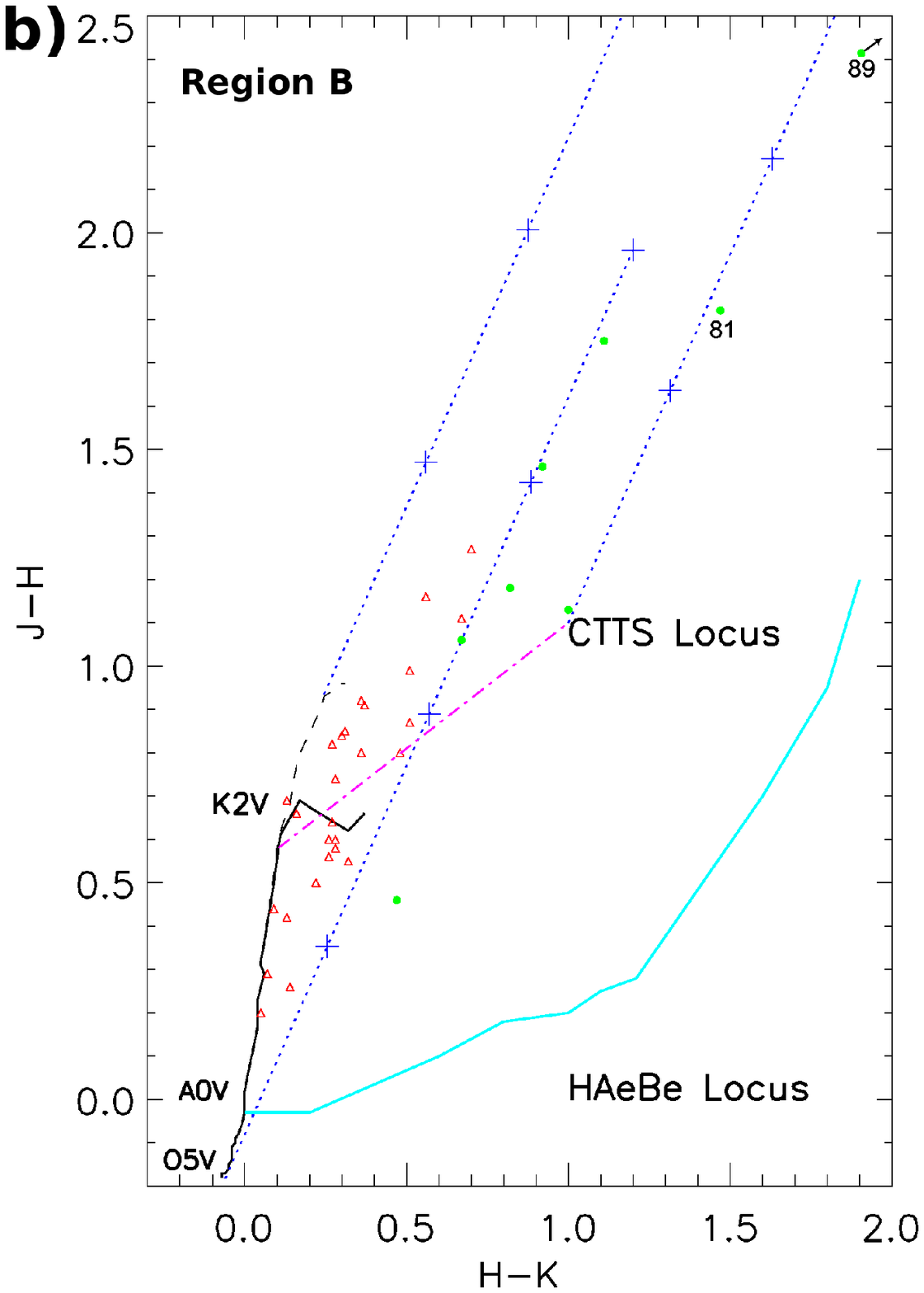}
\centerline{\includegraphics[width=2.4in]{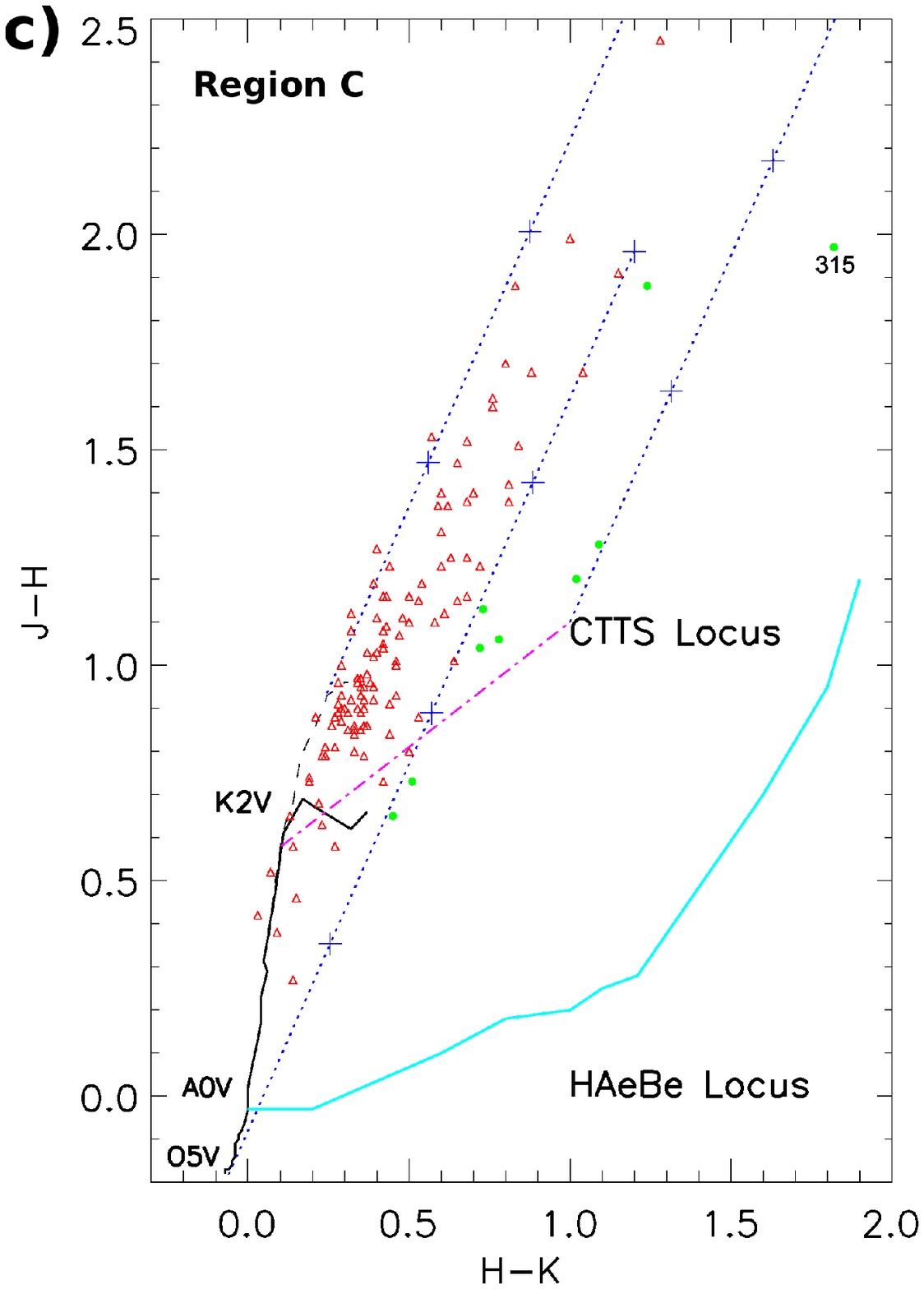}}
\caption{NIR
$J-H$ vs.\ $H-K$ color-color diagram for {\em Chandra} stars with
high-quality $JHK$ photometry (error in both $J-H$ and $H-K$ colors
$<0.1$ mag) in regions A, B, and C as defined in
Figure~\ref{fig:chandra_contour}. The green circles and red triangles
represent sources with and without significant $K$-band excess ($E
(H-K)> 2\sigma (H-K)$) respectively. The black solid and long-dash lines denote the
loci of MS stars and giants, respectively, from Bessell \& Brett
(1988). The purple dash dotted line is the locus for classical T Tauri
stars from Meyer et al. (1997), and the cyan solid line is the locus
for HAeBe stars from Lada \& Adams (1992). The blue dashed lines
represent the standard reddening vector with crosses marking every
$A_V=5$ mag.\label{fig:ch5_ccd}}
\end{figure}
\begin{figure}[h]
\centering 
\includegraphics[width=2.7in]{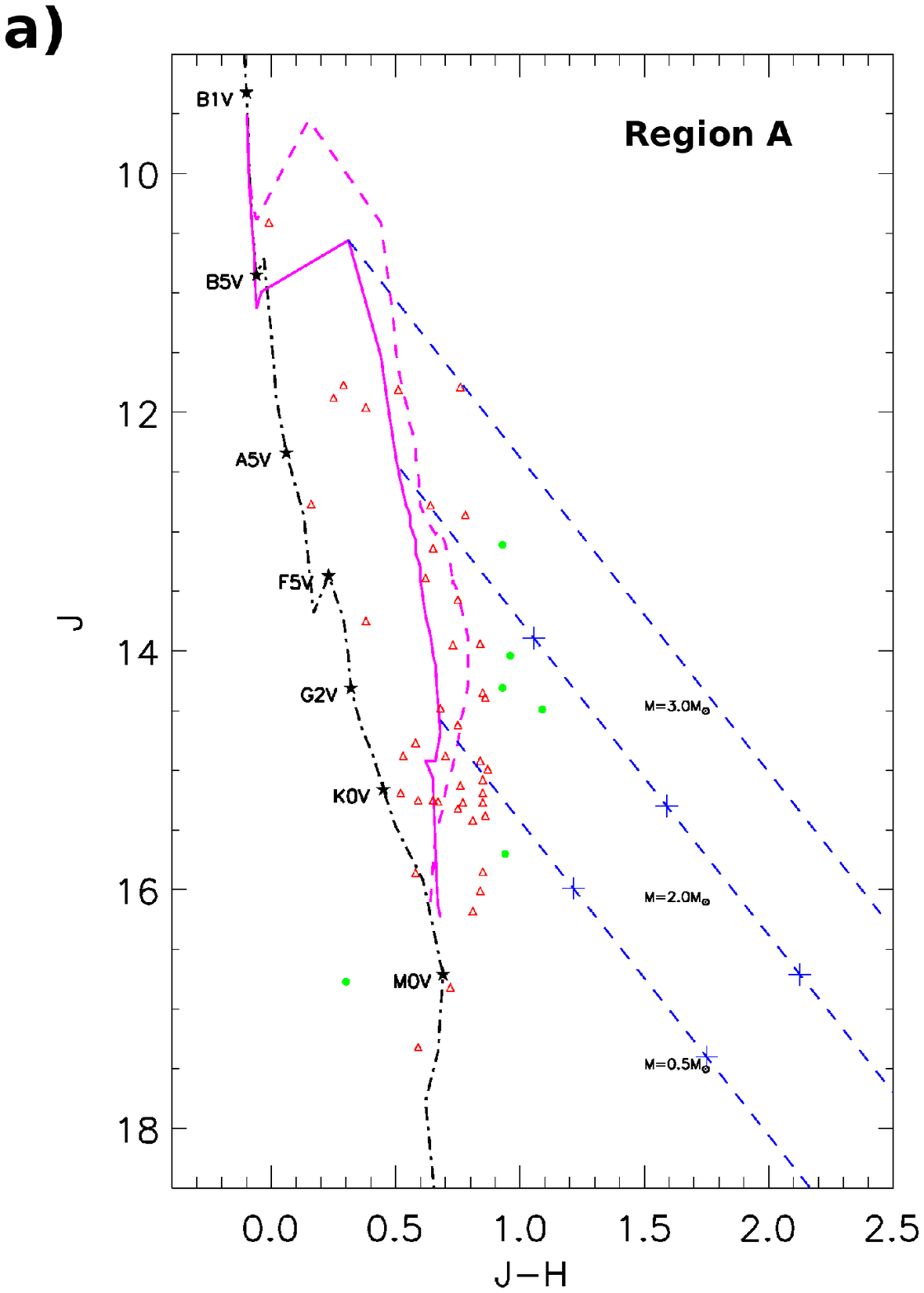}
\includegraphics[width=2.7in]{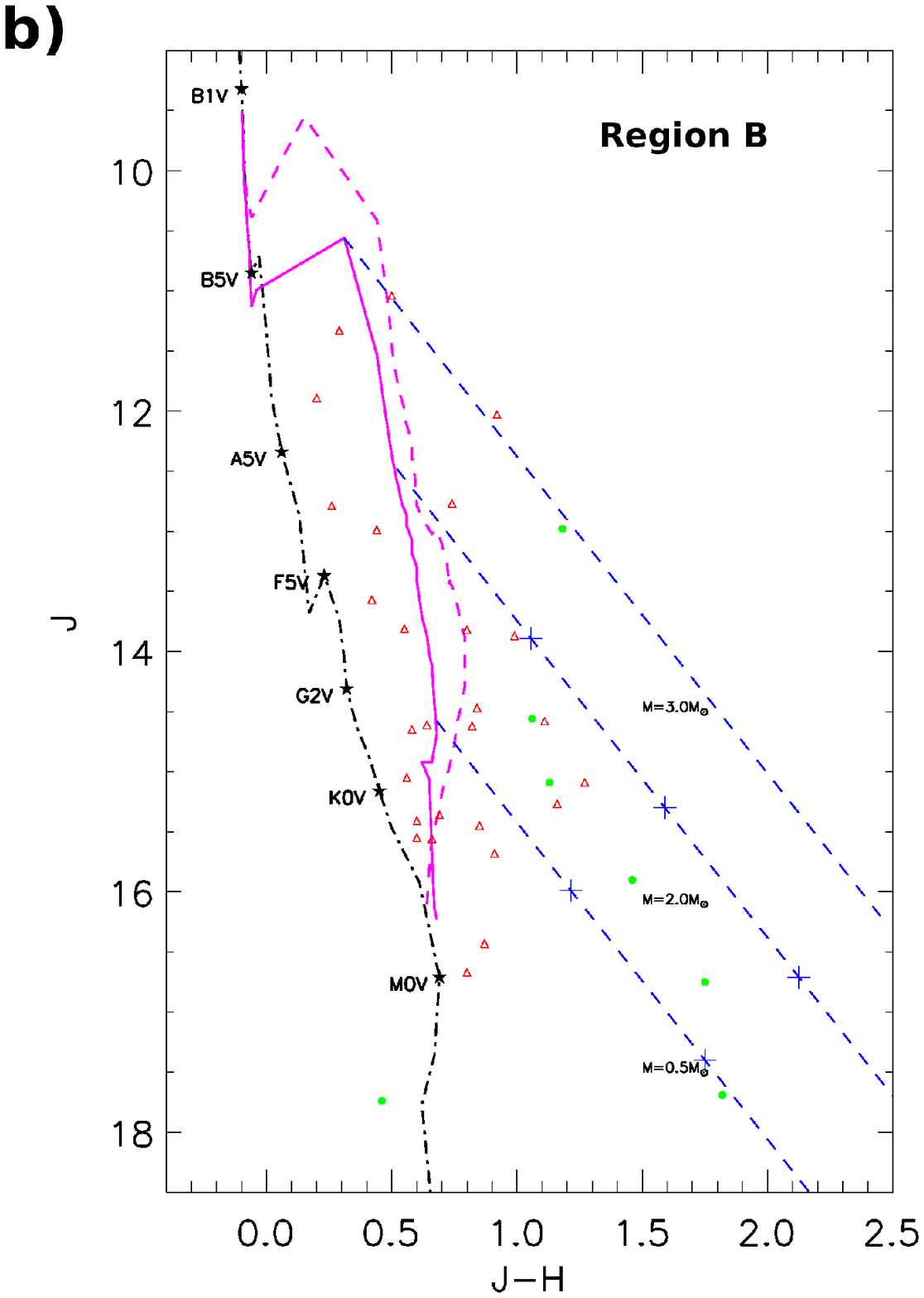}
\centerline{\includegraphics[width=2.7in]{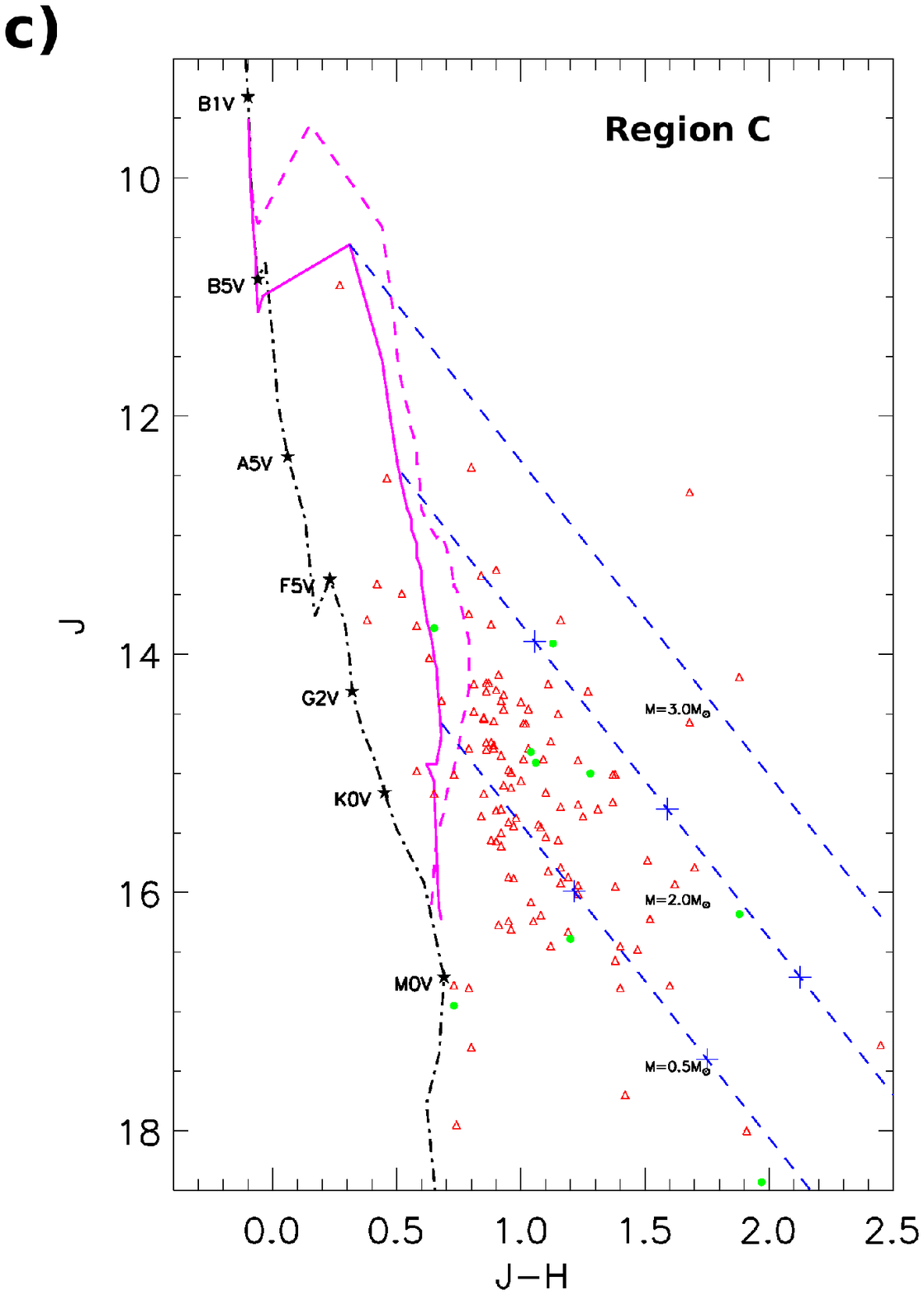}}
\caption{NIR $J$ vs. $J-H$ color-magnitude diagram using the
same sample and symbols as Figure~\ref{fig:ch5_ccd}. The purple solid
line and dashed line is the 2~Myr isochrone and the 1~Myr isochrone
for PMS stars from \citet{Siess00}, respectively. Both isochrones are shown here 
because of the uncertainty in stellar ages. The dash dotted line
marks the location of Zero Age Main Sequence (ZAMS) stars. The blue
dashed lines represent the standard reddening vector with asterisks
marking every $A_V=5$ mag and the corresponding stellar masses are
marked.\label{fig:ch5_cmd}}
\end{figure}
\begin{figure}[h]
\centering 
\plotone{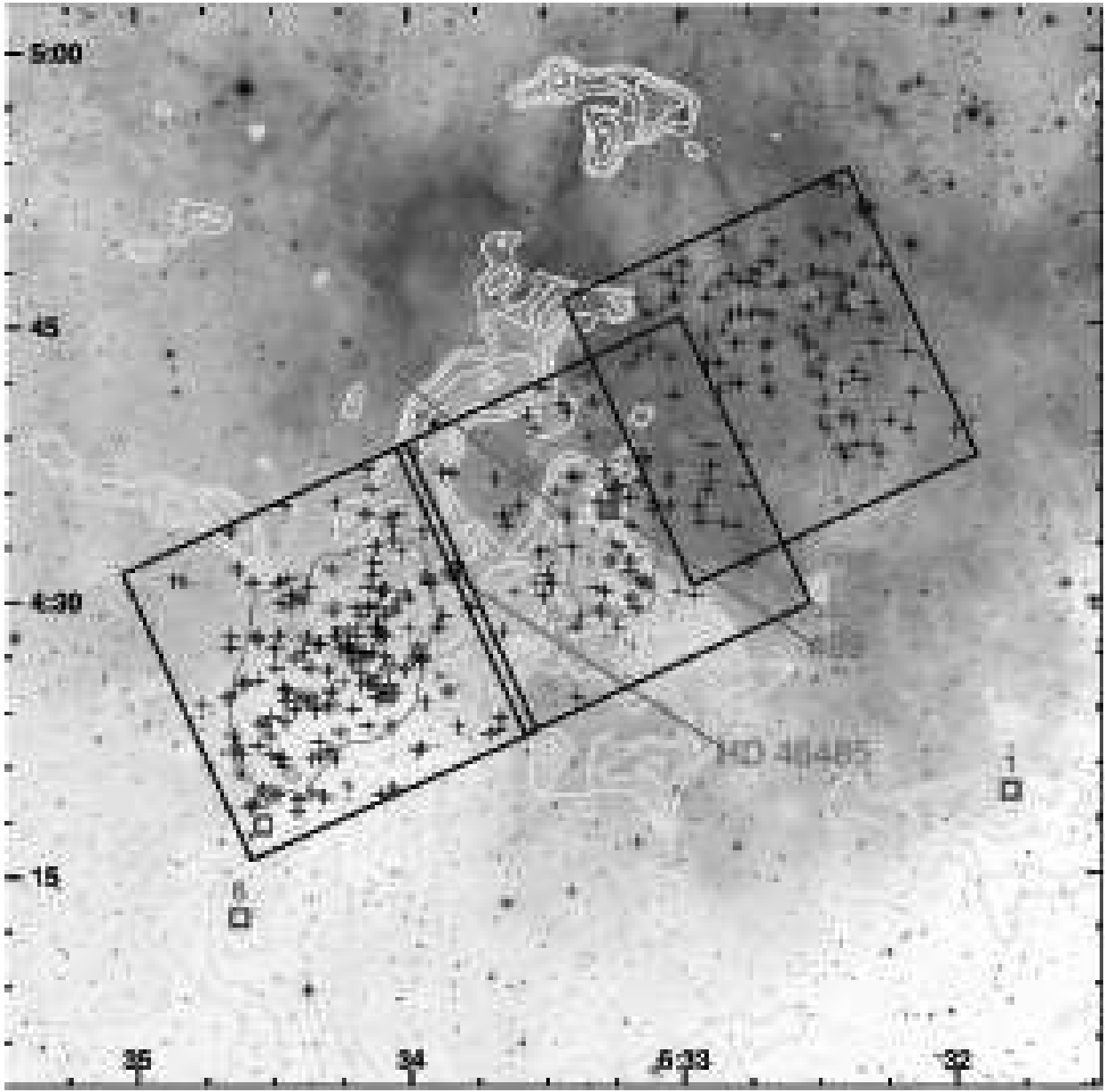}
\caption{Spatial distribution of all the IR counterparts to our
X-ray sources. The dense molecular cloud is outlined by the $^{12}$CO
emission contours \citep{Heyer06}. Overlaid red polygons are the regions A, B, and C
defined by X-ray source densities, and the symbols are the
X-ray-selected stars classified as Class I (magenta circles), Class II
(green diamonds), and Class III (black crosses) based on their NIR
colors. The O7 star and the Class I protostar RMCX \#89 are labeled. The
overall distribution of the Class II/I sources is more confined to
the CO molecular ridge than the Class III stars.\label{fig:co_ir}}
\end{figure}

\end{document}